\documentclass[reprint,
groupedaddress,
 prx,amsmath,amssymb,
 aps]{revtex4-2}
 
\usepackage[normalem]{ulem}
\usepackage{appendix}
\usepackage{blindtext}
\usepackage{multirow}    
\usepackage{graphicx}
\usepackage{mathtools}
\usepackage{dcolumn}
\usepackage{amsthm}
\usepackage{qcircuit}
\usepackage{amsmath}
\usepackage{amsfonts}
\usepackage{amssymb}
\usepackage{ctable}
\usepackage{bm}
\usepackage{bbm}
\usepackage{color}
\usepackage{placeins}
\usepackage{afterpage}
\usepackage{comment}
\usepackage{hyperref}

\makeatletter
\newcommand{\vast}{\bBigg@{4}}
\newcommand{\Vast}{\bBigg@{5}}
\makeatother
\DeclareMathOperator{\Tr}{Tr}
\theoremstyle{definition}

\newtheorem{fact}{Fact}
\newcommand{\CX}{\mathrm{CX}}

\newcommand{\jc}[1]{\textcolor{black}{#1}}
\newcommand{\spuri}[1]{\textcolor{black}{#1}}
\newcommand{\fl}[1]{\textcolor{black}{#1}}

\begin{document}

\title{Estimating the bias of CX gates via character randomized benchmarking}

\author{Jahan Claes}
\author{Shruti Puri}
\affiliation{Department of Applied Physics, Yale University, New Haven, Connecticut 06511, USA\\
Yale Quantum Institute, Yale University, New Haven, Connecticut 06511, USA}

\begin{abstract}
    Recent work has demonstrated that high-threshold quantum error correction is possible for biased-noise qubits, provided one can implement a \jc{controlled-not ($\CX$)} gate that preserves the bias. Bias-preserving $\CX$ gates have been proposed for several biased-noise qubit platforms, most notably Kerr cats. \spuri{However, experimentally measuring the noise bias is challenging as it requires accurately estimating certain low-probability Pauli errors in the presence of much larger state preparation and measurement (SPAM) errors.} In this paper, we introduce bias randomized benchmarking (BRB) as a technique for measuring bias in quantum gates. BRB, like all RB protocols, is highly accurate and immune to SPAM errors. Our first protocol, $\CX$-dihedral BRB, is a straightforward method to measure the bias of the entire $\CX$-dihedral group. Our second protocol, interleaved bias randomized benchmarking (IBRB), is a generalization of interleaved RB tailored to the experimental constraints biased-noise qubits; this is a more involved procedure that directly targets the bias of the $\CX$ gate alone. Our BRB procedures occupy a middle ground between classic RB protocols that only estimate the average fidelity, and tomographic RB protocols that provide more detailed characterization of noise but require more \jc{measurements as well as} experimental capabilities that are not necessarily available in biased-noise qubits.
\end{abstract}

\date{\today}

\maketitle

\section{Introduction}

Achieving scalable quantum computing will require efficient error-correction to counter the effects of noise. Recent work has revealed that there exist error correcting codes that tolerate much higher noise rates~\cite{stephens2013high,tuckett2018ultrahigh,tuckett2019tailoring,tuckett2020fault,ataides2021xzzx,dua2022clifford,claes2022tailored} when the noise is biased, that is, when errors that cause bit flips are suppressed compared to errors that cause only phase flips. \spuri{In order to efficiently detect errors, these proposals rely on using controlled-NOT ($\CX$) gates that are {\it bias-preserving} such that bit-flip noise remain suppressed to leading order~\cite{tuckett2018ultrahigh,tuckett2019tailoring,tuckett2020fault,ataides2021xzzx,darmawan2021practical,dua2022clifford,claes2022tailored,chamberland2022building,chamberland2022universal}.} Without bias-preserving $\CX$ gates, error correction is possible \spuri{but requires more complex circuits}, reducing the effectiveness of the underlying code~\cite{aliferis2008fault}. Recently, $\CX$ gates that preserve the noise bias have been theoretically proposed in Kerr cat~\cite{puri2020bias} and other qubit platforms~\cite{guillaud2019repetition,cong2021hardware} (see also~\cite{xu2021engineering} for improvements to bias-preserving $\CX$ gates in Kerr cat qubits). Experimental efforts towards realizing these proposals are rapidly growing. In order to determine the effectiveness of the bias-tailored codes implemented with the experimentally realized $\CX$ gates, it is necessary to estimate the amount of noise-asymmetry or bias along with the total error probability of these gates. \jc{However, it is challenging to precisely measure the rate of bit flip errors of highly-biased noise channels, as such rates are extremely low.}

A common method for precisely estimating low error probabilities in quantum gates is randomized benchmarking (RB)~\cite{magesan2011scalable,magesan2012characterizing,emerson2005scalable,knill2008randomized} and its derivatives~\cite{magesan2012efficient,carignan2015characterizing,cross2016scalable,helsen2019new,claes2021character,helsen2020general,brown2018randomized,francca2018approximate,harper2017estimating,onorati2019randomized,erhard2019characterizing,chasseur2017hybrid,wallman2015estimating,kimmel2014robust,flammia2020efficient,harper2020efficient,baldwin2020subspace}. In RB, one randomly generates circuits from some set of elementary gates and estimates the average error probability as a function of the circuit depth; the decay rate of the error probability with circuit depth then gives information about the error rate in the circuit. RB protocols typically boast two main advantages over other characterization methods. Primarily, RB protocols \fl{decouple state preparation and measurement (SPAM) errors from gate errors}. In addition, because RB involves evaluating the error rate of circuits composed of many elementary gates, small error probabilities are magnified and may be precisely measured. The accuracy of RB experiments to estimate properties of the error channel can be rigorously guaranteed in a variety of settings~\cite{wallman2018randomized,helsen2019new,claes2021character,merkel2021randomized,wallman2014randomized,helsen2019multiqubit,carignan2018randomized}.

\jc{There exists a zoo of different RB protocols (see~\cite{helsen2020general} for a taxonomy) which differ in whether they characterize a group of gates or a single gate, whether they measure only the average fidelity or additional properties of the noise, and whether they are tailored to specific experimental hardware. The original RB measures the fidelity \fl{averaged over elements of} the Clifford group~\cite{magesan2011scalable,magesan2012characterizing}, but there also exist extensions of RB which measure the average fidelity of other efficiently-simulatable groups~\cite{brown2018randomized,francca2018approximate,cross2016scalable}, most notably character RB~\cite{carignan2015characterizing,helsen2019new,claes2021character} which uses techniques from representation theory to significantly simplify the RB decay functions. In contrast to RB methods characterizing groups of gates, interleaved RB~\cite{magesan2012efficient} and its variants~\cite{harper2017estimating,helsen2019new,xue2019benchmarking,chasseur2017hybrid,onorati2019randomized,erhard2019characterizing} instead measure the average fidelity of a single gate or layer of gates. There also exist RB protocols that measure properties of the noise channel beyond the average fidelity for either groups of gates or specific gates~\cite{wallman2015estimating,feng2016estimating,kimmel2014robust,flammia2020efficient,harper2020efficient,flammia2021averaged}. \jc{Moreover, efforts have been directed towards designing} both group and interleaved RB protocols \jc{that are specifically tailored to} \spuri{the gate-set available in a given hardware}~\cite{baldwin2020subspace,xue2019benchmarking,claes2021character,helsen2022matchgate}.}

In this spirit, here we introduce group and interleaved RB procedures \jc{tailored to extract} information about the total noise as well as noise-asymmetry in biased-noise hardware. Our first procedure is $\CX$-dihedral bias RB (BRB), \spuri{which is} a straightforward modification of $\CX$-dihedral RB~\cite{cross2016scalable,helsen2019new}. \spuri{This procedure} measures the noise bias and average fidelity of the entire $\CX$-dihedral group. Our second procedure, interleaved bias RB (IBRB), is a generalization of interleaved RB~\cite{magesan2012efficient} and the recently introduced 2-for-1 interleaved RB~\cite{helsen2019new,xue2019benchmarking}. \spuri{In particular, the IBRB procedure} is tailored to the available gate-set of biased-noise qubits~\cite{grimm2020stabilization,puri2020bias,puri2017engineering}. This procedure uses interleaved $\CX$ and $\mathrm{Z}$ gates to estimate the bias and fidelity of the $\CX$. \spuri{Both the BRB and IBRB protocols} make heavy use of the recently introduced framework of character RB~\cite{helsen2019new,claes2021character}.

\jc{Our approach to \spuri{biased-noise benchmarking} is notably different from Pauli channel estimation~\cite{flammia2020efficient,harper2020efficient,flammia2021averaged}. Pauli channel estimation is an interleaved RB procedure that uses character RB and randomized compiling to measure the full set of Pauli-diagonal elements of the noise channel associated with any Clifford operation. While Pauli channel estimation would appear to be sufficient for measuring the bias of $\CX$ gates, it has two drawbacks that make it unsuitable for estimating bias in biased-noise qubits. First, Pauli channel estimation demands considerable experimental overhead, as it estimates the probability of each Pauli error rather than the probability of sets of Pauli errors \spuri{such as bit-flips and phase-flips}. More importantly, Pauli channel estimation requires interleaving the full Pauli group between the Clifford operators, and \fl{is only guaranteed to measure} the Clifford's noise when the Pauli group is high-fidelity. This is not suitable for biased-noise qubits, where we generically expect $X$ and $Y$ gates to be of comparable fidelity to $\CX$ gates \spuri{if implemented in a bias-preserving manner}~\cite{puri2020bias}. \spuri{On the other hand, if $X$ and $Y$ gates are not implemented in a bias-preserving manner then it won't be possible to accurately estimate the suppressed bit-flip rate in the $\CX$ gates.} \spuri{In contrast to Pauli channel estimation, our interleaved bias RB is designed specifically to use only interleaved $Z$ gates, which are trivially bias-preserving and can be implemented with high-fidelity~\cite{puri2020bias,darmawan2021practical}}}.

Our paper is organized as follows. In Section \ref{sec:DefineBias}, we define the bias and fidelity of a multi-qubit gate in terms of the Pauli-diagonal part of its noise channel. In Section \ref{sec:IntroduceGroups}, we introduce the Pauli, Z, and $\CX$-dihedral groups that we will use to benchmark the bias. In Section \ref{sec:DefineProcedures}, we give the step-by-step instructions for our bias RB procedures and illustrate them with simulated experiments. The derivations of these procedures, as well as technical details about biased-noise error channels, are relegated to appendices.

\section{Defining the bias}
\label{sec:DefineBias}
Intuitively, the bias of a noise channel $\Lambda$ is a measure of the likelihood that an error will not flip a bit and will instead only apply an erroneous phase. We refer to errors that only apply an erroneous phase as \textbf{dephasing errors}\jc{, while we refer to errors that include a bit-flip as $\textbf{non-dephasing errors}$, even if they also apply erroneous phases. For example, an error $\mathbbm{1}\otimes Z$ is dephasing, while $Y\otimes Z$ is non-dephasing.}

To formally define the bias, we introduce the \textbf{${\pmb\chi}$-matrix} of a quantum channel~\cite{wood2011tensor}. Since the Pauli group on $N$ qubits forms a basis for the set of all operators on $N$ qubits, we can write the action of an arbitrary noise channel as
\begin{equation}
\Lambda(\rho) = \sum_{\substack{\vec\alpha_1,\vec\beta_1\\\vec\alpha_2,\vec\beta_2}}\chi_{\vec\alpha_1\vec\beta_1,\vec\alpha_2\vec\beta_2}\mathrm{X}(\vec\alpha_1)\mathrm{Z}(\vec\beta_1)\rho \mathrm{Z}(\vec\beta_2)\mathrm{X}(\vec\alpha_2)\label{eq:DefnChiMatrix}
\end{equation}
for some coefficients $\chi_{\vec\alpha_1\vec\beta_1,\vec\alpha_2\vec\beta_2}$ \footnote{The usual definition of the $\chi$-matrix differs from this one by factors of $i$ off the diagonal, which are not relevant for defining the bias.}. \jc{Here $\vec\alpha_i,\vec\beta_i\in\mathbbm{Z}_2^N$ are vectors of $0$s and $1$s that index the Pauli group, and} for any single qubit operator $\mathrm{O}$ we use the notation
\begin{equation}
\mathrm{O}(\vec v) := \mathrm{O}_1^{v_1}\otimes\cdots\otimes \mathrm{O}_N^{v_N}.
\end{equation}

The diagonal of the $\chi$-matrix is non-negative and sums to one. As a Pauli error is dephasing if and only if $\vec\alpha=0$, we define the probabilities of dephasing and non-dephasing errors, $p_{\mathrm{D}}$ and $p_{\mathrm{ND}}$, by
\begin{align}
    p_{\mathrm{D}} &:= \sum_{\vec\beta\neq\vec 0} \chi_{0\vec\beta,0\vec\beta}\label{eq:DefnDephasing}\\
    p_{\mathrm{ND}} &:= \sum_{\substack{\vec\alpha\neq\vec 0\\\vec\beta}} \chi_{\vec\alpha\vec\beta,\vec\alpha\vec\beta}.\label{eq:Defnnon-dephasing}
\end{align}
Finally, we define the \textbf{bias} $\eta$ as the ratio of the error probabilities,
\begin{equation}
    \eta := p_{\mathrm{D}}/p_{\mathrm{ND}}.\label{eq:BiasDefinition}
\end{equation}
This definition of the bias for a multi-qubit gate was originally given in~\cite{puri2020bias}.

The probability of no error is given by $\chi_{\vec 0\vec 0,\vec 0\vec 0}$, which is related to $p_{\mathrm{D}}$ and $p_{\mathrm{ND}}$ in the obvious way
\begin{align}
    \chi_{\vec 0\vec 0,\vec 0\vec 0}=1-p_{\mathrm{D}}-p_{\mathrm{ND}}.\label{eq:chi00InTermsOfPs}
\end{align}
The usual measure of gate quality is the average fidelity $F_\Lambda$, which in can be written in terms of the $\chi$-matrix as~\cite{nielsen2002simple,horodecki1999general}
\begin{equation}
    F_\Lambda = \frac{2^{N}\chi_{\vec 0 \vec 0,\vec 0 \vec 0}+1}{2^N+1} \label{eq:fidelityChiMatrixRelation}
\end{equation}
where $N$ is the number of qubits. There exist numerous RB procedures to measure the average fidelity of a group of gates~\cite{helsen2019new,claes2021character,cross2016scalable,helsen2020general,carignan2015characterizing,magesan2011scalable,magesan2012characterizing} or a specific gate~\cite{magesan2012efficient,helsen2019new,xue2019benchmarking,harper2017estimating,chasseur2017hybrid,onorati2019randomized} and therefore determine $\chi_{\vec 0 \vec 0,\vec 0 \vec 0}$.

Note that while $p_{\mathrm{D}}$ and $p_{\mathrm{ND}}$ only directly give information about the diagonal elements of the $\chi$-matrix, complete-positivity of the noise channel $\Lambda$ implies the off-diagonal elements of the $\chi$-matrix are bounded by the diagonal elements as~\cite[Appendix D]{kimmel2014robust}
\begin{equation}
    |\chi_{\vec\alpha_1\vec\beta_1,\vec\alpha_2\vec\beta_2}|\leq\sqrt{\chi_{\vec\alpha_1\vec\beta_1,\vec\alpha_1\vec\beta_1}\chi_{\vec\alpha_2\vec\beta_2,\vec\alpha_2\vec\beta_2}}.
\end{equation}
One can use this bound to show that if two biased noise channels $\Lambda_A$ and $\Lambda_B$ have non-dephasing error probabilities $p_{\mathrm{ND}}^A$ and $p_{\mathrm{ND}}^B$, their composition will still have \begin{equation}
p_{\mathrm{ND}}\lesssim  p_{\mathrm{ND}}^A+p_{\mathrm{ND}}^B +2\sqrt{p_{\mathrm{ND}}^Ap_{\mathrm{ND}}^B}   \label{eq:CombinedNDBound}
\end{equation}
and similar for $p_\mathrm{D}$ (see Appendix \ref{appendix:CompositeChannels}). Thus, $p_\mathrm{ND}$ as defined in Eq.~\ref{eq:Defnnon-dephasing} is a useful characterization of the channel's behavior under composition even if the off-diagonal elements are not negligible. We also note that previous work on quantum error correction has demonstrated that measurement of the stabilizers of an error-correcting code causes the off-diagonal elements of the $\chi$-matrix to rapidly decay \cite{huang2019performance,beale2018quantum,bravyi2018correcting,geller2013efficient}.

\section{The Pauli, Z, and $\CX$-Dihedral quantum groups}
\label{sec:IntroduceGroups}
We will focus specifically on the Pauli, Z, and $\CX$-dihedral~\cite{cross2016scalable,garion2020synthesis} groups. These are all finite subgroups of the full unitary group that can be efficiently simulated. We will consider these groups to be defined modulo overall phases for convenience.

The $N$-qubit Pauli group $\mathcal{P}_N$ is the group generated by the Pauli $\mathrm{X}$ and $\mathrm{Z}$ operators on all $N$ qubits. As $\mathrm{X}\mathrm{Z}=\mathrm{Z}\mathrm{X}$ up to a phase, we may assume all $\mathrm{X}$ operators appear to the left of all $\mathrm{Z}$ operators. We then have
\begin{align}
    \mathcal{P}_N :=& \langle \mathrm{X}_1,\mathrm{Z}_1,...,\mathrm{X}_N,\mathrm{Z}_N\rangle/U(1)\\
    =& \left\{\mathrm{X}(\vec{\alpha})\mathrm{Z}(\vec{\beta}):\vec \alpha\in\mathbb{Z}_2^N,\vec \beta\in\mathbb{Z}_2^N\right\}.
\end{align}
Since $\mathrm{X}(\vec\alpha_1)\mathrm{Z}(\vec\beta_1)\mathrm{X}(\vec\alpha_2)\mathrm{Z}(\vec\beta_2)=\mathrm{X}(\vec\alpha_1+\vec\alpha_2)\mathrm{Z}(\vec\beta_1+\vec\beta_2)$ up to a phase, this group is isomorphic to $\mathbb{Z}^{2N}_2$

The $N$-qubit Z group $\mathcal{Z}_N$ is the commutative subgroup of the Pauli group generated by all single-qubit $Z$:
\begin{align}
    \mathcal{Z}_N :=& \langle \mathrm{Z}_i: 1\leq i\leq N\rangle/U(1)\\
     =& \left\{\mathrm{Z}(\vec \beta):\vec \beta\in \mathbb{Z}_2^N\right\}.
\end{align}
This group is clearly isomorphic to $\mathbb{Z}_2^N$.

Finally, the $N$-qubit $\CX$-Dihedral group $\mathcal{D}_N$ is the group generated by the $N$-qubit dihedral group along with all $\CX$ gates between any pair of qubits~\cite{cross2016scalable}:
\begin{equation}
    \mathcal{D}_N := \langle \mathrm{X}_i,\mathrm{T}_i,\CX_{i,j}: 1\leq i,j\leq N\rangle /U(1).
\end{equation}
\fl{Here, $\mathrm{T}_i:=\exp\{i\pi \mathrm{Z}_i/4\}$ is the T-gate}, while $\CX_{i,j}$ denotes the $\CX$ gate with \spuri{the first index,} $i$, the control and \spuri{the second index,} $j$, the target. Simple presentations of $\mathcal{D}_N$ were introduced in~\cite{cross2016scalable,amy2016finite}, and efficient decompositions of elements of $\mathcal{D}_N$ which minimize the number of two-qubit gates are given in~\cite{garion2020synthesis}.

\section{The bias RB procedures}
\label{sec:DefineProcedures}
In this section, we outline our two bias RB procedures. The first is a simple and scalable protocol for measuring $p_{\mathrm{D}}$ and $p_{\mathrm{ND}}$ for the entire $\CX$-dihedral group, which may be a useful proxy for the performance of the $\CX$ gate when the $\mathrm{X}$ and $\mathrm{T}$ gates are high-fidelity.

%$\mathrm{X}_1\CX_{1,2}\mathrm{X}_1$
The second is a protocol for measuring $p_{\mathrm{D}}$ and $p_{\mathrm{ND}}$ for the $\CX$ gate $\CX_{1,2}$ acting on two qubits, by interleaving with elements of $\mathcal{Z}_2$ \spuri{and randomly swapping between the usual $\CX_{1,2}$ gate and a $|0\rangle$-controlled $\CX$ gate which flips the target qubit if the control qubit is in the $|0\rangle$ state. The $|0\rangle$-controlled $\CX$ gate can be written as a $\CX$ gate conjugated by a Pauli $X$ on the control qubit, $\mathrm{X}_1\CX_{1,2}\mathrm{X}_1$.} This second protocol is specifically designed for \spuri{biased-noise, stabilized-cat qubits}, where the fidelity of diagonal $\mathrm{Z}$ gates is much higher than the fidelity of $\mathrm{X}$, and where we can apply both a bias-preserving $\CX_{1,2}$ and a bias-preserving \spuri{$|0\rangle$-controlled $\CX$} gate by simply \jc{changing the phase of a drive~\cite{puri2020bias,guillaud2019repetition,chamberland2022building}} so that these two operations have similar $p_{\mathrm{D}}$ and $p_{\mathrm{ND}}$. However, we expect this protocol to be generally applicable to other biased-noise architectures, since arbitrary biased-noise architectures will likely have much higher-fidelity diagonal gates than off-diagonal gates \cite{guillaud2019repetition,chamberland2022building}, and $\mathrm{X}_1\CX_{1,2}\mathrm{X}_1$ differs from $\CX_{1,2}$ simply by swapping the roles of $|0\rangle$ and $|1\rangle$ in the control qubit.

\subsection{$\CX$-dihedral bias RB (\spuri{BRB})}

Standard Clifford RB measures the fidelity \fl{averaged over elements of the} Clifford group; similarly, $\CX$-dihedral bias RB measures $p_{\mathrm{D}}$ and $p_{\mathrm{ND}}$ \fl{averaged over elements of the} $\CX$-dihedral group $\mathcal{D}_N$. Our procedure is essentially identical to the previous character RB procedure for estimating the fidelity of the $\CX$-dihedral group~\cite{carignan2015characterizing,helsen2019new}. The only modification is a post-processing step to extract the dephasing and non-dephasing error probabilities instead of just the average fidelity.

For convenience, we make the standard RB assumption of gate-independent noise, so that the noisy implementation of any $U\in\mathcal{D}_N$ is $\Lambda\circ U$ for a noise channel $\Lambda$ independent of $U$. However, like the usual character RB, our procedure works for gate-dependent noise, provided all $U\in\mathcal{D}_N$ are high-fidelity; see~\cite{helsen2019new,claes2021character} for proofs.

\begin{table}
    \centering
    \begin{tabular}{c|c|c|c|c}
        $b$ & $\rho_b$ & $E_b$ & $\chi_b\Big(\mathrm{X}(\vec{\alpha})\mathrm{Z}(\vec{\beta})\Big)$ &$S_b(n)$\\\specialrule{.2em}{0em}{0em}
        1 & $\bigotimes^N|0\rangle\langle 0|$ & $\bigotimes^N \mathrm{Z}$ & $(-1)^{|\vec\alpha|}$ &$A_1\lambda_1^n$ \\\hline
        2 & $\bigotimes^N|+\rangle\langle +|$ & $\bigotimes^N \mathrm{X}$ & $(-1)^{|\vec\beta|}$ & $A_2\lambda_2^n$ \\\hline
    \end{tabular}
    \caption{Initial states, measurements, weighting, and fitting functions for $\CX$-dihedral bias RB. Here, $|\vec\alpha|$ denotes the $\mathbb{Z}_2$-valued sum of all elements of $\vec\alpha$, and similarly for $|\vec\beta|$.}
    \label{tab:MeasurementsAndInitialStatesCNOTD}
\end{table}

The $\CX$-dihedral bias RB procedure is:
\begin{enumerate}
    \item For each $b$ in Table~\ref{tab:MeasurementsAndInitialStatesCNOTD}:
    \begin{enumerate}
        \item For arbitrary $n$, choose unitaries $U_0\in\mathcal{P}_N$ and $U_1,...,U_n\in \mathcal{D}_N$ at random. Set $U_{n+1}=U_1^\dagger\cdots U_n^\dagger$. \label{step:ChoosingUnitariesCNOTD}
        \item Prepare the initial state $\rho_b$ listed in Table~\ref{tab:MeasurementsAndInitialStatesCNOTD}.
        \item Successively apply the gates $(U_1 U_0)$, $U_2$, $U_3$, \dots, $U_{n+1}$. Note that instead of applying $U_0$ and then $U_1$, we compile the product $U_1U_0$ into a single element of $\mathcal{D}_N$.\label{step:GateApplyCNOTD}
        \item Perform a measurement of the observable $E_b$ in Table~\ref{tab:MeasurementsAndInitialStatesCNOTD}. Weight the outcome by $\chi_b^*(U_0)$.\label{step:MeasurementCNOTD}
        \item Repeat steps~\ref{step:ChoosingUnitariesCNOTD}-\ref{step:MeasurementCNOTD} many times to estimate the {\bf character-weighted survival probability}
        \begin{equation}
            S_b(n) :=\quad \smashoperator{\mathop{\mathbb{E}}_{\substack{U_0\in \mathcal{P}_N\\ U_1\cdots U_{n}\in \mathcal{D}_N}}}\quad\left[\chi_b^*(U_0)P_{\{U_i\}}\right]
        \end{equation}
        where $P_{\{U_i\}}$ denotes the expectation value of $E_b$ after applying the gates in step~\ref{step:GateApplyCNOTD} and $\mathbbm{E}[\cdot]$ denotes the average over the choices of gates.\label{step:CharacterWeightingCNOTD}
        \item Repeat steps~\ref{step:ChoosingUnitariesCNOTD}-\ref{step:CharacterWeightingCNOTD} for many different values of $n$ to estimate the whole $S_b(n)$ curve.
        \item Fit \spuri{the decay curve}, $S_b(n)$, to the functional form listed in Table~\ref{tab:MeasurementsAndInitialStatesCNOTD} to estimate $\lambda_b$.
    \end{enumerate}
    \item Estimate the dephasing and non-dephasing error probabilities of the error channel $\Lambda$ as
    \begin{equation}
        \begin{split}
            p_{\mathrm{D}} &=\frac{2^N-1}{4^N}\left[1+\left(2^N-1\right)\lambda_1-2^N\lambda_2\right]\\
            p_{\mathrm{ND}} &= \frac{2^N-1}{2^N}\left[1-\lambda_1\right].
        \end{split} \label{eq:CNOTDProbEstimates}
    \end{equation}
\end{enumerate}

Note that the weights and measurement outcomes satisfy $|\chi_b^*(U_0)P_{\{U_i\}}|\leq 1$, so that if we take $\mathcal{N}_s$ samples to estimate $S_b(n)$ at a specific value of $b$ and $n$, the statistical uncertainty in our estimate is roughly $1/\sqrt{\mathcal{N}_s}$, independent of $n$ or $N$. To achieve a good \emph{relative} uncertainty, then, we simply need the true value of $S_b(n)$ to be sufficiently large. For high-fidelity gates, we show in \jc{our derivation in} Appendix~\ref{sec:DerivingCNOTD} that $A_b\approx 1$, $\lambda_b\approx 1$, so that we can reliably \spuri{fit the decay curve}.

\begin{figure}
    \centering
    \includegraphics[width=\columnwidth]{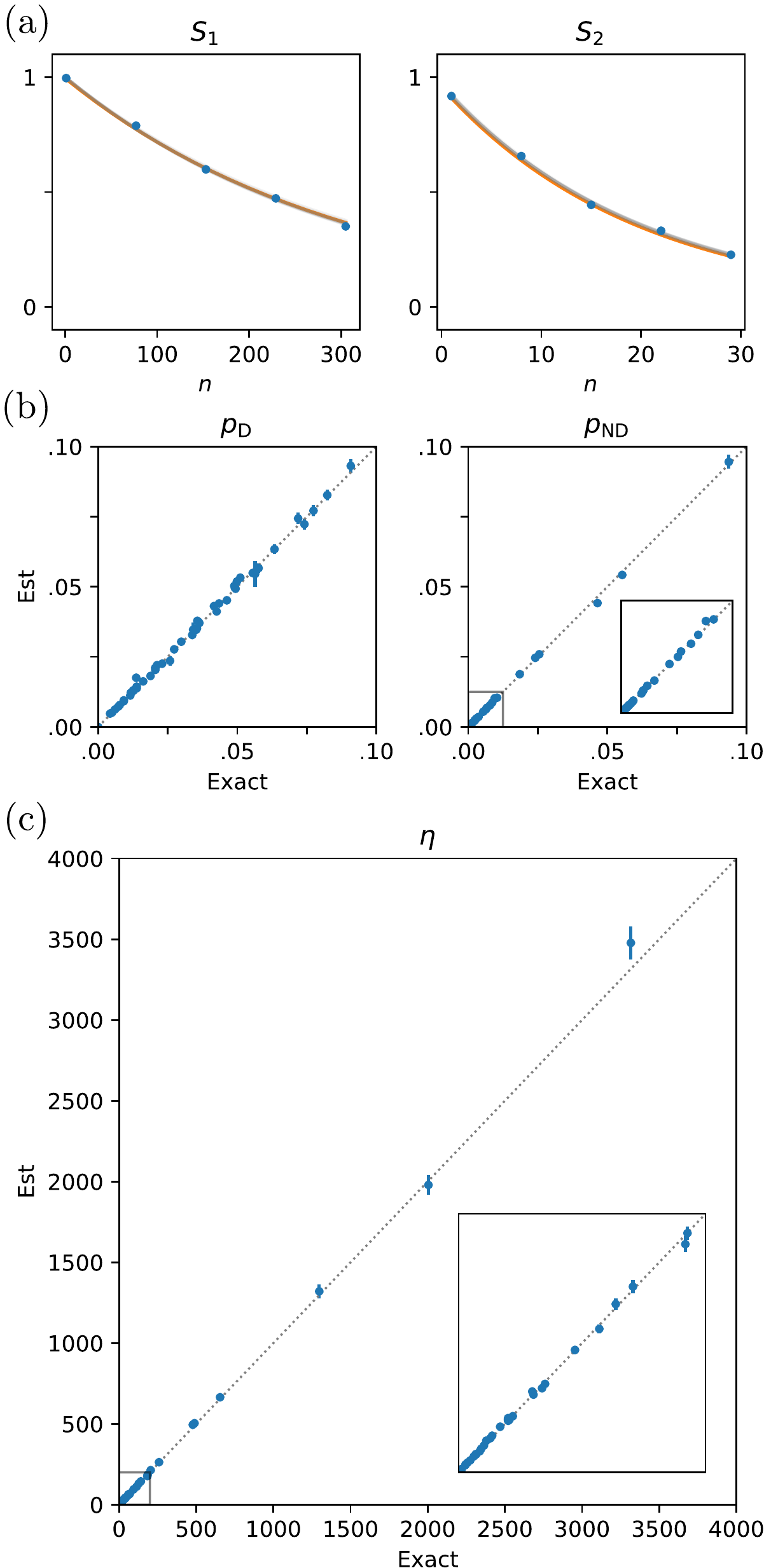}
    \caption{Simulated $\CX$-dihedral bias RB experiments. (a) An example of estimating $S_b(n)$, with the $S_b(n)$ curves plotted in orange and our estimates of $S_b(n)$ given by blue dots. To estimate sensitivity to statistical errors, we generate multiple fits of $S_b(n)$ using bootstrap resampling of our data, which are plotted in grey and generally overlap the exact $S_b(n)$ curves. (b) The probabilities $p_\mathrm{D}$ and $p_{\mathrm{ND}}$ extracted from the fits for $50$ randomly generated error channels. Error bars denote the standard deviation over $50$ resamplings. (c) The estimated bias $\eta$ for each of these error channels. We see that even at very high bias, we can accurately estimate the value of $\eta$.}
    \label{fig:SimulatedDihedralRB}
\end{figure}

We give an example of $\CX$-dihedral BRB in Fig. \ref{fig:SimulatedDihedralRB}. \jc{Here, we generate random error channels $\Lambda$ by generating random sets of Kraus operators, and simulate a $\CX$-dihedral BRB experiment (see Appendix~\ref{sec:RandomErrors} for details on the random error channels).} Fig. \ref{fig:SimulatedDihedralRB}a illustrates the experiment for a single error channel, where we estimate the value of $S_b(n)$ for a few different values of $n$ and fit the data to the functional forms given in Table \ref{tab:MeasurementsAndInitialStatesCNOTD}. In this simulation, each value of $S_b(n)$ is estimated using $5000$ measurements. To demonstrate the effectiveness of our procedure, we repeat this experiment for many different error channels, using Eq. \ref{eq:CNOTDProbEstimates} to estimate $p_{\mathrm{D}}$ and $p_{\mathrm{ND}}$ and finally using these estimates to extract the bias. In Fig. \ref{fig:SimulatedDihedralRB}b we plot our estimated probabilities versus the exact probabilities of the error channel, and in Fig. \ref{fig:SimulatedDihedralRB}c we do the same for the bias. To estimate the error bars, we use bootstrap resampling. Visually, it is clear that we are accurately estimating $p_{\mathrm{D}}$, $p_{\mathrm{ND}}$, and $\eta$ even for very high biases. To verify this, we compute the reduced-$\chi^2$ statistic for our estimates \jc{of $p_\mathrm{D}$, $p_\mathrm{ND}$, and $\eta$}, and find $\chi^2$ between $1$ and $1.3$, indicating that our bootstrapping is accurately estimating the error bars to within a factor of between $1$ and $\sqrt{1.3}$.

\subsection{Interleaved bias RB (\spuri{IBRB})}

A common approach to estimate the fidelity of a single gate is interleaved RB, in which one interleaves the gate of interest with random elements from an interleaving group designed to simplify the error channel. Originally, the gate of interest was required to be from the interleaving group~\cite{magesan2012efficient}, but later work has relaxed this requirement~\cite{harper2017estimating,helsen2019new,xue2019benchmarking,chasseur2017hybrid,onorati2019randomized}. Interleaved RB estimates the fidelity of the combined error channel $\Lambda_U\circ \Lambda_G$ of the gate $U$ and interleaving group $G$; knowing the fidelity of $\Lambda_U\circ\Lambda_G$ and the fidelity of $\Lambda_G$ allows one to bound the fidelity of $\Lambda_U$, with bounds that become tight as the fidelity of $\Lambda_G$ approaches one~\cite{magesan2012efficient,kimmel2014robust,carignan2019bounding}. Thus, we typically require the interleaving group to have high-fidelity. On the other hand, the advent of randomized compiling~\cite{knill2005quantum,viola2005random,wallman2016noise} implies that in some cases it is not necessary to separately estimate the fidelities of $\Lambda_U$ and $\Lambda_G$, as $\Lambda_U\circ\Lambda_G$ is the relevant error channel for a circuit that has been randomly compiled with the group $G$. However, in the case of randomized compiling, it is still necessary for $G$ to be high-fidelity, so that the randomized compilation does not add in significant additional noise.

We develop an interleaved bias RB procedure that directly estimates the bias of the $\CX$ gate in Kerr cat qubits~\cite{grimm2020stabilization,puri2020bias,puri2017engineering} or other biased-noise platforms~\cite{guillaud2019repetition,chamberland2022building,cong2021hardware}. In biased-noise qubits, we generally expect gates that are diagonal in the $\mathrm{Z}$-basis to have much higher fidelity than non-diagonal gates. It is thus desirable to have a protocol that uses $\mathcal{Z}_2$ as the interleaving group, as the fidelity of $X$ and $Y$ gates may be no better than the fidelity of the $\CX$ gate \footnote{we could also include single-qubit $T$ gates or even two-qubit $CZ$ gates in our group, as these gates are also high-fidelity for Kerr cat qubits, but we did not find these additional gates helpful in designing our protocol.}. Restricting to an interleaving group such as $\mathcal{Z}_2$ that is diagonal in the $Z$-basis introduces considerable complications, as we will see below.

For convenience, we define $C:=\CX_{1,2}$ to be a $\CX$ gate on the two qubits. We also define the gate $C':=\mathrm{X}_1\CX_{1,2}\mathrm{X}_1$, which is similar to $C$, except that it applies $\mathrm{X}_2$ when qubit $1$ is in $|0\rangle$ rather than $|1\rangle$. The Kerr cat system can implement bias-preserving versions of both of these gates using similar procedures, and we expect that both $C$ and $C'$ will have similar dephasing and non-dephasing probabilities. These features will likely be shared by any biased-noise system. Define the error channels $\Lambda_C$ and $\Lambda_{C'}$ to be the error channels associated with $C$ and $C'$ respectively, so the noisy implementation of $C$ is $C\circ\Lambda_C$ and similar for $C'$. In addition define $\Lambda_G$ to be the error channel associated with gates $U\in\mathcal{Z}_2$, so the noisy implementation of $U$ is $\Lambda_G\circ U$. Finally, define $\Lambda = \Lambda_C\circ\Lambda_G$ and $\Lambda'=\Lambda_{C'}\circ\Lambda_G$ to be the composed error channels. Our protocol will directly estimate the dephasing and non-dephasing probabilities of the average channel $(\Lambda+\Lambda')/2$. If the dephasing and non-dephasing probabilities of $\Lambda_C$ and $\Lambda_{C'}$ are close and $\Lambda_G\approx\mathbbm{1}$, this is identical to the dephasing and non-dephasing probabilities of $\Lambda_C$ alone. On the other hand, even without these assumptions, we will see below that $(\Lambda+\Lambda')/2$ is the relevant error channel for a certain randomized compilation procedure.

The interleaved bias RB procedure is:

\begin{table}
    \centering
    \begin{tabular}{c|c|c|c|c}
        $b$ & $\rho_b^{(a)}$ & $E_b^{(a)}$ & $\chi_b\big(\mathrm{Z}_1^{\beta_1}\mathrm{Z}_2^{\beta_2}\big)$ &$S_b(n)$\\\specialrule{.2em}{0em}{0em}
        $0+$ & $\frac{1}{2}\mathrm{Z}\otimes|0\rangle\langle 0|$ & $\mathrm{Z}\otimes\mathbbm{1}$ & $1$ & $A_{0+}\lambda_{0+}^n+B_{0+}$\\\hline
        $0-$ & $\frac{1}{2}\mathrm{Z}\otimes|0\rangle\langle 0|$ & $\mathrm{Z}\otimes \mathrm{Z}$& $1$ & $A_{0-}\lambda_{0-}^n+B_{0-}\kappa_{0-}^n$ \\\hline
        \multirow{2}{*}{$1+$} & $\frac{1}{2}\mathbbm{1}\otimes|+\rangle\langle +|$ & $\mathbbm{1}\otimes \mathrm{X}$  & \multirow{2}{*}{$(-1)^{\beta_2}$} & \multirow{2}{*}{$A_{1+}\lambda_{1+}^n+B_{1+}\kappa_{1+}^n$}\\
        &$\frac{1}{2}\mathrm{Z}\otimes|+\rangle\langle +|$ &$\mathrm{Z}\otimes \mathrm{X}$&&\\\hline
        $1-$ & $|+\! i +\! i\rangle\langle +\! i +\! i|$ & $\mathbbm{1}\otimes \mathrm{Y}$ & $(-1)^{\beta_2}$ & $A_{1-}\lambda_{1-}^n+B_{1-}\kappa_{1-}^n$\\\hline
        \multirow{2}{*}{$2+$} & $|+ 0\rangle\langle + 0|$ & $\mathrm{X}\otimes\mathbbm{1}$ & \multirow{2}{*}{$(-1)^{\beta_1}$} & \multirow{2}{*}{$A_{2+}\lambda_{2+}^n+B_{2+}\kappa_{2+}^n$}\\
        &$|+\! i\ 0\rangle\langle +\! i\ 0|$&$\mathrm{Y}\otimes\mathbbm{1}$&&\\\hline
        \multirow{2}{*}{$2-$} & $|+ 0\rangle\langle + 0|$ & $\mathrm{X}\otimes \mathrm{Z}$ & \multirow{2}{*}{$(-1)^{\beta_1+\beta_2}$} & \multirow{2}{*}{$A_{2-}\lambda_{2-}^n+B_{2-}\kappa_{2-}^n$}\\
        &$|+\! i\ 0\rangle\langle +\! i\ 0|$ & $\mathrm{Y}\otimes \mathrm{Z}$& \\\hline
    \end{tabular}
    \caption{Initial states, measurements, weighting, and fitting functions for interleaved bias RB. \fl{In the fitting functions, $\lambda_b$ and $\kappa_b$ are two independent RB decay parameters.} $|\pm\! i\rangle$ denotes eigenstates of the $\mathrm{Y}$ operator.}
    \label{tab:MeasurementsAndInitialStatesZ}
\end{table}

\begin{enumerate}
\item For $b=0\pm$ and $b=1\pm$ \spuri{in Table~\ref{tab:MeasurementsAndInitialStatesZ}}:
\begin{enumerate}
    \item For arbitrary $n$, choose unitaries $U_1,...,U_{n+1}\in \mathcal{Z}_2$ uniformly at random. Also choose $\CX$ gates $C_1,...,C_n\in\{C,C'\}$ uniformly at random. \label{step:ChoosingUnitariesA}
    \item If there is more than one initial state $\rho_b$ listed in Table~\ref{tab:MeasurementsAndInitialStatesZ}, randomly select one of the listed initial states and prepare it.\label{step:ChoosingStateA}
    \item Alternatively apply the gates from $\mathcal{Z}_2$ and the $\CX$ gates as $U_1,C_1,U_2,C_2,...,U_n,C_n,U_{n+1}$. \label{step:GateApplyA}
    \item Perform a measurement of the observable $E_b$ corresponding to the $\rho_b$ selected in step~\ref{step:ChoosingStateA}. Weight the outcome by $\chi_b^*(\{U_i\})\sigma_\pm(\{C_i\})$, where we define
    \begin{align}
        \sigma_\pm(\{C_i\})&:= (\pm 1)^{\delta_{C_1,C'}+\cdots+\delta_{C_n,C'}}\\
        \chi_b^*(\{U_i\})&:= \prod_i\chi_b^*(U_i)
    \end{align}
    and $\chi_b(U_i)$ is given in Table~\ref{tab:MeasurementsAndInitialStatesZ}. If $b=0-$ or $b=1-$ the effect of $\sigma$ is to multiply the measurement outcome by $(-1)$ for each $C'$ we apply, while otherwise $\sigma$ is trivial.
    \label{step:MakingAMeasurementA}
    \item Repeat steps~\ref{step:ChoosingUnitariesA}-\ref{step:MakingAMeasurementA} many times, to estimate the {\bf signed character-weighted survival probability}
    \begin{equation}
       S_b(n) :=\quad \smashoperator{\mathop{\mathbb{E}}_{\substack{\rho_b\\U_0\cdots U_{n+1}\in \mathcal{Z}_2\\C_1,...,C_{n}\in\{C,C'\}}}}\quad\left[\chi_b^*(\{U_i\})\sigma_\pm(\{C_i\})P_{\{C_i\},\{U_i\},\rho_b}\right]
    \end{equation}
    where $P_{\{C_i\},\{U_i\},\rho_b}$ denotes the expectation value of $E_b$ after applying the gates in step~\ref{step:GateApplyA} to $\rho_b$.\label{step:repeatToGetSurvivalA}
    \item Repeat steps~\ref{step:ChoosingUnitariesA}-\ref{step:repeatToGetSurvivalA} for many different values of $n$ to estimate the whole $S_b(n)$ curve.
    \item Fit $S_b(n)$ to the functional form listed in Table~\ref{tab:MeasurementsAndInitialStatesZ} to estimate $\lambda_b$ and $\kappa_b$, where we take the convention $\Re(\lambda_b)>\Re(\kappa_b)$.\label{step:Fit}
\end{enumerate}
\item For $b=2\pm$ in Table ~\ref{tab:MeasurementsAndInitialStatesZ}:
\begin{enumerate}
    \item For arbitrary $n$, choose unitaries $U_1,...,U_{2n+1}\in \mathcal{Z}_2$ uniformly at random. Also choose $\CX$ gates $C_1,...,C_{2n}\in\{C,C'\}$ uniformly at random.\label{step:ChoosingUnitariesB}
    \item If there is more than one initial state $\rho_b$ listed in Table~\ref{tab:MeasurementsAndInitialStatesZ}, randomly select one of the listed initial states and prepare it.\label{step:ChoosingStateB}
    \item Alternatively apply the gates from $\mathcal{Z}_2$ and the $\CX$ gates as $U_1,C_1,U_2,C_2,...,U_{2n},C_{2n},U_{2n+1}$.\label{step:GateApplyB}
    \item Perform a measurement of the observable $E_b$ corresponding to the $\rho_b$ selected in step~\ref{step:ChoosingStateB}. Weight the outcome by $\chi_b^*(\{U_i\})\sigma_\pm(\{C_i\})$, where we define
    \begin{align}
        \sigma_\pm(\{C_i\})&:= (\pm 1)^{\delta_{C_1,C'}+\cdots+\delta_{C_{2n},C'}}\\
        \chi_b^*(\{U_i\})&:= \prod_{i\text{ odd}}\chi_{2+}^*(U_i)\prod_{i\text{ even}}\chi_{2-}^*(U_i)
    \end{align}
    where $\chi_{2\pm}(U_i)$ are given in Table~\ref{tab:MeasurementsAndInitialStatesZ}.    \label{step:MakingAMeasurementB}
    \item Repeat steps~\ref{step:ChoosingUnitariesB}-\ref{step:MakingAMeasurementB} many times, to estimate the {\bf signed character-weighted survival probability}
    \begin{equation}
        S_b(n) :=\quad \smashoperator{\mathop{\mathbb{E}}_{\substack{\rho_b\\U_0\cdots U_{2n+1}\in \mathcal{Z}_2\\C_1,...,C_{2n}\in\{C,C'\}}}}\quad\left[\chi_b^*(\{U_i\})\sigma_\pm(\{C_i\})P_{\{C_i\},\{U_i\},\rho_b}\right]
    \end{equation}
    where $P_{\{C_i\},\{U_i\},\rho_b}$ denotes the expectation value of $E_b$ after applying the gates in step~\ref{step:GateApplyB} to $\rho_b$.\label{step:repeatToGetSurvivalB}
    \item Repeat steps~\ref{step:ChoosingUnitariesB}-\ref{step:repeatToGetSurvivalB} for many different values of $n$ to estimate the whole $S_b(n)$ curve.
    \item Fit $S_b(n)$ to the functional form listed in Table~\ref{tab:MeasurementsAndInitialStatesZ} to estimate $\lambda_b$ and $\kappa_b$, where we take the convention $\Re(\lambda_b)>\Re(\kappa_b)$.\label{step:FitB}
\end{enumerate}
\item Estimate the dephasing and non-dephasing error probabilities of the combined error channel $\Lambda$ as
    \begin{equation}
    \begin{split}
       p_{\mathrm{D}} &=\frac{1}{16}\big[3\lambda_{0+}+3\lambda_{0-}-3\kappa_{0-}-\lambda_{1+}\\
       &\qquad\qquad-\kappa_{1+}-\lambda_{1-}+\kappa_{1-}-\lambda_{2+}\\
       &\qquad\qquad\qquad-\kappa_{2+}-\lambda_{2-}-\kappa_{2-}-1\big]\\
    p_{\mathrm{ND}} &= 1-\frac{1}{4}\left[1+\lambda_{0+}+\lambda_{0-}-\kappa_{0-}\right].
    \end{split}\label{eq:ZProbabilityEstimates}
    \end{equation}
\end{enumerate}

To realize the mixed, non-positive state $\rho=\frac{1}{2}Z\otimes |0\rangle\langle 0|$ we simply prepare either $|00\rangle\langle 00|$ or $|10\rangle\langle 10|$ with equal probability, and weight the resulting measurement by $(-1)$ if we prepare $|10\rangle\langle 10|$. We realize the states $\rho = \frac{1}{2}\mathbbm{1}\otimes |+\rangle\langle +|$ and  $\rho = \frac{1}{2}Z\otimes|+\rangle\langle +|$ similarly.

In this procedure, for $b=0+$ we need to accurately estimate $\lambda_{0+}$ and for $b\neq 0+$ we need to accurately estimate both $\lambda_b$ and $\kappa_b$. To accurately fit these decay parameters, we require the prefactors $A_{0+}$ and $A_b,B_b$ for $b\neq 0+$ to be large. As we will demonstrate in our derivation in Appendix \ref{sec:DerivingIBRB}, for high-fidelity gates we expect $A_{0+}\approx 1$, as well as $A_b\approx 1/2$ and $B_b\approx 1/2$ for $b\neq 0+$, so that we may accurately fit the decay parameters.

In the case of $b=0-$ and $b=1-$, we also need to distinguish between $\lambda_b$ and $\kappa_b$, since they enter into Eq.~\ref{eq:ZProbabilityEstimates} with different signs. As we demonstrate \jc{in Appendix \ref{sec:DerivingIBRB}}, for high-fidelity gates we expect $\lambda_b\approx 1$ and $\kappa_b\approx-1$, so we can define $\lambda_b$ to be the decay constant with the largest real part. In the case of $b=0+$, $b=1+$, and $b=2\pm$ we cannot differentiate between $\lambda_b$ and $\kappa_b$, since they are both $\approx 1$ for high-fidelity gates, but they enter Eq.~\ref{eq:ZProbabilityEstimates} with the same sign and therefore do not need to be distinguished.

While there are nine initial states and measurements listed in Table~\ref{tab:MeasurementsAndInitialStatesZ}, one can use the same experimental data for $b=0+$ and $b=0-$, the two rows of $b=1+$, for the first row of $b=2+$ and the first row of $b=2-$, and for the second row of $b=2+$ and the second row of $b=2-$, \jc{reducing the cost to five distinct pairs of initial states and measurements.}

\begin{figure*}
    \centering
    \includegraphics[width=2\columnwidth]{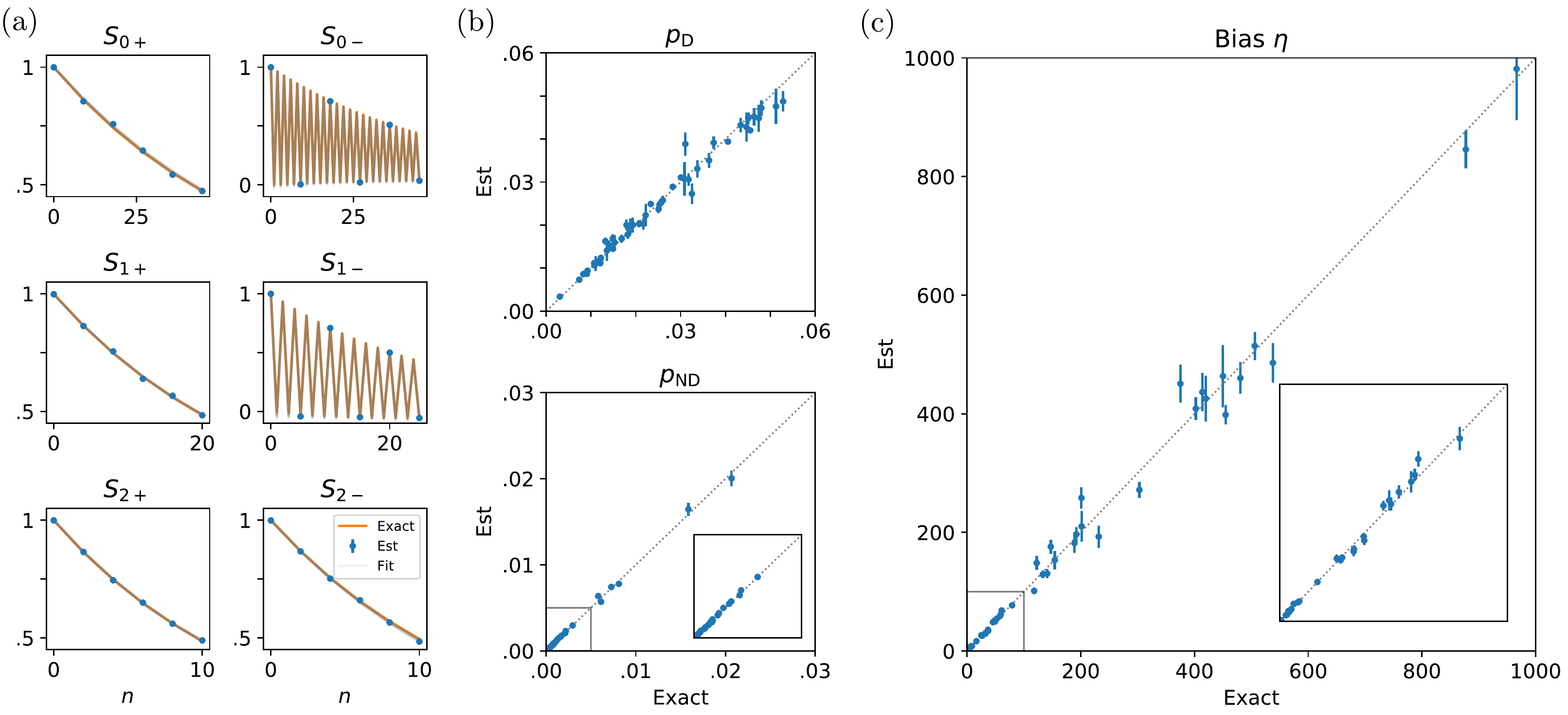}
    \caption{Simulated interleaved bias RB experiments. (a) An example of estimating $S_b(n)$, with the $S_b(n)$ curves plotted in orange and our estimates of $S_b(n)$ given by blue dots. \jc{Note that $S_b(n)$ is oscillatory for $b=0-,1-$, since the decay constants $\lambda_b,\kappa_b$ have opposite sign. To robustly fit $S_b(n)$ in the presence of these oscillations does not require densely sampling data points, since we know the oscillatory period is always $2$; however, it does require taking data at both even and odd sequence lengths, as shown in the figure.} To estimate sensitivity to statistical errors, we generate multiple fits of $S_b(n)$ using bootstrap resampling of our data, which are plotted in grey and generally overlap the exact $S_b(n)$ curves. (b) The probabilities $p_\mathrm{D}$ and $p_{\mathrm{ND}}$ extracted from the fits for $50$ randomly generated error channels. Error bars denote the standard deviation over $50$ resamplings. (c) The estimated bias $\eta$ for each of these error channels. We see that even at very high bias, we can accurately estimate the value of $\eta$.}
    \label{fig:SimulatedInterleavedRB}
\end{figure*}

We give an example of IBRB in Fig. \ref{fig:SimulatedInterleavedRB}. \jc{Here, we generate random error channels $\Lambda_G$, $\Lambda_C$, and $\Lambda_{C'}$ by generating random sets of Kraus operators, and simulate an IBRB experiment (see Appendix~\ref{sec:RandomErrors} for details on the random error channels).} Fig. \ref{fig:SimulatedInterleavedRB}a illustrates the experiment for a single error channel, where we estimate the value of $S_b(n)$ for a few different values of $n$ and fit the data to the functional forms given in Table \ref{tab:MeasurementsAndInitialStatesZ}. Again, each value of $S_b(n)$ is estimated using $5000$ measurements. \fl{Note that for $b=1-,2-$ the function $S_b(n)=A_b\lambda_b^n+B_b\kappa_b^n$ oscillates with period $2$, because $\lambda_b\approx 1$ and $\kappa_b\approx -1$. However, we can still accurately estimate the parameters $A_b,B_b,\lambda_b,\kappa_b$ by taking only a few widely spaced data points, as shown in Fig. \ref{fig:SimulatedInterleavedRB}a, provided we take data at both even and odd sequence lengths $n$. This is because the rapid oscillations are constrained to have period $2$ by the form of $S_b(n)$, so it is not necessary to take fine-grained data at nearby values of $n$ to fit this rapidly oscillating function.} To demonstrate the effectiveness of our procedure, we repeat this experiment for many different error channels, using Eq. \ref{eq:ZProbabilityEstimates} to estimate $p_{\mathrm{D}}$ and $p_{\mathrm{ND}}$ and finally using these estimates to extract the bias. In Fig. \ref{fig:SimulatedInterleavedRB}b we plot our estimated probabilities versus the exact probabilities of the error channel, and in Fig. \ref{fig:SimulatedInterleavedRB}c we do the same for the bias. To estimate the error bars, we again use bootstrap resampling. Visually, it is clear that we are accurately estimating $p_{\mathrm{D}}$, $p_{\mathrm{ND}}$, and $\eta$ even for very high biases. To verify this, we again compute the reduced-$\chi^2$ statistic for our estimates, and find $\chi^2$ between $2$ and $3$, indicating that our bootstrapping is accurately estimating the error bars to within a factor of between $\sqrt{2}$ and $\sqrt{3}$.

\subsection{Randomized compiling for interleaved bias RB}

Our IBRB measures $p_\mathrm{D}$ and $p_\mathrm{ND}$ for the averaged, composite channel $(\Lambda+\Lambda')/2$, with $\Lambda=\Lambda_C\circ\Lambda_G$ and $\Lambda'=\Lambda_{C'}\circ\Lambda_G$. Provided that $p_\mathrm{D}$ and $p_\mathrm{ND}$ for $\Lambda_C$ and $\Lambda_{C'}$ are equal and $\Lambda_G=\mathbbm{1}$, these are also the dephasing and non-dephasing probabilities for $\Lambda_C$ alone. However, in general we would like to avoid assuming the error channels $\Lambda_C$ and $\Lambda_{C'}$ are identical, and we would like to allow for the possibility of $\Lambda_G\neq \mathbbm{1}$.

Previous interleaved RB procedures for determining the fidelity of a gate $C$ have a similar problem; in these procedures one separately estimates the fidelity of $\Lambda_C\circ\Lambda_G$ and $\Lambda_G$, and uses this information to provide bounds on the fidelity of $\Lambda_C$ alone~\cite{magesan2012efficient}. These bounds depend on the fidelity of $\Lambda_G$, with lower fidelity $\Lambda_G$ resulting in looser bounds for the fidelity of $\Lambda_C$. From this point of view, it is important for the interleaving group to be high-fidelity, in order to be able to tightly bound the fidelity of $\Lambda_C$. It is also possible to use this method to bound $p_\mathrm{D}$ and $p_\mathrm{ND}$ of $(\Lambda_C+\Lambda_{C'})/2$, provided we know the corresponding probabilities for $\Lambda_G$ and $\Lambda$ (see Appendix \ref{appendix:InterleavedBounds}). However, benchmarking $p_\mathrm{D}$ and $p_\mathrm{ND}$ of $\Lambda_G$ is challenging for $G=\mathcal{Z}_2$ \jc{as averaging over sequences of $\mathcal{Z}_2$ gates does not randomize error channels enough to guarantee a simple form of the survival probability}.

On the other hand, the technique of randomized compiling~\cite{knill2005quantum,viola2005random,wallman2016noise} provides an alternative interpretation of interleaved RB results. In randomized compiling, one intentionally inserts random elements of a high-fidelity interleaving group between the lower-fidelity gates $C$ in order to eliminate the coherence of the noise. In a circuit that has been randomly compiled, the error channel associated with a gate $C$ is the combined error channel $\Lambda_C\circ\Lambda_G$. From this point of view, we require the interleaving group to be high-fidelity to ensure that randomly inserting elements of the interleaving group into a circuit does not notably increase the errors in the circuit.

\begin{figure}
    \centering
    $$
    \Qcircuit @C=.5em @R=.5em { 
    &\lstick{|+\rangle} & \ctrl{1} & \gate{Z^{\beta_1}} &\gate{X^\alpha} &\ctrl{2}&\gate{X^\alpha} &\ctrl{3}&\ctrl{4}&\measureD{X}\\
    \lstick{} & \qw & \ctrl{-1} & \qw & \qw & \qw & \qw & \qw & \qw & \qw \\
    \lstick{} & \qw & \qw & \gate{Z^{\beta_2}} & \qw & \targ & \qw & \qw & \qw & \qw\\
    \lstick{} & \qw & \qw & \qw & \qw & \qw & \qw & \targ & \qw & \qw\\
    \lstick{} & \qw & \qw & \qw & \qw & \qw & \qw & \qw & \ctrl{-1} & \qw \gategroup{1}{4}{3}{7}{.5em}{--}}
    $$
    $$
    \Qcircuit @C=.5em @R=.5em { 
    &\lstick{|+\rangle} & \ctrl{1} &\ctrl{2} &\ctrl{3}&\ctrl{4}&\gate{Z^{\beta_1+\beta_2}}&\measureD{X}\\
    \lstick{} & \qw & \ctrl{-1} & \qw & \qw & \qw & \qw & \qw \\
    \lstick{} & \qw & \qw  & \targ & \qw & \qw & \gate{Z^{\beta_2}X^\alpha} &\qw\\
    \lstick{} & \qw & \qw & \qw & \targ & \qw & \qw & \qw\\
    \lstick{} & \qw & \qw & \qw & \qw & \ctrl{-1} & \qw & \qw}
    $$
    \caption{Randomizing a single $\CX_{c,t}$ gate in a circuit that measures the stabilizer of the XZZX code, where $c$ denotes the control qubit and $t$ the target. Rather than apply the bare $\CX_{c,t}$ gate, we randomly insert an element of $\mathcal{Z}_2$ before the gate, and randomly apply either $\CX_{c,t}$ or $\mathrm{X}_c\CX_{c,t}\mathrm{X}_c$. This modification of the circuit results in an overall Pauli operator, which can be tracked in software.}
    \label{fig:RandomizedCompiling}
\end{figure}
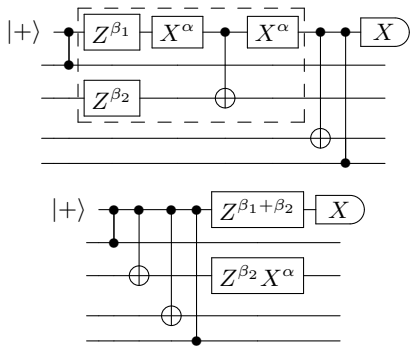

We can do a modified version of randomized compiling for bias-preserving $\CX$ gates. Each time a $\CX$ gate appears in a circuit, we randomly insert an element of $\mathcal{Z}_2$ before it, and with $50\%$ probability replace it with \spuri{$|0\rangle$-controlled $\CX$, $\mathrm{X}_c\CX_{c,t}\mathrm{X}_c$}. These extra Pauli operators can be commuted through the rest of the circuit and their effect can be tracked in software. This is illustrated in Fig.~\ref{fig:RandomizedCompiling} for the example of a circuit that measures the stabilizer of the XZZX surface code~\cite{ataides2021xzzx}.

The error channel for the resulting randomized gate then has the $p_\mathrm{D}$ and $p_\mathrm{ND}$ that we measure in our experiment. Therefore, we can always achieve the error rates $p_\mathrm{D}$ and $p_\mathrm{ND}$ by this randomized compiling procedure.

Like the usual randomized compiling, our biased-noise randomized compilation procedure has the additional benefit of limiting how errors can build up when composing noisy gates. We show in Appendix \ref{appendix:RandomizedCompiling} that if two error channels $\Lambda_A$ and $\Lambda_B$ are randomly compiled as above, their composition $(\Lambda_A\circ\Lambda_B)$ will have a non-dephasing error probability
\begin{equation}
    p_\mathrm{ND}\approx p_\mathrm{ND}^A+p_\mathrm{ND}^B,\label{eq:randomizedCompilingCombined}
\end{equation}
which, \spuri{in contrast to} to Eq. \ref{eq:CombinedNDBound}, says that $p_\mathrm{ND}$ grows linearly rather than quadratically in the \jc{number of composed error channels} when the circuit is randomly compiled by $\mathcal{Z}_N$. This is similar to the behavior of the average infidelity for full randomized compilation, which also may grow quadratically under composition for generic noise channels but grows linearly for randomly compiled circuits \cite{carignan2019bounding}.

\section{Conclusion}

Measuring the bias of a highly-biased gate is a delicate process, as the non-dephasing error probability must be precisely estimated. By using techniques from randomized benchmarking, we can precisely estimate these error probabilities. The essential ingredient in our method is defining efficiently measurable weighted survival probabilities whose decay rates depend only on $p_\mathrm{ND}$. Because we consider variable sequence lengths, our method \fl{estimates gate error rates independently from} SPAM errors, even if \fl{the SPAM} errors are much larger than the non-dephasing \fl{gate} errors. By measuring the weighted survival probabilities for long gate sequences, we can magnify the effect of small non-dephasing errors, allowing us to precisely measure arbitrarily small error probabilities by simply increasing our sequence length.

Our interleaved bias RB in particular is highly tailored to the experimental constraints of biased noise qubits. In general, interleaved RB works because the interleaved gates randomize the error channel while not adding significant additional errors. However, in the case of biased noise qubits, we can only interleave Z gates without introducing additional errors; X and Y gates are generally as error-prone as $\CX$ gates. As a result, we were \jc{motivated} to add additional randomization by swapping between $C$ and $C'$, which allowed for sufficient randomization of the error channel. This is in contrast to standard techniques for estimating Pauli channels, which assume one can freely add Pauli operators to a circuit without adding significant errors \cite{flammia2020efficient,flammia2021averaged,wallman2016noise,harper2020efficient}. We expect our techniques to be highly relevant to near-term experiments, as the numerous proposals for bias-preserving $\CX$ gates \cite{puri2020bias,guillaud2019repetition,cong2021hardware} are realized experimentally.

\acknowledgements

We thank Joel J. Wallman for helpful discussions about randomized compiling and its implications for our procedure and Steven T. Flammia for a critical reading of our manuscript. \jc{This work is supported by the ARO under grant number W911NF18-1-0212}.

\appendix

\section{Derivation of the procedures}
\label{sec:Derivations}
To derive these procedures requires a detour into some background mathematics. To begin, we review a few necessary aspects of representation theory, and define a natural representation for quantum groups, the Liouville representation. Next, we determine the irreducible representations of the Liouville representation of $\mathcal{P}_N$, $\mathcal{Z}_N$, and $\mathcal{D}_N$. Then we derive how the dephasing and non-dephasing probabilities may be written as the trace over invariant subspaces of the Liouville representations. Finally, armed with this background, we derive each of the bias RB procedures.

\subsection{Background: Representation theory}

Given a finite group $G$, a {\bf unitary representation} is a map assigning a unitary matrix to each group element such that group multiplication is preserved:
\begin{align}
\phi:G\rightarrow U(m),\quad \phi(g_1g_2)=\phi(g_1)\phi(g_2).
\end{align}
where $U(m)$ is the group of $m\times m$ unitary matrices acting on $\mathbb{C}^m$. A representation is {\bf irreducible} if the image of $\phi$ doesn't preserve any proper subspace of $\mathbb{C}^m$. Every finite-dimensional representation can be uniquely decomposed as the direct sum of irreducible representations (irreps):
\begin{align}
    \phi&\simeq a_1\phi_1\oplus\cdots\oplus a_I\phi_I\\
    \mathbb{C}^m &\simeq V_1^{a_1}\oplus\cdots\oplus V_I^{a_I}
\end{align}
where $\phi_i:G\rightarrow V_i$ is irreducible and $a_i\phi_i$ is standard shorthand for the direct sum of $a_i$ copies of $\phi_i$. The number $a_i$ is refered to as the {\bf multiplicity} of the $i$th irrep. Finally, the {\bf character} $\chi$ of a representation is defined by
\begin{equation}
    \chi(g) = \Tr\left(\phi(g)\right).
\end{equation}

We will repeatedly use two elementary facts about representations.
\begin{fact}[Projection Formula]
    If $\phi\simeq a_1\phi_1\oplus\cdots\oplus a_I\phi_I$, then the projector $\hat\Pi_i$ onto $V_i^{a_i}\subseteq\mathbb{C}^m$ is given by
    \begin{equation}
        \hat{\Pi}_i=\frac{\text{dim}(V_i)}{|G|}\sum_{g\in G} \chi_i^*(g)\phi(g)\label{eq:Projection}
    \end{equation}
    where $\chi_i$ is the character of $\phi_i$.\label{fact:Projection}
\end{fact}
\begin{fact}[Schur's Lemma]
    If $\phi\simeq a_1\phi_1\oplus\cdots\oplus a_I\phi_I$, then for any matrix $Q\in \mathrm{GL}(m)$ we have
    \begin{equation}
        \frac{1}{|G|}\sum_{g\in G} \phi(g^{-1}) Q \phi(g)\simeq (Q_1\otimes \mathbbm{1}_1)\oplus\cdots\oplus (Q_I\otimes \mathbbm{1}_I)\label{eq:Schur}
    \end{equation}
    where $Q_i$ is some $a_i\times a_i$ matrix, and $Q_i\otimes \mathbbm{1}_i$ is a matrix that acts on $V_i^{a_i}$ by mixing the $a_i$ copies of $V_i$ but acting as the identity on the degrees of freedom within each copy of $V_i$. In particular, if $a_i=1$ for all $i$, we have
    \begin{equation}
        \frac{1}{|G|}\sum_{g\in G} \phi(g^{-1}) Q \phi(g)=\sum_i\frac{\Tr(\hat{\Pi}_i Q)}{\Tr(\hat{\Pi}_i)}\hat{\Pi}_i\label{eq:SchurIrreducible}
    \end{equation}
    where $\hat\Pi_i$ is the projector onto the single copy of $V_i\subseteq \mathbb{C}^m$.
    \label{fact:Schur}
\end{fact}\noindent
See~\cite{fulton2013representation} for proofs of these facts, as well as more details on the representation theory of finite groups.

In this paper, we will be interested in the case where $G\subset U(2^N)$ is a finite subgroup of the unitary group on $N$ qubits. In this case, the standard action of $U\in G$ on a density matrix, $\rho\mapsto U\rho U^\dagger$, is a unitary representation of $G$ on the vector space of density matrices. Choosing a basis for $\mathcal{H}$, we can more conveniently represent a density matrix $\rho=\sum_{i,j} \rho_{ij}|i\rangle\langle j|$ by a vector $|\rho\rangle\rangle := \sum_{i,j}\rho_{ij}|i\rangle|j\rangle$. In terms of this vectorized density matrix, it is simple to see that the action of a unitary $U$ on $|\rho\rangle\rangle$ is given by
\begin{equation}
    |U\rho U^\dagger\rangle\rangle = U\otimes U^*|\rho\rangle\rangle = \hat U |\rho \rangle\rangle
\end{equation}
where we've defined $\hat U:=U\otimes U^*$ to be the matrix representation of the unitary $U$ acting on the space of vectorized density matrices. This representation $\phi: U\mapsto \hat U$ sending a $d\times d$ unitary to a $d^2\times d^2$ unitary is known as the {\bf Liouville representation}.

We can define the Liouville representation of a quantum channel $\Lambda: \rho\mapsto \sum_a K_a\rho K_a^\dagger$ by $\hat\Lambda:= \sum_a K_a\otimes K_a^*$, in which case we have
\begin{equation}
    \big|\Lambda(\rho)\big\rangle\big\rangle = \hat{\Lambda}|\rho\rangle\rangle.
\end{equation}

Finally, we can write the expectation value of an observable $E$ over a state $\rho$ in terms of the Liouville representation as
\begin{equation}
    \langle E\rangle_\rho =\Tr(E^\dagger \rho) = \langle\langle E|\rho\rangle\rangle
\end{equation}

We refer to~\cite{wood2011tensor} for a more detailed treatment of both quantum channels and the Liouville representation.

\subsection{Irreps of the quantum groups}

\begin{table}
\begin{tabular}{c|c|c}
    $i$ & $\hat{\Pi}_i^\mathcal{P}$ & $\chi_i(X^\alpha Z^\beta )$ \\\specialrule{.2em}{0em}{0em}
    $0$ & $\frac{1}{2}|\mathbbm{1}\rangle\rangle\langle\langle\mathbbm{1}|$ & $1$\\\hline
    $1$ & $\frac{1}{2}|Z\rangle\rangle\langle\langle Z|$ & $(-1)^{\alpha}$\\\hline
    $2$ & $\frac{1}{2}|Y\rangle\rangle\langle\langle Y|$ & $(-1)^{\alpha+\beta}$ \\\hline
    $3$ & $\frac{1}{2}|X\rangle\rangle\langle\langle X|$ & $(-1)^\beta$ \\\hline
\end{tabular}
\caption{Irreps, projectors, and characters of the Liouville representation of $\mathcal{P}_1$.}
\label{tab:PauliIrreps}
\begin{tabular}{c|c|c}
    $i$ & $\hat{\Pi}_i^\mathcal{Z}$ & $\chi_i\left(Z_1^{\beta_1}Z_2^{\beta_2}\right)$ \\\specialrule{.2em}{0em}{0em}
    \multirow{4}{*}{$0$} & $\frac{1}{4}|\mathbbm{1}\mathbbm{1}\rangle\rangle\langle\langle\mathbbm{1}\mathbbm{1}|$ & \multirow{4}{*}{$1$}\\
    & $\frac{1}{4}|\mathbbm{1}Z\rangle\rangle\langle\langle\mathbbm{1}Z|$ & \\
    & $\frac{1}{4}|Z\mathbbm{1}\rangle\rangle\langle\langle Z\mathbbm{1}|$ & \\
    & $\frac{1}{4}|ZZ\rangle\rangle\langle\langle ZZ|$ & \\\hline
    \multirow{4}{*}{$1$} & $\frac{1}{4}|\mathbbm{1}X\rangle\rangle\langle\langle\mathbbm{1}X|$ & \multirow{4}{*}{$(-1)^{\beta_2}$}\\
    & $\frac{1}{4}|\mathbbm{1}Y\rangle\rangle\langle\langle\mathbbm{1}Y|$ & \\
    & $\frac{1}{4}|ZX\rangle\rangle\langle\langle ZX|$ & \\
    & $\frac{1}{4}|ZY\rangle\rangle\langle\langle ZY|$ & \\\hline
    \multirow{4}{*}{$2$} & $\frac{1}{4}|X\mathbbm{1}\rangle\rangle\langle\langle X\mathbbm{1}|$ & \multirow{4}{*}{$(-1)^{\beta_1}$}\\
    & $\frac{1}{4}|XZ\rangle\rangle\langle\langle XZ|$ & \\
    & $\frac{1}{4}|Y\mathbbm{1}\rangle\rangle\langle\langle Y\mathbbm{1}|$ & \\
    & $\frac{1}{4}|YZ\rangle\rangle\langle\langle YZ|$ & \\\hline
    \multirow{4}{*}{$3$} & $\frac{1}{4}|XX\rangle\rangle\langle\langle XX|$ & \multirow{4}{*}{$(-1)^{\beta_1+\beta_2}$}\\
    & $\frac{1}{4}|XY\rangle\rangle\langle\langle XY|$ & \\
    & $\frac{1}{4}|YX\rangle\rangle\langle\langle YX|$ & \\
    & $\frac{1}{4}|YY\rangle\rangle\langle\langle YY|$ & \\\hline
\end{tabular}
\caption{Irreps, projectors, and characters of the Liouville representation of $\mathcal{Z}_2$.}
\label{tab:ZIrreps}
\begin{tabular}{c|c}
    $i$ & $\hat{\Pi}_i^\mathcal{D}$ \\\specialrule{.2em}{0em}{0em}
    0 & $\frac{1}{2^N}|\mathbbm{1}\cdots\mathbbm{1}\rangle\rangle\langle\langle\mathbbm{1}\cdots\mathbbm{1}|$\\\hline
    1 & $\frac{1}{2^N}\sum_{P\in\mathcal{Z}_N\setminus\{\mathbbm{1}\cdots\mathbbm{1}\}}|P\rangle\rangle\langle\langle P|$\\\hline
    2 & $\frac{1}{2^N}\sum_{P\in\mathcal{P}_N\setminus\mathcal{Z}_N}|P\rangle\rangle\langle\langle P|$\\\hline
    \end{tabular}
    \caption{Irreps and projectors of the Liouville representation of $\mathcal{D}_N$. For this work, we will not need the characters of $\mathcal{D}_N$.}
    \label{tab:DihedralIrreps}
\end{table}

For the Pauli, Z, and $\CX$-dihedral groups, we will need to understand the decomposition of their Liouville representation into irreps, and the characters of those irreps.

In the case of the $1$-qubit Pauli group, the Liouville representation decomposes into four non-isomorphic irreps, with projectors and characters given in Table~\ref{tab:PauliIrreps}. The Liouville representation of the $N$-qubit Pauli group $\mathcal{P}_N$ then decomposes into $4^N$ non-isomorphic irreps indexed by vectors $\vec i \in\mathbb{Z}_4^N$, with projectors and characters
\begin{align}
    \hat{\Pi}^\mathcal{P}_{\vec{i}} &= \hat{\Pi}^\mathcal{P}_{i_1}\otimes\cdots\otimes\hat{\Pi}_{i_N}^{\mathcal{P}}\label{eq:PauliIrreps}\\
    \chi_{\vec i}\left(X(\vec{\alpha})Z(\vec{\beta})\right)&= \chi_{i_1}(X^{\alpha_1}Z^{\beta_1})\cdots\chi_{i_N}(X^{\alpha_N}Z^{\beta_N}).\nonumber
\end{align}

One can similarly determine a factorized form for the irreps of the Liouville representation of the $\mathrm{Z}$ group $\mathcal{Z}_N$, but we will need only the case $N=2$ here. The Liouville representation of $\mathcal{Z}_2$ decomposes into 16 irreps with projectors and characters given Table~\ref{tab:ZIrreps}. Note that in this case, each irrep has multiplicity $4$.

Finally, the Liouville representation of the $\CX$-dihedral group $\mathcal{D}_N$ decomposes into the three irreps given in Table~\ref{tab:DihedralIrreps}. Note that the number of irreps is independent of $N$. Derivation of these irreps can be found in~\cite{cross2016scalable}.

\subsection{The dephasing/non-dephasing error probabilities in the Liouville representation}

From the definition of the dephasing and non-dephasing error probabilities, Eqs.~\ref{eq:DefnDephasing} and~\ref{eq:Defnnon-dephasing}, we want to express $p_\mathrm{D}$ and $p_\mathrm{ND}$ as the trace of $\hat\Lambda$ over subspaces $\mathcal{Z}_N$ and $\mathcal{P}_N\setminus\mathcal{Z}_N$ of the Pauli group. Note that these subspaces are invariant under the action of any bias-preserving operator. It is straightforward to show
\begin{align}
    \Tr_{\mathcal{Z}_N}(\hat\Lambda) &= \frac{1}{2^N}\sum_{P\in\mathcal{Z}_N}\langle\langle P|\hat\Lambda|P\rangle\rangle\\
    &= 2^N(1-p_{\mathrm{ND}})\\
    \Tr_{\mathcal{P}_N\setminus\mathcal{Z}_N}(\hat\Lambda) &=\frac{1}{2^N}\sum_{P\in\mathcal{P}_N\setminus\mathcal{Z}_N}\langle\langle P|\hat\Lambda|P\rangle\rangle\\
    &=(4^N-2^N)(1-p_{\mathrm{ND}})-4^Np_{\mathrm{D}}
\end{align}
Rearranging this gives formulas for the dephasing and non-dephasing probabilities
\begin{align}
\begin{split}
    p_{\mathrm{D}} &=\frac{1}{4^N}\left[\left(2^N-1\right)\Tr_{\mathcal{Z}_N}(\hat\Lambda)-\Tr_{\mathcal{P}_N\setminus\mathcal{Z}_N}(\hat\Lambda)\right]\\
    p_{\mathrm{ND}} &= 1-\frac{1}{2^N}\Tr_{\mathcal{Z}_N}(\hat\Lambda).
\end{split}\label{eq:dephasingAndnon-dephasingProbabilities}
\end{align}
Our goal in these RB procedures is to determine the two traces $\Tr_{\mathcal{Z}_N}(\hat\Lambda)$ and $\Tr_{\mathcal{P}_N\setminus\mathcal{Z}_N}(\hat\Lambda)$.

\subsection{Deriving $\CX$-dihedral bias RB}
\label{sec:DerivingCNOTD}
The character-weighted survival probability $S_b$ can be written as
\begin{widetext}
\begin{equation}
    S_b(n)=\smashoperator{\mathop{\mathbb{E}}_{\substack{U_0\in \mathcal{P}_N\\ U_1\cdots U_{n}\in \mathcal{D}_N}}}\quad\left[\langle\langle E_b|\hat \Lambda_M \hat \Lambda \hat{U}_{n+1}\hat\Lambda\hat{U}_N\cdots\hat U_2\hat\Lambda\hat U_1\hat U_0\hat\Lambda_P|\rho_b\rangle\rangle\chi_b^*(U_0)\right]
\end{equation}
where $\Lambda$ is the error channel associated with elements of $\mathcal{D}_N$, and we have included unknown preparation and measurement errors $\Lambda_P$ and $\Lambda_M$. We make the standard RB change-of-variables, defining $V_1:=U_1$ and inductively defining $V_{m}:=U_mV_{m-1}$. Note that the expectation value over $\{V_m\}$ is equivalent to the expectation value over $\{U_m\}$. Thus, we can write
\begin{align}
    S_b(n)&=\smashoperator{\mathop{\mathbb{E}}_{\substack{U_0\in \mathcal{P}_N\\ V_1\cdots V_{n}\in \mathcal{D}_N}}}\quad\left[\langle\langle E_b|\hat \Lambda_M \hat \Lambda \hat{V}_{n}^\dagger\hat\Lambda\hat{V}_n\hat{V}_{n-1}\cdots\hat V_2V_1^\dagger\hat\Lambda\hat V_1\hat U_0\hat\Lambda_P|\rho_b\rangle\rangle\chi_b^*(U_0)\right]\\
    &=\langle\langle E_b|\hat \Lambda_M \hat \Lambda \left[\mathop{\mathbb{E}}_{V\in\mathcal{D}_N}\left[\hat V^\dagger\hat\Lambda V\right]\right]^n\mathop{\mathbb{E}}_{U_0\in \mathcal{P}_N}\left[\hat U_0\chi_b^*(U_0)\right]\hat\Lambda_P|\rho_b\rangle\rangle.
\end{align}
\end{widetext}
The two expectation values can be evaluated using Facts~\ref{fact:Projection} and~\ref{fact:Schur}. First, we note from Eq.~\ref{eq:PauliIrreps} that $\chi_1$ is the character function of the irrep of the Liouville representation indexed by $\vec{i}_1=(1,1,\dots,1)$, and $\chi_2$ is the character function of the irrep indexed by $\vec{i}_2 = (3,3,\dots,3)$. Then Fact~\ref{fact:Projection} says that
\begin{equation}
    \mathop{\mathbb{E}}_{U_0\in \mathcal{P}_N}\left[\hat U_0\chi_b^*(U_0)\right] = \hat{\Pi}^\mathcal{P}_{\vec{i}_b}
\end{equation}
where the projectors have the explicit form (see Eq.~\ref{eq:PauliIrreps})
\begin{equation}
\hat{\Pi}_{\vec{i}_b}^\mathcal{P}=\left\{\begin{aligned}
        \frac{1}{2^N}|Z\cdots Z\rangle\rangle\langle\langle Z\cdots Z|, &\quad  b=1\\
        \frac{1}{2^N}|X\cdots X\rangle\rangle\langle\langle X\cdots X|, &\quad  b=2
\end{aligned}\right.\label{eq:PauliProjectorsCNOTD}
\end{equation}
Second, we note that, since the Liouville representation of the dihedral group is multiplicity-free (all $a_i=1$), Fact~\ref{fact:Schur} gives
\begin{equation}
\mathop{\mathbb{E}}_{V\in\mathcal{D}_N}\left[\hat V^\dagger\hat\Lambda V\right] = \sum_i\frac{\Tr(\hat\Pi_i^\mathcal{D}\hat\Lambda)}{\Tr(\hat\Pi_i^\mathcal{D})}\hat\Pi_i^\mathcal{D}.
\end{equation}

Combining these facts allows us to simplify the survival probability as
\begin{align}
    S_b(n)&=\sum_i\langle\langle E_b|\hat \Lambda_M \hat \Lambda \left[\frac{\Tr(\hat\Pi_i^\mathcal{D}\hat\Lambda)}{\Tr(\hat\Pi_i^\mathcal{D})}\right]^n\hat\Pi_i^\mathcal{D}\hat{\Pi}^\mathcal{P}_{\vec{i}_b}\hat\Lambda_P|\rho_b\rangle\rangle.
\end{align}

We have carefully chosen $\chi_b$ to give us a projector $\hat\Pi_{\vec{i}_b}^\mathcal{P}$ such that $\hat{\Pi}_i^\mathcal{D}\hat{\Pi}_{\vec{i}_b}^\mathcal{P}=\delta_{i,b}\hat\Pi_{\vec{i}_b}$, as can be checked from the formulas for $\hat{\Pi}_i^\mathcal{D}$ in Table \ref{tab:DihedralIrreps}. Therefore, all terms in the sum vanish except for $i=b$, and we can write the final form of $S_b(n)$ as
\begin{equation}
    S_b(n)= \underbrace{\langle\langle E_b|\hat \Lambda_M \hat \Lambda \hat{\Pi}^\mathcal{P}_{\vec{i}_b}\hat\Lambda_P|\rho_b\rangle\rangle}_{A_b} \Bigg[\underbrace{\frac{\Tr(\hat\Pi_b^\mathcal{D}\hat\Lambda)}{\Tr(\hat\Pi_b^\mathcal{D})}}_{\lambda_b}\Bigg]^n
\end{equation}
where we have defined $A_b$, $\lambda_b$ to match the form given in Table \ref{tab:MeasurementsAndInitialStatesCNOTD}. Note that $\lambda_b$ only depends on $\Lambda$, and all effects of the SPAM errors $\Lambda_P$ and $\Lambda_M$ are absorbed in $A_b$. Note also that provided we have reasonably high-fidelity preparation, gates, and measurements, we can approximate $A_b$ as
\begin{equation}
    A_b\approx \langle\langle E_b|\hat\Pi_{\vec{i}_b}^\mathcal{P}|\rho_b\rangle\rangle = 1
\end{equation}
where the last equality is found by plugging in the explicit formulas for $E_b$, $\hat\Pi_{\vec i_b}^\mathcal{P}$, and $\rho_b$ given in Eq.~\ref{eq:PauliProjectorsCNOTD} and Table~\ref{tab:MeasurementsAndInitialStatesCNOTD}. Therefore, the prefactor in front of $\lambda_b^n$ is large, and we can accurately fit $\lambda_b$.

To finish, we note that plugging in the explicit formulas from Table~\ref{tab:DihedralIrreps} for $\hat{\Pi}_b^\mathcal{D}$ gives
\begin{align}
    \lambda_1 & = \frac{\Tr_{\mathcal{Z}}(\hat\Lambda)-1}{2^N-1} \\
    \lambda_2 & = \frac{\Tr_{\mathcal{P}\setminus\mathcal{Z}}(\hat\Lambda)}{4^N-2^N}.
\end{align}
Plugging this into Eq.~\ref{eq:dephasingAndnon-dephasingProbabilities} for $p_\mathrm{D}$ and $p_\mathrm{ND}$ gives the estimates in Eq.~\ref{eq:CNOTDProbEstimates}, as desired.
\subsection{Deriving interleaved bias RB}

\label{sec:DerivingIBRB}

\subsubsection{Deriving the survival probabilities for $b=0\pm$ and $b=1\pm$}

We begin by considering the survival probability $S_{0+}$. For convenience, we define the operator $\hat{M}_{0+}$ by
\begin{align}
    \hat M_{0+}&:=\frac{1}{2}\hat{\Pi}_0^\mathcal{Z}(\hat C\hat\Lambda+\hat C'\hat\Lambda')\hat{\Pi}_0^\mathcal{Z}.
\end{align}
Note that $\hat{\Pi}_0^\mathcal{Z}$ is a rank-$4$ projector, so $\hat M_{0+}$ is a rank-$4$ operator in the general case.

We now evaluate the survival probability $S_{0+}(n)$ in terms of $M_{0+}$. We note that Fact~\ref{fact:Projection} implies that $\mathop{\mathbb{E}}_{U\in\mathcal{Z}_2}[\hat{U}]=\hat\Pi_0^\mathcal{Z}$. We then have
\begin{widetext}
\begin{align}
S_{0+}(n)&=\underset{\substack{U_1\cdots U_{n+1} \in \mathcal{Z}_2\\C_1,...,C_n\in \{C,C'\}}}{\mathbb{E}}\left[\langle\langle E_{0+}|\hat{\Lambda}_M \hat\Lambda_G\hat{U}_{n+1} \hat C_n\hat\Lambda_n\hat U_n\hat{C}_{n-1}\hat\Lambda_{n-1}\hat U_{n-1}\cdots \hat{U}_2\hat C_1\hat\Lambda_1\hat U_1\hat\Lambda_P|\rho_{0+}\rangle\rangle\right]\\
&=\underset{C_1,...,C_n\in \{C,C'\}}{\mathbb{E}}\Big[\langle\langle E_{0+}|\hat{\Lambda}_M\hat\Lambda_G \underbrace{\hat{\Pi}^\mathcal{Z}_0 \hat C_n\hat\Lambda_n\hat{\Pi}^\mathcal{Z}_0}_{\hat{M}_{0+}}\underbrace{\hat{\Pi}^\mathcal{Z}_0\hat{C}_{n-1}\hat\Lambda_{n-1}\hat{\Pi}^\mathcal{Z}_0}_{\hat{M}_{0+}}\cdots \underbrace{\hat{\Pi}^\mathcal{Z}_0\hat C_1\hat\Lambda_1\hat{\Pi}^\mathcal{Z}_0}_{\hat{M}_{0+}}\hat\Lambda_P|\rho_{0+}\rangle\rangle\Big]\\
&=\langle\langle E_{0+}|\hat{\Lambda}_M\hat\Lambda_G \hat M_{0+}^n \hat\Lambda_P|\rho_{0+}\rangle\rangle
\end{align}
\end{widetext}
Given that $\hat M_{0+}$ has rank $4$, we can expand it in terms of its eigenvalues $\mu_j$ and corresponding left and right eigenvectors $|\phi_j\rangle\rangle$ and $|\psi_j\rangle\rangle$ as
\begin{equation}
    \hat{M}_{0+}=\sum_{j=1}^4|\psi_j\rangle\rangle\mu_j\langle\langle \phi_j|
\end{equation}
where we have normalized our eigenvectors so that $\langle\langle\psi_j|\phi_k\rangle\rangle=\delta_{j,k}$. In this case, our survival probability becomes
\begin{equation}
S_{0+}(n)=\sum_{j=1}^4\langle\langle E_{0+}|\hat{\Lambda}_M\hat{\Lambda}_G|\psi_j\rangle\rangle\langle\langle\phi_j| \hat\Lambda_P|\rho_{0+}\rangle\rangle\mu_j^n.\label{eq:ExpDecayForm0+}
\end{equation}
Note that again the eigenvalues $\mu_j$ depend only on the gate errors $\Lambda$ and $\Lambda'$, and not on the SPAM errors $\Lambda_P$ and $\Lambda_M$.

In this form, the survival probability is the sum of four exponential decays, which is infeasible to fit to experimental data. However, in the case of high-fidelity gates, we can show that only two exponential decays are relevant using perturbation theory in $\delta\hat \Lambda := (\hat \Lambda-\mathbbm{1})$ and $\delta\hat \Lambda' := (\hat \Lambda'-\mathbbm{1})$. For perfect gates with $\delta\hat\Lambda=\delta\hat\Lambda'=0$, $\hat{M}_{0+}$ has eigenvalues and eigenvectors given by
\begin{align}
|\psi^0_1\rangle\rangle=|\phi^0_1\rangle\rangle &=\frac{1}{2}|\mathbbm{1}\mathbbm{1}\rangle\rangle &\quad &\mu^0_1=1\label{eq:FirstEigenvector1+}\\
|\psi^0_2\rangle\rangle=|\phi^0_2\rangle\rangle  &=\frac{1}{2}|Z\mathbbm{1}\rangle\rangle &\quad &\mu^0_2=1\label{eq:SecondEigenvector1+}\\
|\psi^0_3\rangle\rangle=|\phi^0_3\rangle\rangle &=\frac{1}{2}|\mathbbm{1}Z\rangle\rangle &\quad &\mu^0_3=0\\
|\psi^0_4\rangle\rangle=|\phi^0_4\rangle\rangle  &=\frac{1}{2}|ZZ\rangle\rangle &\quad &\mu^0_4=0.
\end{align}
Then to first order in $\delta\hat\Lambda$ and $\delta\hat\Lambda'$, we have
\begin{align}
\mu_1&=1\\
\mu_2&=\frac{1}{4}\Big\langle\Big\langle Z\mathbbm{1}\Big|\frac{\hat\Lambda+\hat\Lambda'}{2}\Big|Z\mathbbm{1}\Big\rangle\Big\rangle+O(\delta^2)\label{eq:Trace1}\\
\mu_3 &= O(\delta)\\
\mu_4 &= O(\delta).
\end{align}
We therefore see that one eigenvalue is always $1$, and that we may neglect the eigenvalues $\mu_3$ and $\mu_4$, since their contribution to the survival probability $S_{0+}(n)$ is $O(\delta^n)$. We can therefore fit $S_{0+}(n)$ to a single exponential decay plus a constant. In the notation of Table~\ref{tab:MeasurementsAndInitialStatesZ}, $\mu_2$ corresponds to $\lambda_{0+}$.

From Eq.~\ref{eq:ExpDecayForm0+}, the prefactor $A_{0+}$ in front of the exponential decay is given by
\begin{equation}
    A_{0+}=\langle\langle E_{0+}|\hat{\Lambda}_M\hat\Lambda_G|\psi_2\rangle\rangle\langle\langle\phi_2|\hat\Lambda_P|\rho_{0+}\rangle\rangle\label{eq:A0+}
\end{equation}
We can estimate the value of $A_{0+}$ by assuming $\hat\Lambda_G,\hat\Lambda_P,\hat\Lambda_M\approx \mathbbm{1}$ and evaluating Eq.~\ref{eq:A0+} explicitly, provided we can estimate the eigenvectors $|\psi_2\rangle\rangle$, $|\phi_2\rangle\rangle$. Since the eigenvectors at $\delta=0$ (Eqs.~\ref{eq:FirstEigenvector1+} and~\ref{eq:SecondEigenvector1+}) are degenerate, we must use degenerate perturbation theory to find the eigenvectors at $\delta\neq 0$. This means that $|\psi_2\rangle\rangle$ and $|\phi_2\rangle\rangle$ will in general be (up to $O(\delta)$) some linear combination the $\delta=0$ eigenvectors with eigenvalue $1$. Specifically, we have
\begin{align}
    |\psi_2\rangle\rangle & \approx \frac{1}{2}|Z\mathbbm{1}\rangle\rangle\\
    |\phi_2\rangle\rangle & \approx \frac{1}{2}|Z\mathbbm{1}\rangle\rangle+\frac{\gamma}{2}|\mathbbm{1}\mathbbm{1}\rangle\rangle
\end{align}
where $\gamma$ is some constant determined by the specific perturbations $\delta\Lambda$ and $\delta\Lambda'$, and the overall form is restricted by the normalization condition $\langle\langle\phi_i|\psi_j\rangle\rangle=\delta_{i,j}$ and the fact that $|\phi_1\rangle\rangle=\frac{1}{2}|\mathbbm{1}\mathbbm{1}\rangle\rangle$ is a right-eigenvector with eigenvalue $1$ for any trace-preserving map. Using these eigenvectors to evaluate Eq.~\ref{eq:A0+} then gives $A_{0+}\approx 1$, so that we may accurately fit $\lambda_{0+}$.

The case of $b=0-$ and $b=1\pm$ are similar. We define
\begin{align}
    \hat M_{0-}&:=\frac{1}{2}\hat{\Pi}_0(\hat C\hat\Lambda-\hat C'\hat\Lambda')\hat{\Pi}_0\\
    \hat M_{1+}&:=\frac{1}{2}\hat{\Pi}_1(\hat C\hat\Lambda+\hat C'\hat\Lambda')\hat{\Pi}_1\\
    \hat M_{1-}&:=\frac{1}{2}\hat{\Pi}_1(\hat C\hat\Lambda-\hat C'\hat\Lambda')\hat{\Pi}_1
\end{align}
and repeat the above analysis to see that we have for $b=0-$ and $b=1-$
\begin{align}
S_{b}(n)&=\sum_{j=1}^4\langle\langle E_b|\hat{\Lambda}_M\hat{\Lambda}_G|\psi_j\rangle\rangle\langle\langle\phi_j| \hat\Lambda_P|\rho_b\rangle\rangle\mu_j^n,\label{eq:ExpDecayForm1-}
\end{align}
or for $b=1+$
\begin{align}
S_{1+}(n)&=\frac{1}{2}\sum_{j=1}^4\left(\langle\langle E_{1+}^{(1)}|\hat{\Lambda}_M\hat{\Lambda}_G|\psi_j\rangle\rangle\langle\langle\phi_j| \hat\Lambda_P|\rho_{1+}^{(1)}\rangle\rangle\right.\label{eq:ExpDecayForm1+}\\&\qquad\qquad\left.+\langle\langle E_{1+}^{(2)}|\hat{\Lambda}_M\hat{\Lambda}_G|\psi_j\rangle\rangle\langle\langle\phi_j| \hat\Lambda_P|\rho_{1+}^{(2)}\rangle\rangle\right)\mu_j^n,\nonumber
\end{align}
where $\mu_j$, $|\psi_j\rangle\rangle$, and $|\phi_j\rangle\rangle$ are the eigenvalues and eigenvectors of the corresponding $\hat{M}_b$ operator. Performing perturbation theory, we find that the unperturbed $\hat{M}_{0-}$ and $\hat{M}_{1-}$ have eigenvalues $\{\mu_j^0\}=\{1,-1,0,0\}$, while the unperturbed $\hat{M}_{1+}$ has eigenvalues $\{\mu_j^0\}=\{1,1,0,0\}$. We label the two largest-magnitude eigenvalues by $\lambda_b$ and $\kappa_b$, and neglect the remaining eigenvalues that are $O(\delta)$. Working to first order in $\delta\Lambda$ and $\delta\Lambda'$, we find
\begin{align}
\lambda_{0-}-\kappa_{0-}&=\frac{1}{4}\Bigg[\Big\langle\Big\langle\mathbbm{1}Z\Big|\frac{\hat\Lambda+\hat\Lambda'}{2}\Big|\mathbbm{1}Z\Big\rangle\Big\rangle\nonumber\\
&\qquad\qquad+\Big\langle\Big\langle ZZ\Big|\frac{\hat\Lambda+\hat\Lambda'}{2}\Big|ZZ\Big\rangle\Big\rangle\Bigg] \label{eq:Trace2}\\
\lambda_{1+}+\kappa_{1+}&=\frac{1}{4}\Bigg[\Big\langle\Big\langle\mathbbm{1}X\Big|\frac{\hat\Lambda+\hat\Lambda'}{2}\Big|\mathbbm{1}X\Big\rangle\Big\rangle\nonumber\\
&\qquad\qquad+\Big\langle\Big\langle ZX\Big|\frac{\hat\Lambda+\hat\Lambda'}{2}\Big|ZX\Big\rangle\Big\rangle\Bigg]\label{eq:Trace3} \\
\lambda_{1-}-\kappa_{1-}&=\frac{1}{4}\Bigg[\Big\langle\Big\langle\mathbbm{1}Y\Big|\frac{\hat\Lambda+\hat\Lambda'}{2}\Big|\mathbbm{1}Y\Big\rangle\Big\rangle\nonumber\\
&\qquad\qquad+\Big\langle\Big\langle ZY\Big|\frac{\hat\Lambda+\hat\Lambda'}{2}\Big|ZY\Big\rangle\Big\rangle\Bigg].\label{eq:Trace4}
\end{align}

In total, we've demonstrated that for these values of $b$, we can fit $S_b(n)$ to the sum of two exponential decays, as given in Table~\ref{eq:ZProbabilityEstimates}. We can again estimate the prefactors $A_b$ and $B_b$ in front of the decays by assuming $\hat\Lambda_G,\hat \Lambda_P,\hat\Lambda_M\approx \mathbbm{1}$ and evaluating Eqs.~\ref{eq:ExpDecayForm1-} and~\ref{eq:ExpDecayForm1+}. In the case of $b=1+$ we again need to use degenerate perturbation theory, while the cases of $b=0-$ and $b=1-$ are non-degenerate. We find that $A_b,B_b\approx 1/2$, so that we can reliably fit both decay curves. Note that the only reason for averaging over two initial states and final measurements for $b_{1+}$ is to ensure that $A_{1+},B_{1+}\approx 1/2$.

\subsubsection{Deriving the survival probabilities for $b=2\pm$}

We begin by considering the survival probability $S_{2+}$. For convenience, we define the operator $\hat{M}_{2+}$ by
\begin{align}
    \hat M_{2+}&:=\frac{1}{4}\hat{\Pi}_2^\mathcal{Z}(\hat C\hat\Lambda+\hat C'\hat\Lambda')\hat{\Pi}_3^\mathcal{Z}(\hat C\hat\Lambda+\hat C'\hat\Lambda')\hat{\Pi}_2^\mathcal{Z}.
\end{align}
Note that $\hat{\Pi}_2^\mathcal{Z}$ is a rank-$4$ projector, so $\hat M_{2+}$ is a rank-$4$ operator in the general case.

We now evaluate the survival probability $S_{2+}(n)$ in terms of $M_{2+}$. We note that Fact~\ref{fact:Projection} implies that $\mathop{\mathbb{E}}_{U\in\mathcal{Z}_2}[\hat{U}\chi_{2+}^*(U)]=\hat\Pi_2^\mathcal{Z}$ and $\mathop{\mathbb{E}}_{U\in\mathcal{Z}_2}[\hat{U}\chi_{2-}^*(U)]=\hat\Pi_3^\mathcal{Z}$ (re the expression for $\chi_{2\pm}$ in Table~\ref{tab:MeasurementsAndInitialStatesZ} to  the characters of the Liouville representation of $\mathcal{Z}_2$ in Table~\ref{tab:ZIrreps}). We then have
\begin{widetext}
\begin{align}
S_{2+}(n)&=\qquad\smashoperator{\mathop{\mathbb{E}}_{\substack{(a)=(1),(2)\\U_1\cdots U_{2n+1} \in G\\C_1,...,C_{2n}\in \{C,C'\}}}}\quad\ \ \left[\langle\langle E_{2+}^{(a)}|\hat{\Lambda}_M\hat\Lambda_G \hat{U}_{2n+1} \hat C_{2n}\hat\Lambda_{2n}\hat U_{2n}\hat{C}_{2n-1}\hat\Lambda_{n-1}\hat U_{n-1}\cdots \hat{U}_3\hat C_2\hat\Lambda_2 \hat{U}_2\hat C_1\hat\Lambda_1\hat U_1\hat\Lambda_P|\rho_{2+}^{(a)}\rangle\rangle\chi_2^*(\{U_i\})\right]\\
&=\qquad\smashoperator{\mathop{\mathbb{E}}_{\substack{(a)=(1),(2)\\C_1,...,C_{2n}\in \{C,C'\}}}}\quad\ \ \Big[\langle\langle E_{2+}^{(a)}|\hat{\Lambda}_M\hat\Lambda_G \underbrace{\hat{\Pi}_{2}^\mathcal{Z} \hat C_{2n}\hat\Lambda_{2n}\hat{\Pi}^\mathcal{Z}_3\hat{C}_{2n-1}\hat\Lambda_{n-1}\hat{\Pi}^\mathcal{Z}_2}_{\hat{M}_{2+}}\cdots \underbrace{\hat{\Pi}_2^\mathcal{Z}\hat C_2\hat\Lambda_2 \hat{\Pi}_3^\mathcal{Z}\hat C_1\hat\Lambda_1\hat{\Pi}_2^\mathcal{Z}}_{\hat{M}_{2+}}\hat\Lambda_P|\rho_{2+}^{(a)}\rangle\rangle\Big]\\
&=\frac{1}{2}\left(\langle\langle E_{2+}^{(1)}|\hat{\Lambda}_M\hat\Lambda_G \hat M_{2+}^n \hat\Lambda_P|\rho_{2+}^{(1)}\rangle\rangle+\langle\langle E_{2+}^{(2)}|\hat{\Lambda}_M \hat M_{2+}^n \hat\Lambda_P|\rho_{2+}^{(2)}\rangle\rangle\right)\\
&=\sum_{j=1}^4\frac{1}{2}\left(\langle\langle E_{2+}^{(1)}|\hat{\Lambda}_M\hat\Lambda_G|\psi_j\rangle\rangle\langle\langle\phi_j| \hat\Lambda_P|\rho_{2+}^{(1)}\rangle\rangle+\langle\langle E_{2+}^{(2)}|\hat{\Lambda}_M\hat\Lambda_G|\psi_j\rangle\rangle\langle\langle\phi_j| \hat\Lambda_P|\rho_{2+}^{(2)}\rangle\rangle\right)\mu_j^n\label{eq:ExpDecayForm2+}
\end{align}
\end{widetext}
where in the last line, $\mu_j$, $|\psi_j\rangle\rangle$, and $|\phi_j\rangle\rangle$ denote the eigenvalues and eigenvectors of $\hat M_{2+}$.

We again simplify this expression through perturbation theory. When $\delta\hat\Lambda=\delta\hat\Lambda'=0$, $\hat{M}_{2+}$ has eigenvalues and eigenvectors given by
\begin{align}
|\psi^0_1\rangle\rangle=|\phi^0_1\rangle\rangle &=\frac{1}{2}|X\mathbbm{1}\rangle\rangle &\quad &\mu_1^0=1\\
|\psi^0_2\rangle\rangle=|\phi^0_2\rangle\rangle &=\frac{1}{2}|Y\mathbbm{1}\rangle\rangle &\quad &\mu_2^0=1\\
|\psi^0_3\rangle\rangle=|\phi^0_3\rangle\rangle &=\frac{1}{2}|XZ\rangle\rangle &\quad &\mu_3^0=0\\
|\psi^0_4\rangle\rangle=|\phi^0_4\rangle\rangle &=\frac{1}{2}|YZ\rangle\rangle &\quad &\mu_4^0=0.
\end{align}
Then to first order in $\delta\hat\Lambda$ and $\delta\hat\Lambda'$, the eigenvalues satisfy
\begin{align}
\mu_1+\mu_2&=\frac{1}{4}\Bigg[\Big\langle\Big\langle X\mathbbm{1}\Big|\frac{\hat\Lambda+\hat\Lambda'}{2}\Big|X\mathbbm{1}\Big\rangle\Big\rangle\nonumber\\
&\qquad +\Big\langle\Big\langle XX\Big|\frac{\hat\Lambda+\hat\Lambda'}{2}\Big|XX\Big\rangle\Big\rangle\label{eq:Trace5}\\
&\qquad +\Big\langle\Big\langle Y\mathbbm{1}\Big|\frac{\hat\Lambda+\hat\Lambda'}{2}\Big|Y\mathbbm{1}\Big\rangle\Big\rangle\nonumber\\
&\qquad +\Big\langle\Big\langle YX\Big|\frac{\hat\Lambda+\hat\Lambda'}{2}\Big|YX\Big\rangle\Big\rangle\Bigg]-2+O(\delta^2) \nonumber\\
\mu_3 &= O(\delta)\\
\mu_4 &= O(\delta).
\end{align}
We neglect $\mu_3$ and $\mu_4$, and find $S_{2+}$ is the sum of two exponential decays. In the notation of Table~\ref{tab:MeasurementsAndInitialStatesZ}, $\mu_1$ corresponds to $\lambda_{2+}$ and $\mu_2$ corresponds to $\kappa_{2+}$.

Similarly, for $S_{2-}(b)$, we define
\begin{equation}
    \hat M_{2-}:=\frac{1}{4}\hat{\Pi}_2(\hat C\hat\Lambda-\hat C'\hat\Lambda')\hat{\Pi}_3(\hat C\hat\Lambda-\hat C'\hat\Lambda')\hat{\Pi}_2
\end{equation}
in terms of which we can write
\begin{align}
S_{2-}(n)&=\frac{1}{2}\sum_{j=1}^4\left(\langle\langle E_{2-}^{(1)}|\hat{\Lambda}_M\hat\Lambda_G|\psi_j\rangle\rangle\langle\langle\phi_j| \hat\Lambda_P|\rho_{2-}^{(1)}\rangle\rangle\right.\label{eq:ExpDecayForm2-}\\&\qquad\quad\left.+\langle\langle E_{2-}^{(2)}|\hat{\Lambda}_M\hat\Lambda_G|\psi_j\rangle\rangle\langle\langle\phi_j| \hat\Lambda_P|\rho_{2-}^{(2)}\rangle\rangle\right)\mu_j^n,\nonumber
\end{align}
where now $\mu_j$, $|\psi_j\rangle\rangle$, and $|\phi_j\rangle\rangle$ denote the eigenvalues and eigenvectors of $\hat M_{2-}$.

We again use perturbation theory; the unperturbed eigenvalues and eigenvectors are
\begin{align}
|\psi_1^0\rangle\rangle=|\phi_1^0\rangle\rangle &=\frac{1}{2}|XZ\rangle\rangle &\quad &\mu_1^0=1\\
|\psi_2^0\rangle\rangle=|\phi_2^0\rangle\rangle &=\frac{1}{2}|YZ\rangle\rangle &\quad &\mu_2^0=1\\
|\psi_3^0\rangle\rangle=|\phi_3^0\rangle\rangle &=\frac{1}{2}|X\mathbbm{1}\rangle\rangle &\quad &\mu_3^0=0\\
|\psi_4^0\rangle\rangle=|\phi_4^0\rangle\rangle &=\frac{1}{2}|Y\mathbbm{1}\rangle\rangle &\quad &\mu_4^0=0
\end{align}
while to first order in $\delta\hat\Lambda$ and $\delta\hat\Lambda'$, the eigenvectors satisfy
\begin{align}
\mu_1+\mu_2&=\frac{1}{4}\Bigg[\Big\langle\Big\langle XZ\Big|\frac{\hat\Lambda+\hat\Lambda'}{2}\Big|XZ\Big\rangle\Big\rangle\nonumber\\
&\qquad +\Big\langle\Big\langle YY\Big|\frac{\hat\Lambda+\hat\Lambda'}{2}\Big|YY\Big\rangle\Big\rangle\label{eq:Trace6}\\
&\qquad +\Big\langle\Big\langle YZ\Big|\frac{\hat\Lambda+\hat\Lambda'}{2}\Big|YZ\Big\rangle\Big\rangle\nonumber\\
&\qquad +\Big\langle\Big\langle XY\Big|\frac{\hat\Lambda+\hat\Lambda'}{2}\Big|XY\Big\rangle\Big\rangle\Bigg]-2+O(\delta^2) \nonumber\\
\mu_3 &= O(\delta)\\
\mu_4 &= O(\delta).
\end{align}
We again neglect $\mu_3$ and $\mu_4$, and find $S_{2-}$ is also the sum of two exponential decays. In the notation of Table~\ref{tab:MeasurementsAndInitialStatesZ}, $\mu_1$ corresponds to $\lambda_{2-}$ and $\mu_2$ corresponds to $\kappa_{2-}$.

Finally, we can estimate the prefactors $A_{2\pm}$ and $B_{2\pm}$ by assuming $\hat\Lambda_G,\hat\Lambda_P,\hat\Lambda_M\approx\mathbbm{1}$ and evaluating Eqs.~\ref{eq:ExpDecayForm2+} and~\ref{eq:ExpDecayForm2-}. In both cases, we must use degenerate perturbation theory to find the approximate eigenvectors. We again find $A_b,B_b\approx 1/2$.

\subsubsection{Finding the dephasing and non-dephasing probabilities}

We can combine Eqs.~\ref{eq:Trace1} and~\ref{eq:Trace2} to evaluate the trace over $\mathcal{Z}_2$ as
\begin{equation}
    \Tr_{\mathcal{Z}_2}\left(\frac{\hat\Lambda+\hat\Lambda'}{2}\right) = 1+\lambda_{0+}+\lambda_{0-}-\kappa_{0-}\label{eq:ZTrace}
\end{equation}
and we can combine Eqs.~\ref{eq:Trace3},~\ref{eq:Trace4},~\ref{eq:Trace5}, and~\ref{eq:Trace6} to evaluate the trace over $\mathcal{P}_2\setminus\mathcal{Z}_2$ as
\begin{align}
    \Tr_{\mathcal{P}_2\setminus\mathcal{Z}_2}\left(\frac{\hat\Lambda+\hat\Lambda'}{2}\right) &= \lambda_{1+}+\kappa_{1+}+\lambda_{1-}-\kappa_{1-}+\lambda_{2+}\nonumber\\
    &\qquad+\kappa_{2+}+\lambda_{2-}+\kappa_{2-}+4.\label{eq:PmZTrace}
\end{align}
Plugging Eqs.~\ref{eq:ZTrace} and~\ref{eq:PmZTrace} into Eq.~\ref{eq:dephasingAndnon-dephasingProbabilities} for $p_\mathrm{D}$ and $p_\mathrm{ND}$ gives Eq.~\ref{eq:ZProbabilityEstimates} as desired.

\section{Dephasing/non-dephasing probabilities of a composite channel}
\label{appendix:CompositeChannels}
Given two quantum channels
\begin{align}
\Lambda_A(\rho) &= \sum_{\substack{\vec\alpha_1,\vec\beta_1\\\vec\alpha_2,\vec\beta_2}}\chi^A_{\vec\alpha_1\vec\beta_1,\vec\alpha_2\vec\beta_2}\mathrm{X}(\vec\alpha_1)\mathrm{Z}(\vec\beta_1)\rho \mathrm{Z}(\vec\beta_2)\mathrm{X}(\vec\alpha_2)\\
\Lambda_B(\rho) &= \sum_{\substack{\vec\alpha_1,\vec\beta_1\\\vec\alpha_2,\vec\beta_2}}\chi^B_{\vec\alpha_1\vec\beta_1,\vec\alpha_2\vec\beta_2}\mathrm{X}(\vec\alpha_1)\mathrm{Z}(\vec\beta_1)\rho \mathrm{Z}(\vec\beta_2)\mathrm{X}(\vec\alpha_2)
\end{align}
with dephasing and non-dephasing probabilities $p_\mathrm{D}^A$, $p_\mathrm{ND}^A$, $p_\mathrm{D}^B$, and $p_\mathrm{ND}^B$, we want to find bounds on the dephasing/non-dephasing probabilities of the combined channel $(\Lambda_A\circ\Lambda_B)$. We will denote the combined error probabilities by simply $p_\mathrm{D}$ and $p_\mathrm{ND}$.

\subsection{Finding the non-dephasing error probability}
\label{appendix:nondephasingDerivationBound}
From the definition of $p_\mathrm{ND}$, Eq. \ref{eq:Defnnon-dephasing}, we have
\begin{equation}
    p_\mathrm{ND} = \ \ \ \smashoperator{\sum_{\substack{\vec\alpha_1+\vec\gamma_1=\vec\alpha_2+\vec\gamma_2\neq\vec 0\\\vec\beta_1+\vec\delta_1=\vec\beta_2+\vec\delta_2}}}\quad (-1)^{\vec\beta_1\cdot\vec\gamma_1+\vec\beta_2\cdot\vec\gamma_2}\chi_{\vec\alpha_1\vec\beta_1,\vec\alpha_2\vec\beta_2}^A\chi_{\vec\gamma_1\vec\delta_1,\vec\gamma_2\vec\delta_2}^B.\label{eq:CombinedSumND}
\end{equation}
For legibility in what follows, we will denote the diagonal elements $\chi_{\vec\alpha\vec\beta,\vec\alpha\vec\beta}$ of the $\chi$-matrices by simply $\chi_{\vec\alpha\vec\beta}$. Note that the complete-positivity of $\Lambda_A$ and $\Lambda_B$ requires that the $\chi$-matrices are positive semidefinite, which in turn implies $|\chi_{\vec\alpha_1\vec\beta_1,\vec\alpha_2\vec\beta_2}|\leq\sqrt{\chi_{\vec\alpha_1\vec\beta_1}\chi_{\vec\alpha_2\vec\beta_2}}$~\cite{kimmel2014robust}, a fact we will use repeatedly. We will also repeatedly use two elementary inequalities, both of which are versions of the Cauchy-Schwarz inequality:
\begin{align}
    \sum_i \sqrt{a_i b_i}&\leq \sqrt{\sum_i a_i}\sqrt{\sum_i b_i}\\
    \sum_{i=1}^I \sqrt{a_i}&\leq \sqrt{I}\sqrt{\sum_i a_i}. 
\end{align}

The condition $\vec\alpha_1+\vec\gamma_1=\vec\alpha_2+\vec\gamma_2\neq\vec 0$ implies at most two of $\{\vec\alpha_1,\vec\alpha_2,\vec\gamma_1,\vec\gamma_2\}$ can be equal to $\vec 0$. We thus divide the terms in Eq. \ref{eq:CombinedSumND} into several subsets. 
\begin{enumerate}
\item Terms with $\vec\alpha_1=\vec\alpha_2=\vec 0$ (and thus $\vec\gamma_1=\vec\gamma_2\neq\vec 0$).
\item Terms with $\vec\gamma_1=\vec\gamma_2=\vec 0$ (and thus $\vec\alpha_1=\vec\alpha_2\neq\vec 0$).
\item Terms with $\vec\alpha_1=\vec\gamma_2=\vec 0$ (and thus $\vec\alpha_2=\vec\gamma_1\neq \vec 0$).
\item Terms with $\vec\alpha_2=\vec\gamma_1=\vec 0$ (and thus $\vec\alpha_1=\vec\gamma_2\neq \vec 0$).
\item Terms with $\vec\alpha_1=\vec 0$ and $\vec\alpha_2,\vec\gamma_1,\vec\gamma_2\neq \vec 0$ (and thus $\vec\gamma_2=\vec\alpha_2+\vec\gamma_1$).
\item Terms with $\vec\alpha_2=\vec 0$ and $\vec\alpha_1,\vec\gamma_1,\vec\gamma_2\neq \vec 0$ (and $\vec\gamma_2=\vec\alpha_1+\vec\gamma_1$).
\item Terms with $\vec\gamma_1=\vec 0$ and $\vec\alpha_1,\vec\alpha_2,\vec\gamma_2\neq \vec 0$ (and $\vec\gamma_2=\vec\alpha_1+\vec\alpha_2$).
\item Terms with $\vec\gamma_2=\vec 0$ and $\vec\alpha_1,\vec\alpha_2,\vec\gamma_1\neq \vec 0$ (and $\vec\gamma_1=\vec\alpha_1+\vec\alpha_2$).
\item Terms with $\vec\alpha_1,\vec\alpha_2,\vec\gamma_1,\vec\gamma_2\neq\vec 0$ (and $\vec\gamma_2=\vec\alpha_1+\vec\alpha_2+\vec\gamma_1$).
\end{enumerate}

Let's take each of these in turn. We have
\begin{align}
    (1) &= \smashoperator{\sum_{\substack{\vec\gamma_1\neq\vec 0\\\vec\beta_1,\vec\beta_2,\vec\delta_1}}}(-1)^{(\vec\beta_1+\vec\beta_2)\cdot\vec\gamma_1}\chi_{\vec 0\vec\beta_1,\vec 0\vec\beta_2}^A\chi_{\vec\gamma_1\vec\delta_1,\vec\gamma_1(\vec\beta_1+\vec\beta_2+\vec\delta_1)}^B\\
    &=\sum_{\substack{\vec\gamma_1\neq\vec 0\\\vec\delta_1}}\chi_{\vec 0\vec 0,\vec 0\vec 0}^A\chi_{\vec\gamma_1\vec\delta_1,\vec\gamma_1\vec\delta_1}^B\nonumber\\
    &\qquad+\smashoperator{\sum_{\substack{\vec\beta_2,\vec\gamma_1\neq\vec 0\\\vec\delta_1}}}(-1)^{\vec\beta_2\cdot\vec\gamma_1}\chi_{\vec 0\vec 0,\vec 0\vec \beta_2}^A\chi_{\vec\gamma_1\vec\delta_1,\vec\gamma_1(\vec\beta_2+\vec\delta_1)}^B\nonumber\\
    &\qquad +\smashoperator{\sum_{\substack{\vec\beta_1,\vec\gamma_1\neq\vec 0\\\vec\delta_1}}}(-1)^{\vec\beta_1\cdot\vec\gamma_1}\chi_{\vec 0\vec \beta_1,\vec 0\vec 0}^A\chi_{\vec\gamma_1\vec\delta_1,\vec\gamma_1(\vec\beta_1+\vec\delta_1)}^B\\
    &\qquad +\smashoperator{\sum_{\substack{\vec\beta_1,\vec\beta_2,\vec\gamma_1\neq\vec 0\\\vec\delta_1}}}(-1)^{(\vec\beta_1+\vec\beta_2)\cdot\vec\gamma_1}\chi_{\vec 0\vec \beta_1,\vec 0\vec \beta_2}^A\chi_{\vec\gamma_1\vec\delta_1,\vec\gamma_1(\vec\beta_1+\vec\beta_2+\vec\delta_1)}^B\nonumber
\end{align}
The first term is equal to $p_\mathrm{ND}^B$, up to irrelevant higher-order terms. We can bound the magnitude of the remaining terms:
\begin{align}
    \vast|\smashoperator[r]{\sum_{\substack{\vec\beta_2,\vec\gamma_1\neq\vec 0\\\vec\delta_1}}}&(-1)^{\vec\beta_2\cdot\vec\gamma_1}\chi_{\vec 0\vec 0,\vec 0\vec \beta_2}^A\chi_{\vec\gamma_1\vec\delta_1,\vec\gamma_1(\vec\beta_2+\vec\delta_1)}^B\vast|\nonumber\\
    &\leq \smashoperator{\sum_{\vec\beta_2\neq\vec 0}}\sqrt{\chi_{\vec 0\vec 0}^A\chi_{\vec 0\vec\beta_2}^A}\smashoperator{\sum_{\substack{\gamma_1\neq\vec 0\\\vec\delta_1}}}\sqrt{\chi_{\vec\gamma_1\vec\delta_1}^B\chi_{\vec\gamma_1(\vec\beta_2+\vec\delta_1)}^B}\\
    &\leq \smashoperator{\sum_{\vec\beta_2\neq\vec 0}}\sqrt{\chi_{\vec 0\vec\beta_2}^A}\smashoperator{\sum_{\substack{\gamma_1\neq\vec 0\\\vec\delta_1}}}\chi_{\vec\gamma_1\vec\delta_1}^B\\
    &\leq 2^{N/2}\sqrt{\smashoperator{\sum_{\beta_1\neq\vec 0}}\chi_{\vec 0\vec\beta_2}^A} p_\mathrm{ND}^B\\
    &=2^{N/2}\sqrt{p_\mathrm{D}^A}p_\mathrm{ND}^B
\end{align}
\begin{align}
    \vast|\smashoperator[r]{\sum_{\substack{\vec\beta_1,\vec\gamma_1\neq\vec 0\\\vec\delta_1}}}&(-1)^{\vec\beta_1\cdot\vec\gamma_1}\chi_{\vec 0\vec \beta_1,\vec 0\vec 0}^A\chi_{\vec\gamma_1\vec\delta_1,\vec\gamma_1(\vec\beta_1+\vec\delta_1)}^B\vast|\nonumber\\
    &\leq 2^{N/2}\sqrt{p_\mathrm{D}^A}p_\mathrm{ND}^B
\end{align}
\begin{align}
    \vast|&\smashoperator[r]{\sum_{\substack{\vec\beta_1,\vec\beta_2,\vec\gamma_1\neq\vec 0\\\vec\delta_1}}}(-1)^{(\vec\beta_1+\vec\beta_2)\cdot\vec\gamma_1}\chi_{\vec 0\vec \beta_1,\vec 0\vec \beta_2}^A\chi_{\vec\gamma_1\vec\delta_1,\vec\gamma_1(\vec\beta_1+\vec\beta_2+\vec\delta_1)}^B\vast|\nonumber\\
    &\leq \smashoperator{\sum_{\substack{\vec\beta_1,\vec\beta_2\neq\vec 0}}}\sqrt{\chi_{\vec 0\vec\beta_1}^A\chi_{\vec 0\vec\beta_2}^A}\smashoperator{\sum_{\substack{\vec\gamma_1\neq\vec 0\\\vec\delta_1}}}\sqrt{\chi_{\vec\gamma_1\vec\delta_1}^B\chi_{\vec\gamma_1(\vec\beta_1+\vec\beta_2+\vec\delta_1)}^B}\\
    &\leq \smashoperator{\sum_{\substack{\vec\beta_1,\vec\beta_2\neq\vec 0}}}\sqrt{\chi_{\vec 0\vec\beta_1}^A\chi_{\vec 0\vec\beta_2}^A}\smashoperator{\sum_{\substack{\vec\gamma_1\neq\vec 0\\\vec\delta_1}}}\chi_{\vec\gamma_1\vec\delta_1}^B\\   &\leq 2^N\smashoperator{\sum_{\substack{\vec\beta_1\neq\vec 0}}}\chi_{\vec 0\vec\beta_1}^A p_\mathrm{ND}^B\\ 
    &=2^Np_\mathrm{D}^Ap_\mathrm{ND}^B
\end{align}
Thus in total, we have
\begin{equation}
    \left|(1)-p_\mathrm{ND}^B\right|\leq 2^{N/2+1}\sqrt{p_\mathrm{D}^A}p_\mathrm{ND}^B+2^Np_\mathrm{D}^Ap_\mathrm{ND}^B \label{eq:1Bound}
\end{equation}
By symmetry, we similarly have
\begin{equation}
    \left|(2)-p_\mathrm{ND}^A\right|\leq 2^{N/2+1}p_\mathrm{ND}^A\sqrt{p_\mathrm{D}^B}+2^Np_\mathrm{ND}^Ap_\mathrm{D}^B \label{eq:2Bound}
\end{equation}

The remaining terms are only higher-order corrections, and so we bound their magnitudes:
\begin{align}
\left|(3)\right|&\leq \smashoperator{\sum_{\substack{\vec\alpha_2\neq\vec 0\\\vec\beta_1,\vec\beta_2,\vec\delta_2}}}\sqrt{\chi_{\vec 0\vec \beta_1}^A\chi_{\vec \alpha_2\vec\beta_2}^A\chi_{\vec\alpha_2(\vec\beta_1+\vec\beta_2+\vec\delta_2)}^B\chi_{\vec 0\vec\delta_2}^B}\\
&=\smashoperator{\sum_{\substack{\vec\alpha_2\neq\vec 0\\\vec\beta_2}}}\sqrt{\chi_{\vec 0\vec 0}^A\chi_{\vec \alpha_2\vec\beta_2}^A\chi_{\vec\alpha_2\vec\beta_2}^B\chi_{\vec 0\vec 0}^B}\nonumber\\
&\qquad +\smashoperator{\sum_{\substack{\vec\alpha_2,\vec\beta_1\neq\vec 0\\\vec\beta_2}}}\sqrt{\chi_{\vec 0\vec \beta_1}^A\chi_{\vec \alpha_2\vec\beta_2}^A\chi_{\vec\alpha_2(\vec\beta_1+\vec\beta_2)}^B\chi_{\vec 0\vec 0}^B}\nonumber\\
&\qquad +\smashoperator{\sum_{\substack{\vec\alpha_2,\vec\delta_2\neq\vec 0\\\vec\beta_2}}}\sqrt{\chi_{\vec 0\vec 0}^A\chi_{\vec \alpha_2\vec\beta_2}^A\chi_{\vec\alpha_2(\vec\beta_2+\vec\delta_2)}^B\chi_{\vec 0\vec \delta_2}^B}\nonumber\\
&\qquad +\smashoperator{\sum_{\substack{\vec\alpha_2,\vec\beta_1,\vec\delta_2\neq\vec 0\\\vec\beta_2}}}\sqrt{\chi_{\vec 0\vec \beta_1}^A\chi_{\vec \alpha_2\vec\beta_2}^A\chi_{\vec\alpha_2(\vec\beta_1+\vec\beta_2+\vec\delta_2)}^B\chi_{\vec 0\vec \delta_2}^B}.
\end{align}
We can bound each of these four terms as
\begin{align}
\smashoperator{\sum_{\substack{\vec\alpha_2\neq\vec 0\\\vec\beta_2}}}&\sqrt{\chi_{\vec 0\vec 0}^A\chi_{\vec \alpha_2\vec\beta_2}^A\chi_{\vec\alpha_2\vec\beta_2}^B\chi_{\vec 0\vec 0}^B}\nonumber\\
&\qquad\leq \smashoperator{\sum_{\substack{\vec\alpha_2\neq\vec 0\\\vec\beta_2}}}\sqrt{\chi_{\vec \alpha_2\vec\beta_2}^A\chi_{\vec\alpha_2\vec\beta_2}^B}\\
&\qquad\leq \sqrt{\smashoperator{\sum_{\substack{\vec\alpha_2\neq\vec 0\\\vec\beta_2}}}\chi_{\vec \alpha_2\vec\beta_2}^A}\sqrt{\smashoperator{\sum_{\substack{\vec\alpha_2\neq\vec 0\\\vec\beta_2}}}\chi_{\vec\alpha_2\vec\beta_2}^B}\\
&\qquad=\sqrt{p_\mathrm{ND}^Ap_\mathrm{ND}^B}
\end{align}
\begin{align}
\smashoperator{\sum_{\substack{\vec\alpha_2,\vec\beta_1\neq\vec 0\\\vec\beta_2}}}&\sqrt{\chi_{\vec 0\vec \beta_1}^A\chi_{\vec \alpha_2\vec\beta_2}^A\chi_{\vec\alpha_2(\vec\beta_1+\vec\beta_2)}^B\chi_{\vec 0\vec 0}^B}\nonumber\\
&\leq\smashoperator{\sum_{\vec\beta_1\neq\vec 0}}\sqrt{\chi_{\vec 0\vec \beta_1}^A}\smashoperator{\sum_{\substack{\vec\alpha_2\neq \vec 0\\\vec\beta_2}}}\sqrt{\chi_{\vec \alpha_2\vec\beta_2}^A\chi_{\vec\alpha_2(\vec\beta_1+\vec\beta_2)}^B}\\
&\leq\smashoperator{\sum_{\vec\beta_1\neq\vec 0}}\sqrt{\chi_{\vec 0\vec \beta_1}^A}\sqrt{\smashoperator{\sum_{\substack{\vec\alpha_2\neq \vec 0\\\vec\beta_2}}}\chi_{\vec \alpha_2\vec\beta_2}^A}\sqrt{\smashoperator{\sum_{\substack{\vec\alpha_2\neq \vec 0\\\vec\beta_2}}}\chi_{\vec\alpha_2\vec\beta_2}^B}\\
& \leq 2^{N/2}\sqrt{\smashoperator{\sum_{\vec\beta_1\neq\vec 0}}\chi_{\vec 0\vec \beta_1}^A}\sqrt{p_\mathrm{ND}^Ap_\mathrm{ND}^B}\\
& \leq 2^{N/2}\sqrt{p_\mathrm{D}^Ap_\mathrm{ND}^Ap_\mathrm{ND}^B}
\end{align}
\begin{align}
\smashoperator{\sum_{\substack{\vec\alpha_2,\vec\delta_2\neq\vec 0\\\vec\beta_2}}}&\sqrt{\chi_{\vec 0\vec 0}^A\chi_{\vec \alpha_2\vec\beta_2}^A\chi_{\vec\alpha_2(\vec\beta_2+\vec\delta_2)}^B\chi_{\vec 0\vec \delta_2}^B}\leq 2^{N/2}\sqrt{p_\mathrm{D}^Bp_\mathrm{ND}^Ap_\mathrm{ND}^B}
\end{align}
\begin{align}
\smashoperator{\sum_{\substack{\vec\alpha_2,\vec\beta_1,\vec\delta_2\neq\vec 0\\\vec\beta_2}}}&\sqrt{\chi_{\vec 0\vec \beta_1}^A\chi_{\vec \alpha_2\vec\beta_2}^A\chi_{\vec\alpha_2(\vec\beta_1+\vec\beta_2+\vec\delta_2)}^B\chi_{\vec 0\vec \delta_2}^B}\nonumber\\
&=\smashoperator{\sum_{\vec\beta_1\neq \vec 0}}\sqrt{\chi_{\vec 0\vec \beta_1}^A}\smashoperator{\sum_{\vec\delta_2\neq\vec 0}}\sqrt{\chi_{\vec 0\vec \delta_2}^B}\smashoperator{\sum_{\substack{\vec\alpha_2\neq\vec 0\\\vec\beta_2}}}\sqrt{\chi_{\vec \alpha_2\vec\beta_2}^A\chi_{\vec\alpha_2(\vec\beta_1+\vec\beta_2+\vec\delta_2)}^B} \\
&\leq \smashoperator{\sum_{\vec\beta_1\neq \vec 0}}\sqrt{\chi_{\vec 0\vec \beta_1}^A}\smashoperator{\sum_{\vec\delta_2\neq\vec 0}}\sqrt{\chi_{\vec 0\vec \delta_2}^B}\sqrt{\smashoperator{\sum_{\substack{\vec\alpha_2\neq\vec 0\\\vec\beta_2}}}\chi_{\vec \alpha_2\vec\beta_2}^A}\sqrt{\smashoperator{\sum_{\substack{\vec\alpha_2\neq\vec 0\\\vec\beta_2}}}\chi_{\vec\alpha_2\vec\beta_2}^B}\\
&\leq 2^N\sqrt{\smashoperator{\sum_{\vec\beta_1\neq \vec 0}}\chi_{\vec 0\vec \beta_1}^A}\sqrt{\smashoperator{\sum_{\vec\delta_2\neq\vec 0}}\chi_{\vec 0\vec \delta_2}^B}\sqrt{p_\mathrm{ND}^Ap_\mathrm{ND}^B}\\
&\leq 2^N\sqrt{p_\mathrm{D}^Ap_\mathrm{D}^Bp_\mathrm{ND}^Ap_\mathrm{ND}^B}.
\end{align}
Thus, in total we have
\begin{equation}
    |(3)|\leq \sqrt{p_\mathrm{ND}^Ap_\mathrm{ND}^B}\left(1+2^{N/2}\sqrt{p_\mathrm{D}^A}\right)\left(1+2^{N/2}\sqrt{p_\mathrm{D}^B}\right)\label{eq:3Bound}.
\end{equation}

By symmetry, we also have
\begin{equation}
    |(4)|\leq \sqrt{p_\mathrm{ND}^Ap_\mathrm{ND}^B}\left(1+2^{N/2}\sqrt{p_\mathrm{D}^A}\right)\left(1+2^{N/2}\sqrt{p_\mathrm{D}^B}\right)\label{eq:4Bound}.
\end{equation}

Continuing, we have
\begin{align}
    |(5)|&\leq \smashoperator{\sum_{\substack{\vec\alpha_2,\vec\gamma_2\neq\vec 0\\\vec\alpha_2\neq\vec\gamma_2\\\vec\beta_1,\vec\beta_2,\vec\delta_2}}}\sqrt{\chi_{\vec 0\vec \beta_1}^A\chi_{\vec \alpha_2\vec\beta_2}^A\chi_{(\vec\alpha_2+\vec\gamma_2)(\vec\beta_1+\vec\beta_2+\vec\delta_2)}^B\chi_{\vec\gamma_2\vec\delta_2}^B}\\
    &= \smashoperator{\sum_{\substack{\vec\alpha_2,\vec\gamma_2\neq\vec 0\\\vec\alpha_2\neq\vec\gamma_2\\\vec\beta_2,\vec\delta_2}}}\sqrt{\chi_{\vec 0\vec 0}^A\chi_{\vec \alpha_2\vec\beta_2}^A\chi_{(\vec\alpha_2+\vec\gamma_2)(\vec\beta_1+\vec\beta_2+\vec\delta_2)}^B\chi_{\vec\gamma_2\vec\delta_2}^B}\nonumber\\
    &\qquad + \smashoperator{\sum_{\substack{\vec\alpha_2,\vec\beta_1,\vec\gamma_2\neq\vec 0\\\vec\alpha_2\neq\vec\gamma_2\\\vec\beta_2,\vec\delta_2}}}\sqrt{\chi_{\vec 0\vec \beta_1}^A\chi_{\vec \alpha_2\vec\beta_2}^A\chi_{(\vec\alpha_2+\vec\gamma_2)(\vec\beta_1+\vec\beta_2+\vec\delta_2)}^B\chi_{\vec\gamma_2\vec\delta_2}^B}
\end{align}
each of which can be bounded as
\begin{align}
    \smashoperator{\sum_{\substack{\vec\alpha_2,\vec\gamma_2\neq\vec 0\\\vec\alpha_2\neq\vec\gamma_2\\\vec\beta_2,\vec\delta_2}}}&\sqrt{\chi_{\vec 0\vec 0}^A\chi_{\vec \alpha_2\vec\beta_2}^A\chi_{(\vec\alpha_2+\vec\gamma_2)(\vec\beta_1+\vec\beta_2+\vec\delta_2)}^B\chi_{\vec\gamma_2\vec\delta_2}^B}\nonumber\\
    &\leq \smashoperator{\sum_{\substack{\vec\alpha_2\neq\vec 0\\\vec\beta_2}}}\sqrt{\chi_{\vec \alpha_2\vec\beta_2}^A}\smashoperator{\sum_{\substack{\vec\gamma_2\neq\vec 0,\vec\alpha_2\\\vec\delta_2}}}\sqrt{\chi_{(\vec\alpha_2+\vec\gamma_2)(\vec\beta_1+\vec\beta_2+\vec\delta_2)}^B\chi_{\vec\gamma_2\vec\delta_2}^B}\\
    &\leq \smashoperator{\sum_{\substack{\vec\alpha_2\neq\vec 0\\\vec\beta_2}}}\sqrt{\chi_{\vec \alpha_2\vec\beta_2}^A}\smashoperator{\sum_{\substack{\vec\gamma_2\neq\vec 0\\\vec\delta_2}}}\chi_{\vec\alpha_2\vec\delta_2}^B\\
    &\leq 2^N\sqrt{\smashoperator{\sum_{\substack{\vec\alpha_2\neq\vec 0\\\vec\beta_2}}}\chi_{\vec \alpha_2\vec\beta_2}^A}p_\mathrm{ND}^B\\
    &=2^N\sqrt{p_\mathrm{ND}^A}p_\mathrm{ND}^B
\end{align}
\begin{align}
    &\smashoperator{\sum_{\substack{\vec\alpha_2,\vec\beta_1,\vec\gamma_2\neq\vec 0\\\vec\alpha_2\neq\vec\gamma_2\\\vec\beta_2,\vec\delta_2}}}\sqrt{\chi_{\vec 0\vec \beta_1}^A\chi_{\vec \alpha_2\vec\beta_2}^A\chi_{(\vec\alpha_2+\vec\gamma_2)(\vec\beta_1+\vec\beta_2+\vec\delta_2)}^B\chi_{\vec\gamma_2\vec\delta_2}^B}\nonumber\\
    &= \smashoperator{\sum_{\vec\beta_1\neq\vec 0}}\sqrt{\chi_{\vec 0\vec \beta_1}^A}\smashoperator{\sum_{\substack{\vec\alpha_2\neq\vec 0\\\vec\beta_2}}}\sqrt{\chi_{\vec \alpha_2\vec\beta_2}^A}\smashoperator{\sum_{\substack{\vec\gamma_2\neq\vec 0,\vec\alpha_2\\\vec\delta_2}}}\sqrt{\chi_{(\vec\alpha_2+\vec\gamma_2)(\vec\beta_1+\vec\beta_2+\vec\delta_2)}^B\chi_{\vec\gamma_2\vec\delta_2}^B}\\
    &\leq \smashoperator{\sum_{\vec\beta_1\neq\vec 0}}\sqrt{\chi_{\vec 0\vec \beta_1}^A}\smashoperator{\sum_{\substack{\vec\alpha_2\neq\vec 0\\\vec\beta_2}}}\sqrt{\chi_{\vec \alpha_2\vec\beta_2}^A}\smashoperator{\sum_{\substack{\vec\gamma_2\neq\vec 0\\\vec\delta_2}}}\chi_{\vec\gamma_2\vec\delta_2}^B\\
    &\leq 2^{3N/2}\sqrt{\smashoperator{\sum_{\vec\beta_1\neq\vec 0}}\chi_{\vec 0\vec \beta_1}^A}\sqrt{\smashoperator{\sum_{\substack{\vec\alpha_2\neq\vec 0\\\vec\beta_2}}}\chi_{\vec \alpha_2\vec\beta_2}^A}p_\mathrm{ND}^B\\
    &\leq 2^{3N/2}\sqrt{p_\mathrm{ND}^Ap_\mathrm{D}^A}p_\mathrm{ND}^B.
\end{align}
Thus in total,
\begin{align}
    \left|(5)\right|\leq 2^Np_\mathrm{ND}^B\sqrt{p_\mathrm{ND}^A}\left(1+2^{N/2}\sqrt{p_\mathrm{D}^A}\right)\label{eq:5bound}
\end{align}
By symmetry, we also have
\begin{align}
    \left|(6)\right|&\leq 2^Np_\mathrm{ND}^B\sqrt{p_\mathrm{ND}^A}\left(1+2^{N/2}\sqrt{p_\mathrm{D}^A}\right)\label{eq:6bound}\\
    \left|(7)\right|&\leq 2^Np_\mathrm{ND}^A\sqrt{p_\mathrm{ND}^B}\left(1+2^{N/2}\sqrt{p_\mathrm{D}^B}\right)\label{eq:7bound}\\
    \left|(8)\right|&\leq 2^Np_\mathrm{ND}^A\sqrt{p_\mathrm{ND}^B}\left(1+2^{N/2}\sqrt{p_\mathrm{D}^B}\right)\label{eq:8bound}.
\end{align}
Finally,
\begin{align}
    \left|(9)\right|&\leq \smashoperator{\sum_{\substack{\vec\alpha_1,\vec\alpha_2,\vec\gamma_2\neq\vec 0\\\vec\alpha_1+\vec\alpha_2\neq\vec\gamma_2\\\vec\beta_1,\vec\beta_2,\vec\delta_2}}}\sqrt{\chi_{\vec \alpha_1\vec \beta_1}^A\chi_{\vec \alpha_2\vec\beta_2}^A\chi_{(\vec\alpha_1+\vec\alpha_2+\vec\gamma_2)(\vec\beta_1+\vec\beta_2+\vec\delta_2)}^B\chi_{\vec\gamma_2\vec\delta_2}^B}\\
    &= \smashoperator{\sum_{\substack{\vec\alpha_1\neq\vec 0\\\vec\beta_1}}}\sqrt{\chi_{\vec \alpha_1\vec \beta_1}^A}\smashoperator{\sum_{\substack{\vec\alpha_2\neq\vec 0\\\vec\beta_2}}}\sqrt{\chi_{\vec \alpha_2\vec\beta_2}^A}\nonumber\\
    &\qquad\quad\times\smashoperator{\sum_{\substack{\vec\gamma_2\neq\vec 0,(\vec\alpha_1+\vec\alpha_2)\\\vec\delta_2}}}\sqrt{\chi_{(\vec\alpha_1+\vec\alpha_2+\vec\gamma_2)(\vec\beta_1+\vec\beta_2+\vec\delta_2)}^B\chi_{\vec\gamma_2\vec\delta_2}^B}\\
    &\leq \smashoperator{\sum_{\substack{\vec\alpha_1\neq\vec 0\\\vec\beta_1}}}\sqrt{\chi_{\vec \alpha_1\vec \beta_1}^A}\smashoperator{\sum_{\substack{\vec\alpha_2\neq\vec 0\\\vec\beta_2}}}\sqrt{\chi_{\vec \alpha_2\vec\beta_2}^A}\smashoperator{\sum_{\substack{\vec\gamma_2\neq\vec 0\\\vec\delta_2}}}\chi_{\vec\gamma_2\vec\delta_2}^B\\
    &\leq 2^{2N} \smashoperator{\sum_{\substack{\vec\alpha_1\neq\vec 0\\\vec\beta_1}}}\chi_{\vec \alpha_1\vec \beta_1}^A\chi_{\vec \alpha_2\vec\beta_2}^Ap_\mathrm{ND}^B\\
    &\leq 2^{2N}p_\mathrm{ND}^Ap_\mathrm{ND}^B\label{eq:9bound}
\end{align}

Combining the bounds given in Eqs. \ref{eq:1Bound}, \ref{eq:2Bound}, \ref{eq:3Bound}, \ref{eq:4Bound}, \ref{eq:5bound}, \ref{eq:6bound}, \ref{eq:7bound}, \ref{eq:8bound}, and \ref{eq:9bound}, we have that
\begin{equation}
    p_\mathrm{ND}=p_\mathrm{ND}^A+p_\mathrm{ND}^B +\epsilon\label{eq:pNDBoundsAppendix}
\end{equation}
where the error term $\epsilon$ is bounded by
\begin{equation}
    \begin{split}
    |\epsilon|<&2\sqrt{p_\mathrm{ND}^Ap_\mathrm{ND}^B}\\
    &+2^{N/2+1}\left[\sqrt{p_\mathrm{D}^A}p_\mathrm{ND}^B+p_\mathrm{ND}^A\sqrt{p_\mathrm{D}^B}\right.\\&\ \ \qquad\qquad\left.+\sqrt{p_\mathrm{D}^Ap_\mathrm{ND}^Ap_\mathrm{ND}^B}+\sqrt{p_\mathrm{ND}^Ap_\mathrm{D}^Bp_\mathrm{ND}^B}\right]\\
    &+2^{N}\left[p_\mathrm{D}^Ap_\mathrm{ND}^B+p_\mathrm{ND}^Ap_\mathrm{D}^B+2\sqrt{p_\mathrm{D}^Ap_\mathrm{ND}^Ap_\mathrm{D}^Bp_\mathrm{ND}^B}\right.\\&\quad\qquad\qquad\qquad\left.+2p_\mathrm{ND}^A\sqrt{p_\mathrm{D}^B}+2\sqrt{p_\mathrm{D}^A}p_\mathrm{ND}^B\right]\\
    &+2^{3N/2+1}\left[p_\mathrm{ND}^A\sqrt{p_\mathrm{D}^Bp_\mathrm{ND}^B}+\sqrt{p_\mathrm{D}^Ap_\mathrm{ND}^A}p_\mathrm{ND}^B\right]\\
    &+2^{2N}p_\mathrm{ND}^Ap_\mathrm{ND}^B.
    \end{split}
\end{equation}

While the bound on $|\epsilon|$ involves many terms, in the case where $p_\mathrm{D}^{A,B}\ll 2^{-N}$ and $p_\mathrm{ND}^{A,B}/p_\mathrm{D}^{A,B}\ll 2^{-N}$ the bound is essentially just $2\sqrt{p_\mathrm{ND}^Ap_\mathrm{ND}^B}$, which gives Eq. \ref{eq:CombinedNDBound} of the main text. This is the relevant regime for high fidelity, highly-biased channels on a small number of qubits. It is also the relevant regime for error channels in which off-diagonal elements of the $\chi$-matrix are exponentially suppressed in the weight of the Pauli errors, which is likely the case for error-correction circuits as they tend to decohere noise \cite{huang2019performance,beale2018quantum,bravyi2018correcting,geller2013efficient}.

\subsection{Finding the dephasing error probability}

From the definition of $p_\mathrm{D}$, Eq. \ref{eq:DefnDephasing}, we have
\begin{equation}
    p_\mathrm{D} = \ \ \ \smashoperator{\sum_{\substack{\vec\alpha_1+\vec\gamma_1=\vec\alpha_2+\vec\gamma_2=\vec 0\\\vec\beta_1+\vec\delta_1=\vec\beta_2+\vec\delta_2}}}\quad (-1)^{\vec\beta_1\cdot\vec\gamma_1+\vec\beta_2\cdot\vec\gamma_2}\chi_{\vec\alpha_1\vec\beta_1,\vec\alpha_2\vec\beta_2}^A\chi_{\vec\gamma_1\vec\delta_1,\vec\gamma_2\vec\delta_2}^B.
\end{equation}

We could similarly divide the terms in this sum into categories and bound them, as we did for $p_\mathrm{ND}$ above. However, if we assume $p_\mathrm{D}^{A,B}\gg p_\mathrm{ND}^{A,B}$, we can instead approximate $p_\mathrm{D}$ by $p=p_\mathrm{D}+p_\mathrm{ND}$, the total error probability of the channel. We then use the known result \cite[Theorem 1]{carignan2019bounding}
\begin{equation}
    p=p^A+p^B+\epsilon,\qquad |\epsilon|\leq 2\sqrt{p^Ap^B}.\label{eq:pDBoundsAppendix}
\end{equation}

\subsection{Extracting the dephasing and non-dephasing probabilities in IBRB}
\label{appendix:InterleavedBounds}

Given the estimates of $p_\mathrm{D}$ and $p_\mathrm{ND}$ given in Eqs. \ref{eq:pDBoundsAppendix} and \ref{eq:pNDBoundsAppendix}, respectively, we can reverse these equations to try and estimate $p_\mathrm{D}^A$ and $p_\mathrm{ND}^A$ from $p_\mathrm{D}^B$, $p_\mathrm{ND}^B$, $p_\mathrm{D}$, and $p_\mathrm{ND}$. This is relevant if $\Lambda_A$ is the error channel associated to our gate of interest and $\Lambda_B$ is the error channel of our interleaving group. For convenience, we'll assume we're in the regime where we can neglect the error terms proportional to $2^{N/2}$ and higher powers. Rearranging Eqs. \ref{eq:pDBoundsAppendix} and \ref{eq:pNDBoundsAppendix} gives

\begin{align}
p_\mathrm{D}^A&=p_\mathrm{D}+p_\mathrm{D}^B+\epsilon,\qquad &|\epsilon|<2\sqrt{p_\mathrm{D}p_\mathrm{D}^B}\\
p_\mathrm{ND}^A&=p_\mathrm{ND}+p_\mathrm{ND}^B+\epsilon, &|\epsilon|<2\sqrt{p_\mathrm{ND}p_\mathrm{ND}^B}.
\end{align}

We note that in our particular case, it seems difficult to extract $p_\mathrm{D}^B$ and $p_\mathrm{ND}^B$ for the group $\mathcal{Z}_2$, as the Liouville representation of this group has high multiplicity. This is why we prefer the randomized compiling interpretation of $p_\mathrm{D}$ and $p_\mathrm{ND}$ given in the main text.

\section{Randomized compiling with $Z$ randomization}
\label{appendix:RandomizedCompiling}

In this section, we explain the randomized compiling by $\mathcal{Z}_N$ in more detail, and prove that it ensures non-dephasing errors increase linearly under composition (Eq. \ref{eq:randomizedCompilingCombined} of the main text).

Given an $N$-qubit bias-preserving Clifford circuit composed of gates $C_1$, $C_2$, \dots, $C_n$, a noisy implementation of the circuit is given by
\begin{equation}
    \hat C_n\hat\Lambda_n\cdots\hat C_2\hat\Lambda_2\hat C_1\hat\Lambda_1
\end{equation}
where $\hat \Lambda_1$, $\hat\Lambda_2$, \dots, $\hat\Lambda_n$ are the associated error channels. Let's assume that we can implement arbitrary elements of $\mathcal{Z}_N$ with a gate-independent error channel $\hat\Lambda_G$, where $\hat\Lambda_G$ is negligible red to the Clifford errors. We take the convention that the In this case, we can interleave randomly chosen gates $U\in\mathcal{Z}_N$ between the Clifford elements without increasing the error. Note that because the circuit elements are Clifford, the effect of interleaving $U$ can be corrected by an efficiently computable Pauli correction operator. We'll denote the correction operator by $U_{n+1}$. Note that because the Cliffords preserve the bias, we have $U_{n+1}\in\mathcal{Z}_N$.

The resulting noisy circuit is then
\begin{equation}
    \hat\Lambda_G\hat U_{n+1}\hat C_n\hat\Lambda_n\hat\Lambda_G\hat U_n\cdots \hat C_2\hat\Lambda_2\hat\Lambda_G\hat U_2\hat C_1\hat\Lambda_1\hat\Lambda_G\hat U_1.
\end{equation}
Because the Cliffords preserve the bias, commuting any element of $\mathcal{Z}_N$ past a Clifford results in another element of $\mathcal{Z}_N$. We can thus rewrite the circuit as
\begin{equation}
    \hat\Lambda_G \hat C_n\hat V_n^\dagger \hat\Lambda_n\hat\Lambda_G\hat V_n\cdots \hat C_2\hat V_2^\dagger\hat\Lambda_2\hat\Lambda_G\hat V_2\hat C_1\hat V_1^\dagger\hat\Lambda_1\hat\Lambda_G\hat V_1.
\end{equation}
for some $V_i\in\mathcal{Z}_N$ that are also distributed uniformly over $\mathcal{Z}_N$. Taking the expectation value over $\hat V_1,\dots\hat V_n$ results in an effective circuit of the form
\begin{equation}
    \hat\Lambda_G\hat C_n\hat\Lambda_n^T\cdots \hat C_2\hat\Lambda_2^T\hat C_1\hat \Lambda_1^T.
\end{equation}
where the twirled error channel $\Lambda_i^T$ is given by
\begin{equation}
    \hat\Lambda_i^T=\mathop{\mathbbm{E}}_{V\in\mathcal{Z}_N}\left[\hat V^\dagger\hat\Lambda_i\hat\Lambda_G\hat V\right].
\end{equation}

The twirled version of the error channels is highly simplified compared to the original error channel. In terms of the $\chi$-matrix, if the original error channel is given by
\begin{align}
\left(\Lambda_i\circ\Lambda_G\right)(\rho) = \sum_{\substack{\vec\alpha_1,\vec\beta_1\\\vec\alpha_2,\vec\beta_2}}\chi_{\vec\alpha_1\vec\beta_1,\vec\alpha_2\vec\beta_2}\mathrm{X}(\vec\alpha_1)\mathrm{Z}(\vec\beta_1)\rho \mathrm{Z}(\vec\beta_2)\mathrm{X}(\vec\alpha_2)
\end{align}
then the twirled error channel is given by
\begin{align}
\Lambda_i^T(\rho) = \sum_{\substack{\vec\alpha\\\vec\beta_1,\vec\beta_2}}\chi_{\vec\alpha\vec\beta_1,\vec\alpha\vec\beta_2}\mathrm{X}(\vec\alpha)\mathrm{Z}(\vec\beta_1)\rho \mathrm{Z}(\vec\beta_2)\mathrm{X}(\vec\alpha)
\end{align}
where the off-diagonal elements of the $\chi$-matrix with $\vec\alpha_1\neq\vec\alpha_2$ are set to zero.

If we consider the composition of two twirled error channels $(\Lambda_A^T\circ\Lambda_B^T)$, we can again estimate the non-dephasing probability of the composition in terms of the dephasing and non-dephasing probabilities of $\Lambda_A$ and $\Lambda_B$. In evaluating the sum in Section \ref{appendix:nondephasingDerivationBound}, the fact that the error channels are twirled means the only nonzero terms in the sum have $\vec\alpha_1=\vec\alpha_2$ and $\vec\gamma_1=\vec\gamma_2$. These leaves the sum over subsets $(1)$ and $(2)$ unchanged, makes the sum over subsets $(3)$-$(8)$ zero, and lets us reevaluate the sum over subset $(9)$ as

\begin{align}
    \left|(9)\right|&\leq \smashoperator{\sum_{\substack{\vec\alpha,\vec\gamma\neq\vec 0\\\vec\beta_1,\vec\beta_2,\vec\delta_2}}}\sqrt{\chi_{\vec \alpha\vec \beta_1}^A\chi_{\vec \alpha\vec\beta_2}^A\chi_{\vec\gamma(\vec\beta_1+\vec\beta_2+\vec\delta_2)}^B\chi_{\vec\gamma\vec\delta_2}^B}\\
    &= \smashoperator{\sum_{\substack{\vec\alpha\neq\vec 0\\\vec\beta_1}}}\sqrt{\chi_{\vec \alpha\vec \beta_1}^A}\smashoperator{\sum_{\substack{\vec\beta_2}}}\sqrt{\chi_{\vec \alpha\vec\beta_2}^A}\smashoperator{\sum_{\substack{\vec\gamma\neq\vec 0\\\vec\delta_2}}}\sqrt{\chi_{\vec\gamma(\vec\beta_1+\vec\beta_2+\vec\delta_2)}^B\chi_{\vec\gamma\vec\delta_2}^B}\\
    &\leq \smashoperator{\sum_{\substack{\vec\alpha\neq\vec 0\\\vec\beta_1}}}\sqrt{\chi_{\vec \alpha\vec \beta_1}^A}\smashoperator{\sum_{\substack{\vec\beta_2}}}\sqrt{\chi_{\vec \alpha\vec\beta_2}^A}\smashoperator{\sum_{\substack{\vec\gamma\neq\vec 0\\\vec\delta_2}}}\chi_{\vec\gamma\vec\delta_2}^B\\
    &\leq 2^{N}\smashoperator{\sum_{\substack{\vec\alpha\neq\vec 0}}}\sqrt{\smashoperator{\sum_{\vec\beta_1}}\chi_{\vec \alpha\vec \beta_1}^A}\sqrt{\smashoperator{\sum_{\vec\beta_1}}\chi_{\vec \alpha\vec\beta_2}^A}p_\mathrm{ND}^B\\
    &\leq 2^{N}\smashoperator{\sum_{\substack{\vec\alpha\neq\vec 0\\\vec\beta_1}}}\chi_{\vec \alpha\vec \beta_1}^Ap_\mathrm{ND}^B\\
    &\leq 2^N p_\mathrm{ND}^Ap_\mathrm{ND}^B.\label{eq:9boundNew}
\end{align}

Thus by combining Eqs. \ref{eq:1Bound}, \ref{eq:2Bound}, and \ref{eq:9boundNew}, we have
\begin{align}
    p_\mathrm{ND}&=p_\mathrm{ND}^A+p_\mathrm{ND}^B+\epsilon\\
    |\epsilon|&\leq 2^{N/2+1}\left[\sqrt{p_\mathrm{D}^A}p_\mathrm{ND}^B+p_\mathrm{ND}^A\sqrt{p_\mathrm{D}^B}\right]\nonumber\\
    &\qquad +2^N\left[p_\mathrm{ND}^Ap_\mathrm{D}^B+p_\mathrm{D}^Ap_\mathrm{ND}^B+p_\mathrm{ND}^Ap_\mathrm{ND}^B\right].
\end{align}
In the regime of high-fidelity gates on a small number of qubits, $|\epsilon|$ is negligible, which gives Eq. \ref{eq:randomizedCompilingCombined} in the main text.

In contrast, randomized $\mathcal{Z}_2$ compiling produces no notable improvement to the bounds on $p_\mathrm{D}$ for a highly biased noise channel, as in this case the dominant uncertainty in $p_\mathrm{D}$ comes from off-diagonal elements of the $\chi$-matrices $\chi^A$ and $\chi^B$ with $\vec\alpha_1=\vec\alpha_2=\vec\gamma_1=\vec\gamma_2=\vec 0$, which are not affected by twirling.

\section{Generating random biased-noise error channels}
\label{sec:RandomErrors}
Here, we give more details on our procedure to generate random biased-noise error channels to use in our simulation data. We make no claim that this procedure is optimal or generates all realistic error channels; our goal was simply to generate error channels that were both biased and not Pauli-diagonal to illustrate the power of our BRB methods.

We first randomly choose the number $d=1,...,4^N$ of Kraus operators to include, as well as the approximate target probabilities $p_\mathrm{D}$ and $p_\mathrm{ND}$ for the channel. For $i=1,\dots,(d-1)$, we choose each Kraus operator to be either dephasing or non-dephasing with probability $1/2$, and set it to
\begin{equation}
    K_i := \left\{\begin{aligned}
     &\sqrt{\frac{10p_\mathrm{D}}{d}}\sum_{\vec\beta} c_{\vec\beta}\mathrm{Z}(\vec\beta),
     & \ \text{dephasing}\\
     &\sqrt{\frac{10p_\mathrm{ND}}{d}}\sum_{\substack{\vec\alpha\neq\vec 0\\\vec\beta}} c_{\vec\alpha,\vec\beta}\mathrm{X}(\vec\alpha)\mathrm{Z}(\vec\beta),
     & \ \text{non-dephasing}
     \end{aligned}\right.
\end{equation}
where each $c$ is generated by choosing a uniform $r\in[0,1]$, $\theta\in[0,2\pi]$, and setting $c=re^{i\theta}$. The factor of $10$ was inserted ``by hand" to make the resulting error channel approximately have the desired dephasing/non-dephasing probabilities.

Finally, having defined $K_i$ for $i=1,\dots,(d-1)$, we define $K_d$ to be a matrix satisfying
\begin{align}
    K_d^\dagger K_d = \mathbbm{1}-\sum_{i=1}^{d-1}K_i^\dagger K_i
\end{align}
to ensure the overall channel is trace-preserving. While this description of $K_d$ is not unique, one matrix satisfying this equation is given by the Cholesky decomposition of $(\mathbbm{1}-\sum_{i}K_i^\dagger K_i)$ \cite{golub2013matrix}, which is what we used in our simulations.

\bibliography{thebibliography.bib}

%apsrev4-2.bst 2019-01-14 (MD) hand-edited version of apsrev4-1.bst
%Control: key (0)
%Control: author (8) initials jnrlst
%Control: editor formatted (1) identically to author
%Control: production of article title (0) allowed
%Control: page (0) single
%Control: year (1) truncated
%Control: production of eprint (0) enabled
\begin{thebibliography}{64}%
\makeatletter
\providecommand \@ifxundefined [1]{%
 \@ifx{#1\undefined}
}%
\providecommand \@ifnum [1]{%
 \ifnum #1\expandafter \@firstoftwo
 \else \expandafter \@secondoftwo
 \fi
}%
\providecommand \@ifx [1]{%
 \ifx #1\expandafter \@firstoftwo
 \else \expandafter \@secondoftwo
 \fi
}%
\providecommand \natexlab [1]{#1}%
\providecommand \enquote  [1]{``#1''}%
\providecommand \bibnamefont  [1]{#1}%
\providecommand \bibfnamefont [1]{#1}%
\providecommand \citenamefont [1]{#1}%
\providecommand \href@noop [0]{\@secondoftwo}%
\providecommand \href [0]{\begingroup \@sanitize@url \@href}%
\providecommand \@href[1]{\@@startlink{#1}\@@href}%
\providecommand \@@href[1]{\endgroup#1\@@endlink}%
\providecommand \@sanitize@url [0]{\catcode `\\12\catcode `\$12\catcode
  `\&12\catcode `\#12\catcode `\^12\catcode `\_12\catcode `\%12\relax}%
\providecommand \@@startlink[1]{}%
\providecommand \@@endlink[0]{}%
\providecommand \url  [0]{\begingroup\@sanitize@url \@url }%
\providecommand \@url [1]{\endgroup\@href {#1}{\urlprefix }}%
\providecommand \urlprefix  [0]{URL }%
\providecommand \Eprint [0]{\href }%
\providecommand \doibase [0]{https://doi.org/}%
\providecommand \selectlanguage [0]{\@gobble}%
\providecommand \bibinfo  [0]{\@secondoftwo}%
\providecommand \bibfield  [0]{\@secondoftwo}%
\providecommand \translation [1]{[#1]}%
\providecommand \BibitemOpen [0]{}%
\providecommand \bibitemStop [0]{}%
\providecommand \bibitemNoStop [0]{.\EOS\space}%
\providecommand \EOS [0]{\spacefactor3000\relax}%
\providecommand \BibitemShut  [1]{\csname bibitem#1\endcsname}%
\let\auto@bib@innerbib\@empty
%</preamble>
\bibitem [{\citenamefont {Stephens}\ \emph {et~al.}(2013)\citenamefont
  {Stephens}, \citenamefont {Munro},\ and\ \citenamefont
  {Nemoto}}]{stephens2013high}%
  \BibitemOpen
  \bibfield  {author} {\bibinfo {author} {\bibfnamefont {A.~M.}\ \bibnamefont
  {Stephens}}, \bibinfo {author} {\bibfnamefont {W.~J.}\ \bibnamefont
  {Munro}},\ and\ \bibinfo {author} {\bibfnamefont {K.}~\bibnamefont
  {Nemoto}},\ }\bibfield  {title} {\bibinfo {title} {High-threshold topological
  quantum error correction against biased noise},\ }\href@noop {} {\bibfield
  {journal} {\bibinfo  {journal} {Phys. Rev. A}\ }\textbf {\bibinfo {volume}
  {88}},\ \bibinfo {pages} {060301} (\bibinfo {year} {2013})}\BibitemShut
  {NoStop}%
\bibitem [{\citenamefont {Tuckett}\ \emph {et~al.}(2018)\citenamefont
  {Tuckett}, \citenamefont {Bartlett},\ and\ \citenamefont
  {Flammia}}]{tuckett2018ultrahigh}%
  \BibitemOpen
  \bibfield  {author} {\bibinfo {author} {\bibfnamefont {D.~K.}\ \bibnamefont
  {Tuckett}}, \bibinfo {author} {\bibfnamefont {S.~D.}\ \bibnamefont
  {Bartlett}},\ and\ \bibinfo {author} {\bibfnamefont {S.~T.}\ \bibnamefont
  {Flammia}},\ }\bibfield  {title} {\bibinfo {title} {Ultrahigh error threshold
  for surface codes with biased noise},\ }\href@noop {} {\bibfield  {journal}
  {\bibinfo  {journal} {Phys. Rev. Lett.}\ }\textbf {\bibinfo {volume} {120}},\
  \bibinfo {pages} {050505} (\bibinfo {year} {2018})}\BibitemShut {NoStop}%
\bibitem [{\citenamefont {Tuckett}\ \emph {et~al.}(2019)\citenamefont
  {Tuckett}, \citenamefont {Darmawan}, \citenamefont {Chubb}, \citenamefont
  {Bravyi}, \citenamefont {Bartlett},\ and\ \citenamefont
  {Flammia}}]{tuckett2019tailoring}%
  \BibitemOpen
  \bibfield  {author} {\bibinfo {author} {\bibfnamefont {D.~K.}\ \bibnamefont
  {Tuckett}}, \bibinfo {author} {\bibfnamefont {A.~S.}\ \bibnamefont
  {Darmawan}}, \bibinfo {author} {\bibfnamefont {C.~T.}\ \bibnamefont {Chubb}},
  \bibinfo {author} {\bibfnamefont {S.}~\bibnamefont {Bravyi}}, \bibinfo
  {author} {\bibfnamefont {S.~D.}\ \bibnamefont {Bartlett}},\ and\ \bibinfo
  {author} {\bibfnamefont {S.~T.}\ \bibnamefont {Flammia}},\ }\bibfield
  {title} {\bibinfo {title} {Tailoring surface codes for highly biased noise},\
  }\href@noop {} {\bibfield  {journal} {\bibinfo  {journal} {Phys. Rev. X}\
  }\textbf {\bibinfo {volume} {9}},\ \bibinfo {pages} {041031} (\bibinfo {year}
  {2019})}\BibitemShut {NoStop}%
\bibitem [{\citenamefont {Tuckett}\ \emph {et~al.}(2020)\citenamefont
  {Tuckett}, \citenamefont {Bartlett}, \citenamefont {Flammia},\ and\
  \citenamefont {Brown}}]{tuckett2020fault}%
  \BibitemOpen
  \bibfield  {author} {\bibinfo {author} {\bibfnamefont {D.~K.}\ \bibnamefont
  {Tuckett}}, \bibinfo {author} {\bibfnamefont {S.~D.}\ \bibnamefont
  {Bartlett}}, \bibinfo {author} {\bibfnamefont {S.~T.}\ \bibnamefont
  {Flammia}},\ and\ \bibinfo {author} {\bibfnamefont {B.~J.}\ \bibnamefont
  {Brown}},\ }\bibfield  {title} {\bibinfo {title} {Fault-tolerant thresholds
  for the surface code in excess of 5\% under biased noise},\ }\href@noop {}
  {\bibfield  {journal} {\bibinfo  {journal} {Phys. Rev. Lett.}\ }\textbf
  {\bibinfo {volume} {124}},\ \bibinfo {pages} {130501} (\bibinfo {year}
  {2020})}\BibitemShut {NoStop}%
\bibitem [{\citenamefont {{Bonilla Ataides}}\ \emph {et~al.}(2021)\citenamefont
  {{Bonilla Ataides}}, \citenamefont {Tuckett}, \citenamefont {Bartlett},
  \citenamefont {Flammia},\ and\ \citenamefont {Brown}}]{ataides2021xzzx}%
  \BibitemOpen
  \bibfield  {author} {\bibinfo {author} {\bibfnamefont {J.~P.}\ \bibnamefont
  {{Bonilla Ataides}}}, \bibinfo {author} {\bibfnamefont {D.~K.}\ \bibnamefont
  {Tuckett}}, \bibinfo {author} {\bibfnamefont {S.~D.}\ \bibnamefont
  {Bartlett}}, \bibinfo {author} {\bibfnamefont {S.~T.}\ \bibnamefont
  {Flammia}},\ and\ \bibinfo {author} {\bibfnamefont {B.~J.}\ \bibnamefont
  {Brown}},\ }\bibfield  {title} {\bibinfo {title} {The {XZZX} surface code},\
  }\href@noop {} {\bibfield  {journal} {\bibinfo  {journal} {Nat. Commun.}\
  }\textbf {\bibinfo {volume} {12}},\ \bibinfo {pages} {1} (\bibinfo {year}
  {2021})}\BibitemShut {NoStop}%
\bibitem [{\citenamefont {Dua}\ \emph {et~al.}(2022)\citenamefont {Dua},
  \citenamefont {Kubica}, \citenamefont {Jiang}, \citenamefont {Flammia},\ and\
  \citenamefont {Gullans}}]{dua2022clifford}%
  \BibitemOpen
  \bibfield  {author} {\bibinfo {author} {\bibfnamefont {A.}~\bibnamefont
  {Dua}}, \bibinfo {author} {\bibfnamefont {A.}~\bibnamefont {Kubica}},
  \bibinfo {author} {\bibfnamefont {L.}~\bibnamefont {Jiang}}, \bibinfo
  {author} {\bibfnamefont {S.~T.}\ \bibnamefont {Flammia}},\ and\ \bibinfo
  {author} {\bibfnamefont {M.~J.}\ \bibnamefont {Gullans}},\ }\bibfield
  {title} {\bibinfo {title} {Clifford-deformed surface codes},\ }\href@noop {}
  {\bibfield  {journal} {\bibinfo  {journal} {arXiv preprint arXiv:2201.07802}\
  } (\bibinfo {year} {2022})}\BibitemShut {NoStop}%
\bibitem [{\citenamefont {Claes}\ \emph {et~al.}(2022)\citenamefont {Claes},
  \citenamefont {Bourassa},\ and\ \citenamefont {Puri}}]{claes2022tailored}%
  \BibitemOpen
  \bibfield  {author} {\bibinfo {author} {\bibfnamefont {J.}~\bibnamefont
  {Claes}}, \bibinfo {author} {\bibfnamefont {J.~E.}\ \bibnamefont
  {Bourassa}},\ and\ \bibinfo {author} {\bibfnamefont {S.}~\bibnamefont
  {Puri}},\ }\bibfield  {title} {\bibinfo {title} {Tailored cluster states with
  high threshold under biased noise},\ }\href@noop {} {\bibfield  {journal}
  {\bibinfo  {journal} {arXiv preprint arXiv:2201.10566}\ } (\bibinfo {year}
  {2022})}\BibitemShut {NoStop}%
\bibitem [{\citenamefont {Darmawan}\ \emph {et~al.}(2021)\citenamefont
  {Darmawan}, \citenamefont {Brown}, \citenamefont {Grimsmo}, \citenamefont
  {Tuckett},\ and\ \citenamefont {Puri}}]{darmawan2021practical}%
  \BibitemOpen
  \bibfield  {author} {\bibinfo {author} {\bibfnamefont {A.~S.}\ \bibnamefont
  {Darmawan}}, \bibinfo {author} {\bibfnamefont {B.~J.}\ \bibnamefont {Brown}},
  \bibinfo {author} {\bibfnamefont {A.~L.}\ \bibnamefont {Grimsmo}}, \bibinfo
  {author} {\bibfnamefont {D.~K.}\ \bibnamefont {Tuckett}},\ and\ \bibinfo
  {author} {\bibfnamefont {S.}~\bibnamefont {Puri}},\ }\bibfield  {title}
  {\bibinfo {title} {Practical quantum error correction with the {XZZX} code
  and {K}err-cat qubits},\ }\href@noop {} {\bibfield  {journal} {\bibinfo
  {journal} {PRX Quantum}\ }\textbf {\bibinfo {volume} {2}},\ \bibinfo {pages}
  {030345} (\bibinfo {year} {2021})}\BibitemShut {NoStop}%
\bibitem [{\citenamefont {Chamberland}\ \emph {et~al.}(2022)\citenamefont
  {Chamberland}, \citenamefont {Noh}, \citenamefont {Arrangoiz-Arriola},
  \citenamefont {Campbell}, \citenamefont {Hann}, \citenamefont {Iverson},
  \citenamefont {Putterman}, \citenamefont {Bohdanowicz}, \citenamefont
  {Flammia}, \citenamefont {Keller} \emph {et~al.}}]{chamberland2022building}%
  \BibitemOpen
  \bibfield  {author} {\bibinfo {author} {\bibfnamefont {C.}~\bibnamefont
  {Chamberland}}, \bibinfo {author} {\bibfnamefont {K.}~\bibnamefont {Noh}},
  \bibinfo {author} {\bibfnamefont {P.}~\bibnamefont {Arrangoiz-Arriola}},
  \bibinfo {author} {\bibfnamefont {E.~T.}\ \bibnamefont {Campbell}}, \bibinfo
  {author} {\bibfnamefont {C.~T.}\ \bibnamefont {Hann}}, \bibinfo {author}
  {\bibfnamefont {J.}~\bibnamefont {Iverson}}, \bibinfo {author} {\bibfnamefont
  {H.}~\bibnamefont {Putterman}}, \bibinfo {author} {\bibfnamefont {T.~C.}\
  \bibnamefont {Bohdanowicz}}, \bibinfo {author} {\bibfnamefont {S.~T.}\
  \bibnamefont {Flammia}}, \bibinfo {author} {\bibfnamefont {A.}~\bibnamefont
  {Keller}}, \emph {et~al.},\ }\bibfield  {title} {\bibinfo {title} {Building a
  fault-tolerant quantum computer using concatenated cat codes},\ }\href@noop
  {} {\bibfield  {journal} {\bibinfo  {journal} {PRX Quantum}\ }\textbf
  {\bibinfo {volume} {3}},\ \bibinfo {pages} {010329} (\bibinfo {year}
  {2022})}\BibitemShut {NoStop}%
\bibitem [{\citenamefont {Chamberland}\ and\ \citenamefont
  {Campbell}(2022)}]{chamberland2022universal}%
  \BibitemOpen
  \bibfield  {author} {\bibinfo {author} {\bibfnamefont {C.}~\bibnamefont
  {Chamberland}}\ and\ \bibinfo {author} {\bibfnamefont {E.~T.}\ \bibnamefont
  {Campbell}},\ }\bibfield  {title} {\bibinfo {title} {Universal quantum
  computing with twist-free and temporally encoded lattice surgery},\
  }\href@noop {} {\bibfield  {journal} {\bibinfo  {journal} {PRX Quantum}\
  }\textbf {\bibinfo {volume} {3}},\ \bibinfo {pages} {010331} (\bibinfo {year}
  {2022})}\BibitemShut {NoStop}%
\bibitem [{\citenamefont {Aliferis}\ and\ \citenamefont
  {Preskill}(2008)}]{aliferis2008fault}%
  \BibitemOpen
  \bibfield  {author} {\bibinfo {author} {\bibfnamefont {P.}~\bibnamefont
  {Aliferis}}\ and\ \bibinfo {author} {\bibfnamefont {J.}~\bibnamefont
  {Preskill}},\ }\bibfield  {title} {\bibinfo {title} {Fault-tolerant quantum
  computation against biased noise},\ }\href@noop {} {\bibfield  {journal}
  {\bibinfo  {journal} {Phys. Rev. A}\ }\textbf {\bibinfo {volume} {78}},\
  \bibinfo {pages} {052331} (\bibinfo {year} {2008})}\BibitemShut {NoStop}%
\bibitem [{\citenamefont {Puri}\ \emph {et~al.}(2020)\citenamefont {Puri},
  \citenamefont {St-Jean}, \citenamefont {Gross}, \citenamefont {Grimm},
  \citenamefont {Frattini}, \citenamefont {Iyer}, \citenamefont {Krishna},
  \citenamefont {Touzard}, \citenamefont {Jiang}, \citenamefont {Blais} \emph
  {et~al.}}]{puri2020bias}%
  \BibitemOpen
  \bibfield  {author} {\bibinfo {author} {\bibfnamefont {S.}~\bibnamefont
  {Puri}}, \bibinfo {author} {\bibfnamefont {L.}~\bibnamefont {St-Jean}},
  \bibinfo {author} {\bibfnamefont {J.~A.}\ \bibnamefont {Gross}}, \bibinfo
  {author} {\bibfnamefont {A.}~\bibnamefont {Grimm}}, \bibinfo {author}
  {\bibfnamefont {N.~E.}\ \bibnamefont {Frattini}}, \bibinfo {author}
  {\bibfnamefont {P.~S.}\ \bibnamefont {Iyer}}, \bibinfo {author}
  {\bibfnamefont {A.}~\bibnamefont {Krishna}}, \bibinfo {author} {\bibfnamefont
  {S.}~\bibnamefont {Touzard}}, \bibinfo {author} {\bibfnamefont
  {L.}~\bibnamefont {Jiang}}, \bibinfo {author} {\bibfnamefont
  {A.}~\bibnamefont {Blais}}, \emph {et~al.},\ }\bibfield  {title} {\bibinfo
  {title} {Bias-preserving gates with stabilized cat qubits},\ }\href@noop {}
  {\bibfield  {journal} {\bibinfo  {journal} {Sci. Adv.}\ }\textbf {\bibinfo
  {volume} {6}},\ \bibinfo {pages} {5901} (\bibinfo {year} {2020})}\BibitemShut
  {NoStop}%
\bibitem [{\citenamefont {Guillaud}\ and\ \citenamefont
  {Mirrahimi}(2019)}]{guillaud2019repetition}%
  \BibitemOpen
  \bibfield  {author} {\bibinfo {author} {\bibfnamefont {J.}~\bibnamefont
  {Guillaud}}\ and\ \bibinfo {author} {\bibfnamefont {M.}~\bibnamefont
  {Mirrahimi}},\ }\bibfield  {title} {\bibinfo {title} {Repetition cat qubits
  for fault-tolerant quantum computation},\ }\href@noop {} {\bibfield
  {journal} {\bibinfo  {journal} {Phys. Rev. X}\ }\textbf {\bibinfo {volume}
  {9}},\ \bibinfo {pages} {041053} (\bibinfo {year} {2019})}\BibitemShut
  {NoStop}%
\bibitem [{\citenamefont {Cong}\ \emph {et~al.}(2021)\citenamefont {Cong},
  \citenamefont {Wang}, \citenamefont {Levine}, \citenamefont {Keesling},\ and\
  \citenamefont {Lukin}}]{cong2021hardware}%
  \BibitemOpen
  \bibfield  {author} {\bibinfo {author} {\bibfnamefont {I.}~\bibnamefont
  {Cong}}, \bibinfo {author} {\bibfnamefont {S.-T.}\ \bibnamefont {Wang}},
  \bibinfo {author} {\bibfnamefont {H.}~\bibnamefont {Levine}}, \bibinfo
  {author} {\bibfnamefont {A.}~\bibnamefont {Keesling}},\ and\ \bibinfo
  {author} {\bibfnamefont {M.~D.}\ \bibnamefont {Lukin}},\ }\bibfield  {title}
  {\bibinfo {title} {Hardware-efficient, fault-tolerant quantum computation
  with {R}ydberg atoms},\ }\href@noop {} {\bibfield  {journal} {\bibinfo
  {journal} {arXiv preprint arXiv:2105.13501}\ } (\bibinfo {year}
  {2021})}\BibitemShut {NoStop}%
\bibitem [{\citenamefont {Xu}\ \emph {et~al.}(2021)\citenamefont {Xu},
  \citenamefont {Iverson}, \citenamefont {Brandao},\ and\ \citenamefont
  {Jiang}}]{xu2021engineering}%
  \BibitemOpen
  \bibfield  {author} {\bibinfo {author} {\bibfnamefont {Q.}~\bibnamefont
  {Xu}}, \bibinfo {author} {\bibfnamefont {J.~K.}\ \bibnamefont {Iverson}},
  \bibinfo {author} {\bibfnamefont {F.~G.}\ \bibnamefont {Brandao}},\ and\
  \bibinfo {author} {\bibfnamefont {L.}~\bibnamefont {Jiang}},\ }\bibfield
  {title} {\bibinfo {title} {Engineering fast bias-preserving gates on
  stabilized cat qubits},\ }\href@noop {} {\bibfield  {journal} {\bibinfo
  {journal} {arXiv preprint arXiv:2105.13908}\ } (\bibinfo {year}
  {2021})}\BibitemShut {NoStop}%
\bibitem [{\citenamefont {Magesan}\ \emph {et~al.}(2011)\citenamefont
  {Magesan}, \citenamefont {Gambetta},\ and\ \citenamefont
  {Emerson}}]{magesan2011scalable}%
  \BibitemOpen
  \bibfield  {author} {\bibinfo {author} {\bibfnamefont {E.}~\bibnamefont
  {Magesan}}, \bibinfo {author} {\bibfnamefont {J.~M.}\ \bibnamefont
  {Gambetta}},\ and\ \bibinfo {author} {\bibfnamefont {J.}~\bibnamefont
  {Emerson}},\ }\bibfield  {title} {\bibinfo {title} {Scalable and robust
  randomized benchmarking of quantum processes},\ }\href@noop {} {\bibfield
  {journal} {\bibinfo  {journal} {Phys. Rev. Lett.}\ }\textbf {\bibinfo
  {volume} {106}},\ \bibinfo {pages} {180504} (\bibinfo {year}
  {2011})}\BibitemShut {NoStop}%
\bibitem [{\citenamefont {Magesan}\ \emph
  {et~al.}(2012{\natexlab{a}})\citenamefont {Magesan}, \citenamefont
  {Gambetta},\ and\ \citenamefont {Emerson}}]{magesan2012characterizing}%
  \BibitemOpen
  \bibfield  {author} {\bibinfo {author} {\bibfnamefont {E.}~\bibnamefont
  {Magesan}}, \bibinfo {author} {\bibfnamefont {J.~M.}\ \bibnamefont
  {Gambetta}},\ and\ \bibinfo {author} {\bibfnamefont {J.}~\bibnamefont
  {Emerson}},\ }\bibfield  {title} {\bibinfo {title} {Characterizing quantum
  gates via randomized benchmarking},\ }\href@noop {} {\bibfield  {journal}
  {\bibinfo  {journal} {Phys. Rev. A}\ }\textbf {\bibinfo {volume} {85}},\
  \bibinfo {pages} {042311} (\bibinfo {year} {2012}{\natexlab{a}})}\BibitemShut
  {NoStop}%
\bibitem [{\citenamefont {Emerson}\ \emph {et~al.}(2005)\citenamefont
  {Emerson}, \citenamefont {Alicki},\ and\ \citenamefont
  {{\.Z}yczkowski}}]{emerson2005scalable}%
  \BibitemOpen
  \bibfield  {author} {\bibinfo {author} {\bibfnamefont {J.}~\bibnamefont
  {Emerson}}, \bibinfo {author} {\bibfnamefont {R.}~\bibnamefont {Alicki}},\
  and\ \bibinfo {author} {\bibfnamefont {K.}~\bibnamefont {{\.Z}yczkowski}},\
  }\bibfield  {title} {\bibinfo {title} {Scalable noise estimation with random
  unitary operators},\ }\href@noop {} {\bibfield  {journal} {\bibinfo
  {journal} {Journal of Optics B: Quantum and Semiclassical Optics}\ }\textbf
  {\bibinfo {volume} {7}},\ \bibinfo {pages} {S347} (\bibinfo {year}
  {2005})}\BibitemShut {NoStop}%
\bibitem [{\citenamefont {Knill}\ \emph {et~al.}(2008)\citenamefont {Knill},
  \citenamefont {Leibfried}, \citenamefont {Reichle}, \citenamefont {Britton},
  \citenamefont {Blakestad}, \citenamefont {Jost}, \citenamefont {Langer},
  \citenamefont {Ozeri}, \citenamefont {Seidelin},\ and\ \citenamefont
  {Wineland}}]{knill2008randomized}%
  \BibitemOpen
  \bibfield  {author} {\bibinfo {author} {\bibfnamefont {E.}~\bibnamefont
  {Knill}}, \bibinfo {author} {\bibfnamefont {D.}~\bibnamefont {Leibfried}},
  \bibinfo {author} {\bibfnamefont {R.}~\bibnamefont {Reichle}}, \bibinfo
  {author} {\bibfnamefont {J.}~\bibnamefont {Britton}}, \bibinfo {author}
  {\bibfnamefont {R.~B.}\ \bibnamefont {Blakestad}}, \bibinfo {author}
  {\bibfnamefont {J.~D.}\ \bibnamefont {Jost}}, \bibinfo {author}
  {\bibfnamefont {C.}~\bibnamefont {Langer}}, \bibinfo {author} {\bibfnamefont
  {R.}~\bibnamefont {Ozeri}}, \bibinfo {author} {\bibfnamefont
  {S.}~\bibnamefont {Seidelin}},\ and\ \bibinfo {author} {\bibfnamefont
  {D.~J.}\ \bibnamefont {Wineland}},\ }\bibfield  {title} {\bibinfo {title}
  {Randomized benchmarking of quantum gates},\ }\href@noop {} {\bibfield
  {journal} {\bibinfo  {journal} {Physical Review A}\ }\textbf {\bibinfo
  {volume} {77}},\ \bibinfo {pages} {012307} (\bibinfo {year}
  {2008})}\BibitemShut {NoStop}%
\bibitem [{\citenamefont {Magesan}\ \emph
  {et~al.}(2012{\natexlab{b}})\citenamefont {Magesan}, \citenamefont
  {Gambetta}, \citenamefont {Johnson}, \citenamefont {Ryan}, \citenamefont
  {Chow}, \citenamefont {Merkel}, \citenamefont {Da~Silva}, \citenamefont
  {Keefe}, \citenamefont {Rothwell}, \citenamefont {Ohki} \emph
  {et~al.}}]{magesan2012efficient}%
  \BibitemOpen
  \bibfield  {author} {\bibinfo {author} {\bibfnamefont {E.}~\bibnamefont
  {Magesan}}, \bibinfo {author} {\bibfnamefont {J.~M.}\ \bibnamefont
  {Gambetta}}, \bibinfo {author} {\bibfnamefont {B.~R.}\ \bibnamefont
  {Johnson}}, \bibinfo {author} {\bibfnamefont {C.~A.}\ \bibnamefont {Ryan}},
  \bibinfo {author} {\bibfnamefont {J.~M.}\ \bibnamefont {Chow}}, \bibinfo
  {author} {\bibfnamefont {S.~T.}\ \bibnamefont {Merkel}}, \bibinfo {author}
  {\bibfnamefont {M.~P.}\ \bibnamefont {Da~Silva}}, \bibinfo {author}
  {\bibfnamefont {G.~A.}\ \bibnamefont {Keefe}}, \bibinfo {author}
  {\bibfnamefont {M.~B.}\ \bibnamefont {Rothwell}}, \bibinfo {author}
  {\bibfnamefont {T.~A.}\ \bibnamefont {Ohki}}, \emph {et~al.},\ }\bibfield
  {title} {\bibinfo {title} {Efficient measurement of quantum gate error by
  interleaved randomized benchmarking},\ }\href@noop {} {\bibfield  {journal}
  {\bibinfo  {journal} {Phys. Rev. Lett.}\ }\textbf {\bibinfo {volume} {109}},\
  \bibinfo {pages} {080505} (\bibinfo {year} {2012}{\natexlab{b}})}\BibitemShut
  {NoStop}%
\bibitem [{\citenamefont {Carignan-Dugas}\ \emph {et~al.}(2015)\citenamefont
  {Carignan-Dugas}, \citenamefont {Wallman},\ and\ \citenamefont
  {Emerson}}]{carignan2015characterizing}%
  \BibitemOpen
  \bibfield  {author} {\bibinfo {author} {\bibfnamefont {A.}~\bibnamefont
  {Carignan-Dugas}}, \bibinfo {author} {\bibfnamefont {J.~J.}\ \bibnamefont
  {Wallman}},\ and\ \bibinfo {author} {\bibfnamefont {J.}~\bibnamefont
  {Emerson}},\ }\bibfield  {title} {\bibinfo {title} {Characterizing universal
  gate sets via dihedral benchmarking},\ }\href@noop {} {\bibfield  {journal}
  {\bibinfo  {journal} {Phys. Rev. A}\ }\textbf {\bibinfo {volume} {92}},\
  \bibinfo {pages} {060302} (\bibinfo {year} {2015})}\BibitemShut {NoStop}%
\bibitem [{\citenamefont {Cross}\ \emph {et~al.}(2016)\citenamefont {Cross},
  \citenamefont {Magesan}, \citenamefont {Bishop}, \citenamefont {Smolin},\
  and\ \citenamefont {Gambetta}}]{cross2016scalable}%
  \BibitemOpen
  \bibfield  {author} {\bibinfo {author} {\bibfnamefont {A.~W.}\ \bibnamefont
  {Cross}}, \bibinfo {author} {\bibfnamefont {E.}~\bibnamefont {Magesan}},
  \bibinfo {author} {\bibfnamefont {L.~S.}\ \bibnamefont {Bishop}}, \bibinfo
  {author} {\bibfnamefont {J.~A.}\ \bibnamefont {Smolin}},\ and\ \bibinfo
  {author} {\bibfnamefont {J.~M.}\ \bibnamefont {Gambetta}},\ }\bibfield
  {title} {\bibinfo {title} {Scalable randomised benchmarking of non-{Clifford}
  gates},\ }\href@noop {} {\bibfield  {journal} {\bibinfo  {journal} {npj
  Quantum Inf.}\ }\textbf {\bibinfo {volume} {2}},\ \bibinfo {pages} {1}
  (\bibinfo {year} {2016})}\BibitemShut {NoStop}%
\bibitem [{\citenamefont {Helsen}\ \emph
  {et~al.}(2019{\natexlab{a}})\citenamefont {Helsen}, \citenamefont {Xue},
  \citenamefont {Vandersypen},\ and\ \citenamefont {Wehner}}]{helsen2019new}%
  \BibitemOpen
  \bibfield  {author} {\bibinfo {author} {\bibfnamefont {J.}~\bibnamefont
  {Helsen}}, \bibinfo {author} {\bibfnamefont {X.}~\bibnamefont {Xue}},
  \bibinfo {author} {\bibfnamefont {L.~M.~K.}\ \bibnamefont {Vandersypen}},\
  and\ \bibinfo {author} {\bibfnamefont {S.}~\bibnamefont {Wehner}},\
  }\bibfield  {title} {\bibinfo {title} {A new class of efficient randomized
  benchmarking protocols},\ }\href@noop {} {\bibfield  {journal} {\bibinfo
  {journal} {npj Quantum Inf.}\ }\textbf {\bibinfo {volume} {5}},\ \bibinfo
  {pages} {1} (\bibinfo {year} {2019}{\natexlab{a}})}\BibitemShut {NoStop}%
\bibitem [{\citenamefont {Claes}\ \emph {et~al.}(2021)\citenamefont {Claes},
  \citenamefont {Rieffel},\ and\ \citenamefont {Wang}}]{claes2021character}%
  \BibitemOpen
  \bibfield  {author} {\bibinfo {author} {\bibfnamefont {J.}~\bibnamefont
  {Claes}}, \bibinfo {author} {\bibfnamefont {E.}~\bibnamefont {Rieffel}},\
  and\ \bibinfo {author} {\bibfnamefont {Z.}~\bibnamefont {Wang}},\ }\bibfield
  {title} {\bibinfo {title} {Character randomized benchmarking for
  non-multiplicity-free groups with applications to subspace, leakage, and
  matchgate randomized benchmarking},\ }\href@noop {} {\bibfield  {journal}
  {\bibinfo  {journal} {PRX Quantum}\ }\textbf {\bibinfo {volume} {2}},\
  \bibinfo {pages} {010351} (\bibinfo {year} {2021})}\BibitemShut {NoStop}%
\bibitem [{\citenamefont {Helsen}\ \emph {et~al.}(2020)\citenamefont {Helsen},
  \citenamefont {Roth}, \citenamefont {Onorati}, \citenamefont {Werner},\ and\
  \citenamefont {Eisert}}]{helsen2020general}%
  \BibitemOpen
  \bibfield  {author} {\bibinfo {author} {\bibfnamefont {J.}~\bibnamefont
  {Helsen}}, \bibinfo {author} {\bibfnamefont {I.}~\bibnamefont {Roth}},
  \bibinfo {author} {\bibfnamefont {E.}~\bibnamefont {Onorati}}, \bibinfo
  {author} {\bibfnamefont {A.~H.}\ \bibnamefont {Werner}},\ and\ \bibinfo
  {author} {\bibfnamefont {J.}~\bibnamefont {Eisert}},\ }\bibfield  {title}
  {\bibinfo {title} {A general framework for randomized benchmarking},\
  }\href@noop {} {\bibfield  {journal} {\bibinfo  {journal} {arXiv preprint
  arXiv:2010.07974}\ } (\bibinfo {year} {2020})}\BibitemShut {NoStop}%
\bibitem [{\citenamefont {Brown}\ and\ \citenamefont
  {Eastin}(2018)}]{brown2018randomized}%
  \BibitemOpen
  \bibfield  {author} {\bibinfo {author} {\bibfnamefont {W.~G.}\ \bibnamefont
  {Brown}}\ and\ \bibinfo {author} {\bibfnamefont {B.}~\bibnamefont {Eastin}},\
  }\bibfield  {title} {\bibinfo {title} {Randomized benchmarking with
  restricted gate sets},\ }\href@noop {} {\bibfield  {journal} {\bibinfo
  {journal} {Phys. Rev. A}\ }\textbf {\bibinfo {volume} {97}},\ \bibinfo
  {pages} {062323} (\bibinfo {year} {2018})}\BibitemShut {NoStop}%
\bibitem [{\citenamefont {Fran{\c{c}}a}\ and\ \citenamefont
  {Hashagen}(2018)}]{francca2018approximate}%
  \BibitemOpen
  \bibfield  {author} {\bibinfo {author} {\bibfnamefont {D.~S.}\ \bibnamefont
  {Fran{\c{c}}a}}\ and\ \bibinfo {author} {\bibfnamefont {A.}~\bibnamefont
  {Hashagen}},\ }\bibfield  {title} {\bibinfo {title} {Approximate randomized
  benchmarking for finite groups},\ }\href@noop {} {\bibfield  {journal}
  {\bibinfo  {journal} {J. Phys. A}\ }\textbf {\bibinfo {volume} {51}},\
  \bibinfo {pages} {395302} (\bibinfo {year} {2018})}\BibitemShut {NoStop}%
\bibitem [{\citenamefont {Harper}\ and\ \citenamefont
  {Flammia}(2017)}]{harper2017estimating}%
  \BibitemOpen
  \bibfield  {author} {\bibinfo {author} {\bibfnamefont {R.}~\bibnamefont
  {Harper}}\ and\ \bibinfo {author} {\bibfnamefont {S.~T.}\ \bibnamefont
  {Flammia}},\ }\bibfield  {title} {\bibinfo {title} {Estimating the fidelity
  of {T} gates using standard interleaved randomized benchmarking},\
  }\href@noop {} {\bibfield  {journal} {\bibinfo  {journal} {Quantum Sci.
  Technol.}\ }\textbf {\bibinfo {volume} {2}},\ \bibinfo {pages} {015008}
  (\bibinfo {year} {2017})}\BibitemShut {NoStop}%
\bibitem [{\citenamefont {Onorati}\ \emph {et~al.}(2019)\citenamefont
  {Onorati}, \citenamefont {Werner},\ and\ \citenamefont
  {Eisert}}]{onorati2019randomized}%
  \BibitemOpen
  \bibfield  {author} {\bibinfo {author} {\bibfnamefont {E.}~\bibnamefont
  {Onorati}}, \bibinfo {author} {\bibfnamefont {A.}~\bibnamefont {Werner}},\
  and\ \bibinfo {author} {\bibfnamefont {J.}~\bibnamefont {Eisert}},\
  }\bibfield  {title} {\bibinfo {title} {Randomized benchmarking for individual
  quantum gates},\ }\href@noop {} {\bibfield  {journal} {\bibinfo  {journal}
  {Phys. Rev. Lett.}\ }\textbf {\bibinfo {volume} {123}},\ \bibinfo {pages}
  {060501} (\bibinfo {year} {2019})}\BibitemShut {NoStop}%
\bibitem [{\citenamefont {Erhard}\ \emph {et~al.}(2019)\citenamefont {Erhard},
  \citenamefont {Wallman}, \citenamefont {Postler}, \citenamefont {Meth},
  \citenamefont {Stricker}, \citenamefont {Martinez}, \citenamefont
  {Schindler}, \citenamefont {Monz}, \citenamefont {Emerson},\ and\
  \citenamefont {Blatt}}]{erhard2019characterizing}%
  \BibitemOpen
  \bibfield  {author} {\bibinfo {author} {\bibfnamefont {A.}~\bibnamefont
  {Erhard}}, \bibinfo {author} {\bibfnamefont {J.~J.}\ \bibnamefont {Wallman}},
  \bibinfo {author} {\bibfnamefont {L.}~\bibnamefont {Postler}}, \bibinfo
  {author} {\bibfnamefont {M.}~\bibnamefont {Meth}}, \bibinfo {author}
  {\bibfnamefont {R.}~\bibnamefont {Stricker}}, \bibinfo {author}
  {\bibfnamefont {E.~A.}\ \bibnamefont {Martinez}}, \bibinfo {author}
  {\bibfnamefont {P.}~\bibnamefont {Schindler}}, \bibinfo {author}
  {\bibfnamefont {T.}~\bibnamefont {Monz}}, \bibinfo {author} {\bibfnamefont
  {J.}~\bibnamefont {Emerson}},\ and\ \bibinfo {author} {\bibfnamefont
  {R.}~\bibnamefont {Blatt}},\ }\bibfield  {title} {\bibinfo {title}
  {Characterizing large-scale quantum computers via cycle benchmarking},\
  }\href@noop {} {\bibfield  {journal} {\bibinfo  {journal} {Nat. Commun.}\
  }\textbf {\bibinfo {volume} {10}},\ \bibinfo {pages} {1} (\bibinfo {year}
  {2019})}\BibitemShut {NoStop}%
\bibitem [{\citenamefont {Chasseur}\ \emph {et~al.}(2017)\citenamefont
  {Chasseur}, \citenamefont {Reich}, \citenamefont {Koch},\ and\ \citenamefont
  {Wilhelm}}]{chasseur2017hybrid}%
  \BibitemOpen
  \bibfield  {author} {\bibinfo {author} {\bibfnamefont {T.}~\bibnamefont
  {Chasseur}}, \bibinfo {author} {\bibfnamefont {D.~M.}\ \bibnamefont {Reich}},
  \bibinfo {author} {\bibfnamefont {C.~P.}\ \bibnamefont {Koch}},\ and\
  \bibinfo {author} {\bibfnamefont {F.~K.}\ \bibnamefont {Wilhelm}},\
  }\bibfield  {title} {\bibinfo {title} {Hybrid benchmarking of arbitrary
  quantum gates},\ }\href@noop {} {\bibfield  {journal} {\bibinfo  {journal}
  {Phys. Rev. A}\ }\textbf {\bibinfo {volume} {95}},\ \bibinfo {pages} {062335}
  (\bibinfo {year} {2017})}\BibitemShut {NoStop}%
\bibitem [{\citenamefont {Wallman}\ \emph {et~al.}(2015)\citenamefont
  {Wallman}, \citenamefont {Granade}, \citenamefont {Harper},\ and\
  \citenamefont {Flammia}}]{wallman2015estimating}%
  \BibitemOpen
  \bibfield  {author} {\bibinfo {author} {\bibfnamefont {J.}~\bibnamefont
  {Wallman}}, \bibinfo {author} {\bibfnamefont {C.}~\bibnamefont {Granade}},
  \bibinfo {author} {\bibfnamefont {R.}~\bibnamefont {Harper}},\ and\ \bibinfo
  {author} {\bibfnamefont {S.~T.}\ \bibnamefont {Flammia}},\ }\bibfield
  {title} {\bibinfo {title} {Estimating the coherence of noise},\ }\href@noop
  {} {\bibfield  {journal} {\bibinfo  {journal} {New J. Phys}\ }\textbf
  {\bibinfo {volume} {17}},\ \bibinfo {pages} {113020} (\bibinfo {year}
  {2015})}\BibitemShut {NoStop}%
\bibitem [{\citenamefont {Kimmel}\ \emph {et~al.}(2014)\citenamefont {Kimmel},
  \citenamefont {da~Silva}, \citenamefont {Ryan}, \citenamefont {Johnson},\
  and\ \citenamefont {Ohki}}]{kimmel2014robust}%
  \BibitemOpen
  \bibfield  {author} {\bibinfo {author} {\bibfnamefont {S.}~\bibnamefont
  {Kimmel}}, \bibinfo {author} {\bibfnamefont {M.~P.}\ \bibnamefont
  {da~Silva}}, \bibinfo {author} {\bibfnamefont {C.~A.}\ \bibnamefont {Ryan}},
  \bibinfo {author} {\bibfnamefont {B.~R.}\ \bibnamefont {Johnson}},\ and\
  \bibinfo {author} {\bibfnamefont {T.}~\bibnamefont {Ohki}},\ }\bibfield
  {title} {\bibinfo {title} {Robust extraction of tomographic information via
  randomized benchmarking},\ }\href@noop {} {\bibfield  {journal} {\bibinfo
  {journal} {Phys. Rev. X}\ }\textbf {\bibinfo {volume} {4}},\ \bibinfo {pages}
  {011050} (\bibinfo {year} {2014})}\BibitemShut {NoStop}%
\bibitem [{\citenamefont {Flammia}\ and\ \citenamefont
  {Wallman}(2020)}]{flammia2020efficient}%
  \BibitemOpen
  \bibfield  {author} {\bibinfo {author} {\bibfnamefont {S.~T.}\ \bibnamefont
  {Flammia}}\ and\ \bibinfo {author} {\bibfnamefont {J.~J.}\ \bibnamefont
  {Wallman}},\ }\bibfield  {title} {\bibinfo {title} {Efficient estimation of
  {P}auli channels},\ }\href@noop {} {\bibfield  {journal} {\bibinfo  {journal}
  {ACM Trans. Quantum Comput.}\ }\textbf {\bibinfo {volume} {1}},\ \bibinfo
  {pages} {1} (\bibinfo {year} {2020})}\BibitemShut {NoStop}%
\bibitem [{\citenamefont {Harper}\ \emph {et~al.}(2020)\citenamefont {Harper},
  \citenamefont {Flammia},\ and\ \citenamefont
  {Wallman}}]{harper2020efficient}%
  \BibitemOpen
  \bibfield  {author} {\bibinfo {author} {\bibfnamefont {R.}~\bibnamefont
  {Harper}}, \bibinfo {author} {\bibfnamefont {S.~T.}\ \bibnamefont
  {Flammia}},\ and\ \bibinfo {author} {\bibfnamefont {J.~J.}\ \bibnamefont
  {Wallman}},\ }\bibfield  {title} {\bibinfo {title} {Efficient learning of
  quantum noise},\ }\href@noop {} {\bibfield  {journal} {\bibinfo  {journal}
  {Nat. Phys.}\ }\textbf {\bibinfo {volume} {16}},\ \bibinfo {pages} {1184}
  (\bibinfo {year} {2020})}\BibitemShut {NoStop}%
\bibitem [{\citenamefont {Baldwin}\ \emph {et~al.}(2020)\citenamefont
  {Baldwin}, \citenamefont {Bjork}, \citenamefont {Gaebler}, \citenamefont
  {Hayes},\ and\ \citenamefont {Stack}}]{baldwin2020subspace}%
  \BibitemOpen
  \bibfield  {author} {\bibinfo {author} {\bibfnamefont {C.}~\bibnamefont
  {Baldwin}}, \bibinfo {author} {\bibfnamefont {B.}~\bibnamefont {Bjork}},
  \bibinfo {author} {\bibfnamefont {J.}~\bibnamefont {Gaebler}}, \bibinfo
  {author} {\bibfnamefont {D.}~\bibnamefont {Hayes}},\ and\ \bibinfo {author}
  {\bibfnamefont {D.}~\bibnamefont {Stack}},\ }\bibfield  {title} {\bibinfo
  {title} {Subspace benchmarking high-fidelity entangling operations with
  trapped ions},\ }\href@noop {} {\bibfield  {journal} {\bibinfo  {journal}
  {Physical Review Research}\ }\textbf {\bibinfo {volume} {2}},\ \bibinfo
  {pages} {013317} (\bibinfo {year} {2020})}\BibitemShut {NoStop}%
\bibitem [{\citenamefont {Wallman}(2018)}]{wallman2018randomized}%
  \BibitemOpen
  \bibfield  {author} {\bibinfo {author} {\bibfnamefont {J.~J.}\ \bibnamefont
  {Wallman}},\ }\bibfield  {title} {\bibinfo {title} {Randomized benchmarking
  with gate-dependent noise},\ }\href@noop {} {\bibfield  {journal} {\bibinfo
  {journal} {Quantum}\ }\textbf {\bibinfo {volume} {2}},\ \bibinfo {pages} {47}
  (\bibinfo {year} {2018})}\BibitemShut {NoStop}%
\bibitem [{\citenamefont {Merkel}\ \emph {et~al.}(2021)\citenamefont {Merkel},
  \citenamefont {Pritchett},\ and\ \citenamefont
  {Fong}}]{merkel2021randomized}%
  \BibitemOpen
  \bibfield  {author} {\bibinfo {author} {\bibfnamefont {S.~T.}\ \bibnamefont
  {Merkel}}, \bibinfo {author} {\bibfnamefont {E.~J.}\ \bibnamefont
  {Pritchett}},\ and\ \bibinfo {author} {\bibfnamefont {B.~H.}\ \bibnamefont
  {Fong}},\ }\bibfield  {title} {\bibinfo {title} {Randomized benchmarking as
  convolution: Fourier analysis of gate dependent errors},\ }\href@noop {}
  {\bibfield  {journal} {\bibinfo  {journal} {Quantum}\ }\textbf {\bibinfo
  {volume} {5}},\ \bibinfo {pages} {581} (\bibinfo {year} {2021})}\BibitemShut
  {NoStop}%
\bibitem [{\citenamefont {Wallman}\ and\ \citenamefont
  {Flammia}(2014)}]{wallman2014randomized}%
  \BibitemOpen
  \bibfield  {author} {\bibinfo {author} {\bibfnamefont {J.~J.}\ \bibnamefont
  {Wallman}}\ and\ \bibinfo {author} {\bibfnamefont {S.~T.}\ \bibnamefont
  {Flammia}},\ }\bibfield  {title} {\bibinfo {title} {Randomized benchmarking
  with confidence},\ }\href@noop {} {\bibfield  {journal} {\bibinfo  {journal}
  {New J. Phys}\ }\textbf {\bibinfo {volume} {16}},\ \bibinfo {pages} {103032}
  (\bibinfo {year} {2014})}\BibitemShut {NoStop}%
\bibitem [{\citenamefont {Helsen}\ \emph
  {et~al.}(2019{\natexlab{b}})\citenamefont {Helsen}, \citenamefont {Wallman},
  \citenamefont {Flammia},\ and\ \citenamefont
  {Wehner}}]{helsen2019multiqubit}%
  \BibitemOpen
  \bibfield  {author} {\bibinfo {author} {\bibfnamefont {J.}~\bibnamefont
  {Helsen}}, \bibinfo {author} {\bibfnamefont {J.~J.}\ \bibnamefont {Wallman}},
  \bibinfo {author} {\bibfnamefont {S.~T.}\ \bibnamefont {Flammia}},\ and\
  \bibinfo {author} {\bibfnamefont {S.}~\bibnamefont {Wehner}},\ }\bibfield
  {title} {\bibinfo {title} {Multiqubit randomized benchmarking using few
  samples},\ }\href@noop {} {\bibfield  {journal} {\bibinfo  {journal} {Phys.
  Rev. A}\ }\textbf {\bibinfo {volume} {100}},\ \bibinfo {pages} {032304}
  (\bibinfo {year} {2019}{\natexlab{b}})}\BibitemShut {NoStop}%
\bibitem [{\citenamefont {Carignan-Dugas}\ \emph {et~al.}(2018)\citenamefont
  {Carignan-Dugas}, \citenamefont {Boone}, \citenamefont {Wallman},\ and\
  \citenamefont {Emerson}}]{carignan2018randomized}%
  \BibitemOpen
  \bibfield  {author} {\bibinfo {author} {\bibfnamefont {A.}~\bibnamefont
  {Carignan-Dugas}}, \bibinfo {author} {\bibfnamefont {K.}~\bibnamefont
  {Boone}}, \bibinfo {author} {\bibfnamefont {J.~J.}\ \bibnamefont {Wallman}},\
  and\ \bibinfo {author} {\bibfnamefont {J.}~\bibnamefont {Emerson}},\
  }\bibfield  {title} {\bibinfo {title} {From randomized benchmarking
  experiments to gate-set circuit fidelity: how to interpret randomized
  benchmarking decay parameters},\ }\href@noop {} {\bibfield  {journal}
  {\bibinfo  {journal} {New J. Phys}\ }\textbf {\bibinfo {volume} {20}},\
  \bibinfo {pages} {092001} (\bibinfo {year} {2018})}\BibitemShut {NoStop}%
\bibitem [{\citenamefont {Xue}\ \emph {et~al.}(2019)\citenamefont {Xue},
  \citenamefont {Watson}, \citenamefont {Helsen}, \citenamefont {Ward},
  \citenamefont {Savage}, \citenamefont {Lagally}, \citenamefont {Coppersmith},
  \citenamefont {Eriksson}, \citenamefont {Wehner},\ and\ \citenamefont
  {Vandersypen}}]{xue2019benchmarking}%
  \BibitemOpen
  \bibfield  {author} {\bibinfo {author} {\bibfnamefont {X.}~\bibnamefont
  {Xue}}, \bibinfo {author} {\bibfnamefont {T.}~\bibnamefont {Watson}},
  \bibinfo {author} {\bibfnamefont {J.}~\bibnamefont {Helsen}}, \bibinfo
  {author} {\bibfnamefont {D.~R.}\ \bibnamefont {Ward}}, \bibinfo {author}
  {\bibfnamefont {D.~E.}\ \bibnamefont {Savage}}, \bibinfo {author}
  {\bibfnamefont {M.~G.}\ \bibnamefont {Lagally}}, \bibinfo {author}
  {\bibfnamefont {S.~N.}\ \bibnamefont {Coppersmith}}, \bibinfo {author}
  {\bibfnamefont {M.}~\bibnamefont {Eriksson}}, \bibinfo {author}
  {\bibfnamefont {S.}~\bibnamefont {Wehner}},\ and\ \bibinfo {author}
  {\bibfnamefont {L.}~\bibnamefont {Vandersypen}},\ }\bibfield  {title}
  {\bibinfo {title} {Benchmarking gate fidelities in a si/sige two-qubit
  device},\ }\href@noop {} {\bibfield  {journal} {\bibinfo  {journal} {Phys.
  Rev. X}\ }\textbf {\bibinfo {volume} {9}},\ \bibinfo {pages} {021011}
  (\bibinfo {year} {2019})}\BibitemShut {NoStop}%
\bibitem [{\citenamefont {Feng}\ \emph {et~al.}(2016)\citenamefont {Feng},
  \citenamefont {Wallman}, \citenamefont {Buonacorsi}, \citenamefont {Cho},
  \citenamefont {Park}, \citenamefont {Xin}, \citenamefont {Lu}, \citenamefont
  {Baugh},\ and\ \citenamefont {Laflamme}}]{feng2016estimating}%
  \BibitemOpen
  \bibfield  {author} {\bibinfo {author} {\bibfnamefont {G.}~\bibnamefont
  {Feng}}, \bibinfo {author} {\bibfnamefont {J.~J.}\ \bibnamefont {Wallman}},
  \bibinfo {author} {\bibfnamefont {B.}~\bibnamefont {Buonacorsi}}, \bibinfo
  {author} {\bibfnamefont {F.~H.}\ \bibnamefont {Cho}}, \bibinfo {author}
  {\bibfnamefont {D.~K.}\ \bibnamefont {Park}}, \bibinfo {author}
  {\bibfnamefont {T.}~\bibnamefont {Xin}}, \bibinfo {author} {\bibfnamefont
  {D.}~\bibnamefont {Lu}}, \bibinfo {author} {\bibfnamefont {J.}~\bibnamefont
  {Baugh}},\ and\ \bibinfo {author} {\bibfnamefont {R.}~\bibnamefont
  {Laflamme}},\ }\bibfield  {title} {\bibinfo {title} {Estimating the coherence
  of noise in quantum control of a solid-state qubit},\ }\href@noop {}
  {\bibfield  {journal} {\bibinfo  {journal} {Phys. Rev. Lett.}\ }\textbf
  {\bibinfo {volume} {117}},\ \bibinfo {pages} {260501} (\bibinfo {year}
  {2016})}\BibitemShut {NoStop}%
\bibitem [{\citenamefont {Flammia}(2021)}]{flammia2021averaged}%
  \BibitemOpen
  \bibfield  {author} {\bibinfo {author} {\bibfnamefont {S.~T.}\ \bibnamefont
  {Flammia}},\ }\bibfield  {title} {\bibinfo {title} {Averaged circuit
  eigenvalue sampling},\ }\href@noop {} {\bibfield  {journal} {\bibinfo
  {journal} {arXiv preprint arXiv:2108.05803}\ } (\bibinfo {year}
  {2021})}\BibitemShut {NoStop}%
\bibitem [{\citenamefont {Helsen}\ \emph {et~al.}(2022)\citenamefont {Helsen},
  \citenamefont {Nezami}, \citenamefont {Reagor},\ and\ \citenamefont
  {Walter}}]{helsen2022matchgate}%
  \BibitemOpen
  \bibfield  {author} {\bibinfo {author} {\bibfnamefont {J.}~\bibnamefont
  {Helsen}}, \bibinfo {author} {\bibfnamefont {S.}~\bibnamefont {Nezami}},
  \bibinfo {author} {\bibfnamefont {M.}~\bibnamefont {Reagor}},\ and\ \bibinfo
  {author} {\bibfnamefont {M.}~\bibnamefont {Walter}},\ }\bibfield  {title}
  {\bibinfo {title} {Matchgate benchmarking: Scalable benchmarking of a
  continuous family of many-qubit gates},\ }\href@noop {} {\bibfield  {journal}
  {\bibinfo  {journal} {Quantum}\ }\textbf {\bibinfo {volume} {6}},\ \bibinfo
  {pages} {657} (\bibinfo {year} {2022})}\BibitemShut {NoStop}%
\bibitem [{\citenamefont {Grimm}\ \emph {et~al.}(2020)\citenamefont {Grimm},
  \citenamefont {Frattini}, \citenamefont {Puri}, \citenamefont {Mundhada},
  \citenamefont {Touzard}, \citenamefont {Mirrahimi}, \citenamefont {Girvin},
  \citenamefont {Shankar},\ and\ \citenamefont
  {Devoret}}]{grimm2020stabilization}%
  \BibitemOpen
  \bibfield  {author} {\bibinfo {author} {\bibfnamefont {A.}~\bibnamefont
  {Grimm}}, \bibinfo {author} {\bibfnamefont {N.~E.}\ \bibnamefont {Frattini}},
  \bibinfo {author} {\bibfnamefont {S.}~\bibnamefont {Puri}}, \bibinfo {author}
  {\bibfnamefont {S.~O.}\ \bibnamefont {Mundhada}}, \bibinfo {author}
  {\bibfnamefont {S.}~\bibnamefont {Touzard}}, \bibinfo {author} {\bibfnamefont
  {M.}~\bibnamefont {Mirrahimi}}, \bibinfo {author} {\bibfnamefont {S.~M.}\
  \bibnamefont {Girvin}}, \bibinfo {author} {\bibfnamefont {S.}~\bibnamefont
  {Shankar}},\ and\ \bibinfo {author} {\bibfnamefont {M.~H.}\ \bibnamefont
  {Devoret}},\ }\bibfield  {title} {\bibinfo {title} {Stabilization and
  operation of a {K}err-cat qubit},\ }\href@noop {} {\bibfield  {journal}
  {\bibinfo  {journal} {Nature}\ }\textbf {\bibinfo {volume} {584}},\ \bibinfo
  {pages} {205} (\bibinfo {year} {2020})}\BibitemShut {NoStop}%
\bibitem [{\citenamefont {Puri}\ \emph {et~al.}(2017)\citenamefont {Puri},
  \citenamefont {Boutin},\ and\ \citenamefont {Blais}}]{puri2017engineering}%
  \BibitemOpen
  \bibfield  {author} {\bibinfo {author} {\bibfnamefont {S.}~\bibnamefont
  {Puri}}, \bibinfo {author} {\bibfnamefont {S.}~\bibnamefont {Boutin}},\ and\
  \bibinfo {author} {\bibfnamefont {A.}~\bibnamefont {Blais}},\ }\bibfield
  {title} {\bibinfo {title} {Engineering the quantum states of light in a
  {K}err-nonlinear resonator by two-photon driving},\ }\href@noop {} {\bibfield
   {journal} {\bibinfo  {journal} {npj Quantum Inf.}\ }\textbf {\bibinfo
  {volume} {3}},\ \bibinfo {pages} {1} (\bibinfo {year} {2017})}\BibitemShut
  {NoStop}%
\bibitem [{\citenamefont {Wood}\ \emph {et~al.}(2011)\citenamefont {Wood},
  \citenamefont {Biamonte},\ and\ \citenamefont {Cory}}]{wood2011tensor}%
  \BibitemOpen
  \bibfield  {author} {\bibinfo {author} {\bibfnamefont {C.~J.}\ \bibnamefont
  {Wood}}, \bibinfo {author} {\bibfnamefont {J.~D.}\ \bibnamefont {Biamonte}},\
  and\ \bibinfo {author} {\bibfnamefont {D.~G.}\ \bibnamefont {Cory}},\
  }\bibfield  {title} {\bibinfo {title} {Tensor networks and graphical calculus
  for open quantum systems},\ }\href@noop {} {\bibfield  {journal} {\bibinfo
  {journal} {arXiv preprint arXiv:1111.6950}\ } (\bibinfo {year}
  {2011})}\BibitemShut {NoStop}%
\bibitem [{Note1()}]{Note1}%
  \BibitemOpen
  \bibinfo {note} {The usual definition of the $\chi $-matrix differs from this
  one by factors of $i$ off the diagonal, which are not relevant for defining
  the bias.}\BibitemShut {Stop}%
\bibitem [{\citenamefont {Nielsen}(2002)}]{nielsen2002simple}%
  \BibitemOpen
  \bibfield  {author} {\bibinfo {author} {\bibfnamefont {M.~A.}\ \bibnamefont
  {Nielsen}},\ }\bibfield  {title} {\bibinfo {title} {A simple formula for the
  average gate fidelity of a quantum dynamical operation},\ }\href@noop {}
  {\bibfield  {journal} {\bibinfo  {journal} {Phys. Lett. A}\ }\textbf
  {\bibinfo {volume} {303}},\ \bibinfo {pages} {249} (\bibinfo {year}
  {2002})}\BibitemShut {NoStop}%
\bibitem [{\citenamefont {Horodecki}\ \emph {et~al.}(1999)\citenamefont
  {Horodecki}, \citenamefont {Horodecki},\ and\ \citenamefont
  {Horodecki}}]{horodecki1999general}%
  \BibitemOpen
  \bibfield  {author} {\bibinfo {author} {\bibfnamefont {M.}~\bibnamefont
  {Horodecki}}, \bibinfo {author} {\bibfnamefont {P.}~\bibnamefont
  {Horodecki}},\ and\ \bibinfo {author} {\bibfnamefont {R.}~\bibnamefont
  {Horodecki}},\ }\bibfield  {title} {\bibinfo {title} {General teleportation
  channel, singlet fraction, and quasidistillation},\ }\href@noop {} {\bibfield
   {journal} {\bibinfo  {journal} {Phys. Rev. A}\ }\textbf {\bibinfo {volume}
  {60}},\ \bibinfo {pages} {1888} (\bibinfo {year} {1999})}\BibitemShut
  {NoStop}%
\bibitem [{\citenamefont {Huang}\ \emph {et~al.}(2019)\citenamefont {Huang},
  \citenamefont {Doherty},\ and\ \citenamefont
  {Flammia}}]{huang2019performance}%
  \BibitemOpen
  \bibfield  {author} {\bibinfo {author} {\bibfnamefont {E.}~\bibnamefont
  {Huang}}, \bibinfo {author} {\bibfnamefont {A.~C.}\ \bibnamefont {Doherty}},\
  and\ \bibinfo {author} {\bibfnamefont {S.}~\bibnamefont {Flammia}},\
  }\bibfield  {title} {\bibinfo {title} {Performance of quantum error
  correction with coherent errors},\ }\href@noop {} {\bibfield  {journal}
  {\bibinfo  {journal} {Phys. Rev. A}\ }\textbf {\bibinfo {volume} {99}},\
  \bibinfo {pages} {022313} (\bibinfo {year} {2019})}\BibitemShut {NoStop}%
\bibitem [{\citenamefont {Beale}\ \emph {et~al.}(2018)\citenamefont {Beale},
  \citenamefont {Wallman}, \citenamefont {Guti{\'e}rrez}, \citenamefont
  {Brown},\ and\ \citenamefont {Laflamme}}]{beale2018quantum}%
  \BibitemOpen
  \bibfield  {author} {\bibinfo {author} {\bibfnamefont {S.~J.}\ \bibnamefont
  {Beale}}, \bibinfo {author} {\bibfnamefont {J.~J.}\ \bibnamefont {Wallman}},
  \bibinfo {author} {\bibfnamefont {M.}~\bibnamefont {Guti{\'e}rrez}}, \bibinfo
  {author} {\bibfnamefont {K.~R.}\ \bibnamefont {Brown}},\ and\ \bibinfo
  {author} {\bibfnamefont {R.}~\bibnamefont {Laflamme}},\ }\bibfield  {title}
  {\bibinfo {title} {Quantum error correction decoheres noise},\ }\href@noop {}
  {\bibfield  {journal} {\bibinfo  {journal} {Phys. Rev. Lett.}\ }\textbf
  {\bibinfo {volume} {121}},\ \bibinfo {pages} {190501} (\bibinfo {year}
  {2018})}\BibitemShut {NoStop}%
\bibitem [{\citenamefont {Bravyi}\ \emph {et~al.}(2018)\citenamefont {Bravyi},
  \citenamefont {Englbrecht}, \citenamefont {K{\"o}nig},\ and\ \citenamefont
  {Peard}}]{bravyi2018correcting}%
  \BibitemOpen
  \bibfield  {author} {\bibinfo {author} {\bibfnamefont {S.}~\bibnamefont
  {Bravyi}}, \bibinfo {author} {\bibfnamefont {M.}~\bibnamefont {Englbrecht}},
  \bibinfo {author} {\bibfnamefont {R.}~\bibnamefont {K{\"o}nig}},\ and\
  \bibinfo {author} {\bibfnamefont {N.}~\bibnamefont {Peard}},\ }\bibfield
  {title} {\bibinfo {title} {Correcting coherent errors with surface codes},\
  }\href@noop {} {\bibfield  {journal} {\bibinfo  {journal} {npj Quantum Inf.}\
  }\textbf {\bibinfo {volume} {4}},\ \bibinfo {pages} {1} (\bibinfo {year}
  {2018})}\BibitemShut {NoStop}%
\bibitem [{\citenamefont {Geller}\ and\ \citenamefont
  {Zhou}(2013)}]{geller2013efficient}%
  \BibitemOpen
  \bibfield  {author} {\bibinfo {author} {\bibfnamefont {M.~R.}\ \bibnamefont
  {Geller}}\ and\ \bibinfo {author} {\bibfnamefont {Z.}~\bibnamefont {Zhou}},\
  }\bibfield  {title} {\bibinfo {title} {Efficient error models for
  fault-tolerant architectures and the {P}auli twirling approximation},\
  }\href@noop {} {\bibfield  {journal} {\bibinfo  {journal} {Phys. Rev. A}\
  }\textbf {\bibinfo {volume} {88}},\ \bibinfo {pages} {012314} (\bibinfo
  {year} {2013})}\BibitemShut {NoStop}%
\bibitem [{\citenamefont {Garion}\ and\ \citenamefont
  {Cross}(2020)}]{garion2020synthesis}%
  \BibitemOpen
  \bibfield  {author} {\bibinfo {author} {\bibfnamefont {S.}~\bibnamefont
  {Garion}}\ and\ \bibinfo {author} {\bibfnamefont {A.~W.}\ \bibnamefont
  {Cross}},\ }\bibfield  {title} {\bibinfo {title} {Synthesis of
  {CNOT}-dihedral circuits with optimal number of two qubit gates},\
  }\href@noop {} {\bibfield  {journal} {\bibinfo  {journal} {Quantum}\ }\textbf
  {\bibinfo {volume} {4}},\ \bibinfo {pages} {369} (\bibinfo {year}
  {2020})}\BibitemShut {NoStop}%
\bibitem [{\citenamefont {Amy}\ \emph {et~al.}(2016)\citenamefont {Amy},
  \citenamefont {Chen},\ and\ \citenamefont {Ross}}]{amy2016finite}%
  \BibitemOpen
  \bibfield  {author} {\bibinfo {author} {\bibfnamefont {M.}~\bibnamefont
  {Amy}}, \bibinfo {author} {\bibfnamefont {J.}~\bibnamefont {Chen}},\ and\
  \bibinfo {author} {\bibfnamefont {N.~J.}\ \bibnamefont {Ross}},\ }\bibfield
  {title} {\bibinfo {title} {A finite presentation of {CNOT}-dihedral
  operators},\ }\href@noop {} {\bibfield  {journal} {\bibinfo  {journal} {arXiv
  preprint arXiv:1701.00140}\ } (\bibinfo {year} {2016})}\BibitemShut {NoStop}%
\bibitem [{\citenamefont {Carignan-Dugas}\ \emph {et~al.}(2019)\citenamefont
  {Carignan-Dugas}, \citenamefont {Wallman},\ and\ \citenamefont
  {Emerson}}]{carignan2019bounding}%
  \BibitemOpen
  \bibfield  {author} {\bibinfo {author} {\bibfnamefont {A.}~\bibnamefont
  {Carignan-Dugas}}, \bibinfo {author} {\bibfnamefont {J.~J.}\ \bibnamefont
  {Wallman}},\ and\ \bibinfo {author} {\bibfnamefont {J.}~\bibnamefont
  {Emerson}},\ }\bibfield  {title} {\bibinfo {title} {Bounding the average gate
  fidelity of composite channels using the unitarity},\ }\href@noop {}
  {\bibfield  {journal} {\bibinfo  {journal} {New J. Phys}\ }\textbf {\bibinfo
  {volume} {21}},\ \bibinfo {pages} {053016} (\bibinfo {year}
  {2019})}\BibitemShut {NoStop}%
\bibitem [{\citenamefont {Knill}(2005)}]{knill2005quantum}%
  \BibitemOpen
  \bibfield  {author} {\bibinfo {author} {\bibfnamefont {E.}~\bibnamefont
  {Knill}},\ }\bibfield  {title} {\bibinfo {title} {Quantum computing with
  realistically noisy devices},\ }\href@noop {} {\bibfield  {journal} {\bibinfo
   {journal} {Nature}\ }\textbf {\bibinfo {volume} {434}},\ \bibinfo {pages}
  {39} (\bibinfo {year} {2005})}\BibitemShut {NoStop}%
\bibitem [{\citenamefont {Viola}\ and\ \citenamefont
  {Knill}(2005)}]{viola2005random}%
  \BibitemOpen
  \bibfield  {author} {\bibinfo {author} {\bibfnamefont {L.}~\bibnamefont
  {Viola}}\ and\ \bibinfo {author} {\bibfnamefont {E.}~\bibnamefont {Knill}},\
  }\bibfield  {title} {\bibinfo {title} {Random decoupling schemes for quantum
  dynamical control and error suppression},\ }\href@noop {} {\bibfield
  {journal} {\bibinfo  {journal} {Physical review letters}\ }\textbf {\bibinfo
  {volume} {94}},\ \bibinfo {pages} {060502} (\bibinfo {year}
  {2005})}\BibitemShut {NoStop}%
\bibitem [{\citenamefont {Wallman}\ and\ \citenamefont
  {Emerson}(2016)}]{wallman2016noise}%
  \BibitemOpen
  \bibfield  {author} {\bibinfo {author} {\bibfnamefont {J.~J.}\ \bibnamefont
  {Wallman}}\ and\ \bibinfo {author} {\bibfnamefont {J.}~\bibnamefont
  {Emerson}},\ }\bibfield  {title} {\bibinfo {title} {Noise tailoring for
  scalable quantum computation via randomized compiling},\ }\href@noop {}
  {\bibfield  {journal} {\bibinfo  {journal} {Phys. Rev. A}\ }\textbf {\bibinfo
  {volume} {94}},\ \bibinfo {pages} {052325} (\bibinfo {year}
  {2016})}\BibitemShut {NoStop}%
\bibitem [{Note2()}]{Note2}%
  \BibitemOpen
  \bibinfo {note} {We could also include single-qubit $T$ gates or even
  two-qubit $CZ$ gates in our group, as these gates are also high-fidelity for
  Kerr cat qubits, but we did not find these additional gates helpful in
  designing our protocol.}\BibitemShut {Stop}%
\bibitem [{\citenamefont {Fulton}\ and\ \citenamefont
  {Harris}(2013)}]{fulton2013representation}%
  \BibitemOpen
  \bibfield  {author} {\bibinfo {author} {\bibfnamefont {W.}~\bibnamefont
  {Fulton}}\ and\ \bibinfo {author} {\bibfnamefont {J.}~\bibnamefont
  {Harris}},\ }\href@noop {} {\emph {\bibinfo {title} {Representation theory: a
  first course}}},\ Vol.\ \bibinfo {volume} {129}\ (\bibinfo  {publisher}
  {Springer Science \& Business Media},\ \bibinfo {year} {2013})\BibitemShut
  {NoStop}%
\bibitem [{\citenamefont {Golub}\ and\ \citenamefont
  {Van~Loan}(2013)}]{golub2013matrix}%
  \BibitemOpen
  \bibfield  {author} {\bibinfo {author} {\bibfnamefont {G.~H.}\ \bibnamefont
  {Golub}}\ and\ \bibinfo {author} {\bibfnamefont {C.~F.}\ \bibnamefont
  {Van~Loan}},\ }\href@noop {} {\emph {\bibinfo {title} {Matrix
  computations}}}\ (\bibinfo  {publisher} {JHU press},\ \bibinfo {year}
  {2013})\BibitemShut {NoStop}%
\end{thebibliography}%

\end{document}